%% file: main.tex
\documentclass[9pt]{extarticle}

\ifdefined\pdfminorversion\pdfminorversion=7\fi

\usepackage{microtype}
\usepackage[utf8]{inputenc}

\usepackage{eurosym}
\usepackage{url}
\usepackage{amssymb}
\usepackage{amsmath,amsfonts}
\usepackage{pifont}
\usepackage[square,numbers]{natbib}
\usepackage{colortbl}
\usepackage{multirow}
\usepackage{amsthm}
\usepackage{mathtools}
\usepackage{mathrsfs}
\usepackage{booktabs}
\usepackage{tabularx,longtable,array}
\newcolumntype{Y}{>{\raggedright\arraybackslash}X}
\usepackage{textcomp}
\usepackage{manyfoot}
\usepackage[table,xcdraw]{xcolor}
\usepackage{fancyhdr}
\usepackage{wrapfig}

\definecolor{cai_primary}{HTML}{4C9A99}
\definecolor{cai_secondary}{HTML}{307FE2}
\definecolor{cai_accent}{HTML}{1D8348}
\definecolor{cai_dark}{HTML}{3F4444}
\definecolor{human_color}{HTML}{173C47}

\newsavebox{\evfigbox}  %

\definecolor{hf_face}{HTML}{FFD21E}
\definecolor{hf_hand}{HTML}{FF9D00}
\definecolor{hf_finger}{HTML}{E8890C}
\definecolor{hf_ink}{HTML}{3A3B45}
\DeclareRobustCommand{\hfemoji}{\texorpdfstring{%
  \begin{tikzpicture}[baseline=-0.30ex,x=1.18ex,y=1.18ex,line join=round,line cap=round]
    \foreach \s in {-1,1}{%
      \begin{scope}[shift={({\s*1.00},-0.18)},rotate=\s*-18]
        \fill[hf_hand] (0,0) ellipse [x radius=0.40, y radius=0.50];
        \foreach \d in {-0.17,0,0.17}{\draw[hf_finger,line width=0.055ex] (\d,0.40) -- (\d,-0.26);}
      \end{scope}}
    \fill[hf_face] (0,0) circle [radius=0.98];
    \fill[hf_ink] (-0.38,0.22) circle [radius=0.155];
    \fill[hf_ink] ( 0.38,0.22) circle [radius=0.155];
    \draw[hf_ink,line width=0.15ex] (-0.40,-0.26) .. controls (-0.20,-0.60) and (0.20,-0.60) .. (0.40,-0.26);
  \end{tikzpicture}}{}}
\newcommand{\ctzerourl}{https://huggingface.co/aliasrobotics/alias3-micro-1026}

\DeclareRobustCommand{\ctzero}{\texorpdfstring{\texttt{CT0}}{CT0}}
\DeclareRobustCommand{\aliasmicro}{\texorpdfstring{\href{\ctzerourl}{\textcolor{fig_ct0}{\texttt{alias3-micro}}}}{alias3-micro}}
\DeclareRobustCommand{\aliasmini}{\texorpdfstring{\href{https://aliasrobotics.com/aliasLLMs.php}{\textcolor{fig_mini}{\texttt{alias3-mini}}}}{alias3-mini}}
\DeclareRobustCommand{\basemodel}{\texttt{Qwen3.5-4B}}

\usepackage{graphicx}
\usepackage{subcaption}
\usepackage{float}
\usepackage{placeins}
\usepackage{stfloats}
\usepackage{hyperref}
\hypersetup{
    pdftitle={CT0: Cybersecurity Transformer},
    pdfauthor={Víctor Mayoral-Vilches},
    pdfsubject={A compact dual-mode cybersecurity transformer},
    colorlinks=true,
    urlcolor=cai_secondary,
    linkcolor=cai_secondary,
    filecolor=cai_accent,
    citecolor=cai_secondary,
}
\makeatletter
\g@addto@macro{\UrlBreaks}{\do\/\do\-\do\_\do\.\do\=\do\?\do\&}
\makeatother

\usepackage{tikz}
\usetikzlibrary{arrows.meta,positioning,shapes.geometric,calc,patterns,shadows,decorations.pathreplacing,fit,backgrounds}
\usepackage{pgfplots}
\pgfplotsset{compat=1.16}
\input{tex/figure_palette.tex}
\usepgfplotslibrary{groupplots}
\newcommand{\lat}[1]{\textcolor{fig_time}{#1}}
\newcommand{\mk}[1]{\textcolor{fig_s1}{\ensuremath{^{\boldsymbol{#1}}}}}  %
\DeclareRobustCommand{\badge}[2][fig_s1]{\tikz[baseline=(b.base)]\node[circle,fill=#1,text=white,inner sep=0pt,minimum size=1.05em,font=\sffamily\bfseries\tiny](b){#2};}  %

\usepackage{enumitem}
\usepackage{placeins}
\usepackage[breakable,skins]{tcolorbox}
\usepackage{algorithm}
\usepackage{algorithmicx}
\usepackage{algpseudocode}
\usepackage{listings}

\usepackage{geometry}
\renewcommand{\headrulewidth}{0.4pt}
\renewcommand{\footrulewidth}{0.4pt}
\renewcommand{\headrule}{\hbox to\headwidth{\color{cai_primary}\leaders\hrule height \headrulewidth\hfill}}
\renewcommand{\footrule}{\hbox to\headwidth{\color{human_color}\leaders\hrule height \footrulewidth\hfill}}
\usepackage{caption,setspace}
\usepackage{titlesec}
\titleformat{\section}{\normalfont\Large\bfseries\color{cai_primary}}{\thesection}{1em}{}[\titlerule]
\titleformat{\subsection}{\normalfont\large\bfseries\color{human_color}}{\thesubsection}{1em}{}
\titleformat{\subsubsection}{\normalfont\normalsize\bfseries\color{cai_dark}}{\thesubsubsection}{1em}{}

\usepackage{authblk}

\renewcommand\Affilfont{\small\normalfont}
\makeatletter
\renewcommand\AB@affilsepx{\\\protect\Affilfont}
\let\orig@maketitle\maketitle
\renewcommand{\maketitle}{%
  \orig@maketitle%
  \vspace{-1.5em}%
  {\color{cai_primary!30}\hrule height 0.5pt}%
  \vspace{1em}%
}
\makeatother

\makeatletter
\renewenvironment{abstract}{\small\noindent\ignorespaces}{\par}
\makeatother

\begin{document}

\title{\LARGE\textcolor{cai_primary}{\textbf{\ctzero{}: Cybersecurity Transformer}}}

\author[1,$\dagger$]{V\'ictor Mayoral-Vilches}

\affil[1]{\small Alias Robotics, Vitoria-Gasteiz, \'Alava, Spain}

\date{}
\twocolumn[
\maketitle
\vspace{0.3em}

\begin{abstract}
General-purpose frontier models are large, expensive to run, and increasingly reluctant to engage with authorised defensive- and offensive-security work. We ask the opposite question: how small can a \emph{cybersecurity-specialist} transformer be while remaining genuinely useful to a security practitioner? We introduce \texttt{CT0}, a Cybersecurity Transformer architecture, and \aliasmicro{}, its first incarnation: a compact ($\sim$4\,B-parameter) model built on the open \basemodel{} base, whose backbone is specialised by two-phase on-policy distillation on our Cybersecurity AI (CAI) dataset. \ctzero{} serves two modes from one set of weights: System~1 answers typed security decisions (is this file suspicious, is this request an attack, did the model just refuse) in a single forward pass without generating a token, System~2 is the usual tool-using agent, and a confidence gate spends generation only where the fast path is unsure. Specialisation lifts the agentic path: \aliasmicro{} solves $21.1\%$ of \texttt{cybench-subtask} against $7.6\%$ for its base, above every open model of comparable footprint and the densest cybersecurity capability per parameter we measured. The fast path is what the architecture buys: \aliasmicro{} screens eighty files per second for indicators of compromise, catches its own refusals at $96.5\%$ accuracy in $47$\,ms where the model judging itself reaches $76.5\%$ in $254$\,ms and misses half of them, flags jailbreak requests at AUROC $0.966$ in $8$\,ms, and, generating only for the facts it is unsure about, extracts vulnerability facts more accurately than either mode alone at a fifth of the generative cost. All of it is built in about $314$ RTX PRO 6000 GPU-hours and served from one workstation GPU, with no policy layer between operator and target. Across eighteen benchmarks and use cases run with one checkpoint and head, we report where the design helps, where it does not, and what it still lacks.
\end{abstract}
\vspace{0.7em}

\begin{center}
{\small\hfemoji~\textbf{Open weights:}~\url{\ctzerourl}}
\end{center}
\vspace{0.45em}

\begin{center}
\begin{minipage}{0.56\textwidth}
  \centering
  \resizebox{\linewidth}{!}{\input{tex/fig_hero.tex}}
\end{minipage}\hfill
\begin{minipage}{0.42\textwidth}
  \centering
  {\fontsize{7}{9.5}\selectfont
  \setlength{\tabcolsep}{2pt}
  \renewcommand{\arraystretch}{1.23}
  \resizebox{\linewidth}{!}{\input{tex/model_reference_rows.tex}}}
  \par\vspace{5pt}
  {\scriptsize\raggedright
  \mk{\dagger}Estimated load memory with a 20\% allowance; context cache extra. Gemma: Google's 17.9\,GB (8B total incl.\ embeddings, 4.5B effective)~\cite{google2026gemma4memory,google2026gemma4card}; Bonsai: 5.95\,GB PTQ1\_0 text weights, 7.90\,GB with the optional Q8 vision pack~\cite{bonsai2_modelcard2026}; other rows: nominal weights $\times1.2$. \mk{*}Speculative sizes, not vendor disclosures (plausible range: Opus 1--6T). \mk{\S}Cybench Subtask: $40$ capture-the-flag (CTF) challenges split into $171$ under one fixed budget, hence reproducible and comparable, unlike best-of-$k$ model-card figures (Section~\ref{sec:protocol}). \mk{\P}Kolibri-1: open $78$B MoE, $3.5$B active, FP8, same budget; a transcript audit attributes its score to non-converging exploration and to the scaffold's answer parser (Section~\ref{sec:protocol}).
  \par}
\end{minipage}
\end{center}
\vspace{0.3em}
\captionsetup{hypcap=false}
\captionof{figure}{\textbf{\ctzero{} and larger references on Cybench (subtask) under a fixed budget.} The shaded row is \ctzero{}, this paper's target. Under the reproducible protocol of Section~\ref{sec:protocol} (one attempt, five steps, no extended thinking), the $4$\,B \aliasmicro{} solves $21.1\%$ of the subtasks: $2.8\times$ its own base and above the other open models of comparable compute, including a $78$\,B mixture-of-experts with $3.5$\,B active parameters (Kolibri-1, $5.3\%$) and a $27$\,B model compressed to ternary weights (Bonsai~2, $8.2\%$). The alias series then scales to $40.9\%$ (mini) and $56.1\%$ (max), ahead of Mistral Large 4 ($54.4\%$) and Opus-5.5 ($29.2\%$), each measured in one API campaign with the same scaffold; switching extended thinking on moves the two largest references by at most one subtask (Appendix~\ref{app:thinking}).}\label{fig:hero}
\vspace{1.0em}
]
\renewcommand{\thefootnote}{$\dagger$}
\setcounter{footnote}{0}
\footnotetext{Corresponding author: \texttt{victor@aliasrobotics.com}}
\renewcommand{\thefootnote}{\arabic{footnote}}
\setcounter{footnote}{0}

\section{Introduction}\label{sec:introduction}
\input{tex/introduction.tex}

\section{Related Work}\label{sec:related}
\input{tex/related_work.tex}

\section{\ctzero{}: Architecture and Post-training}\label{sec:architecture}
\input{tex/architecture.tex}

\section{Benchmarks and Use Cases}\label{sec:eval}
\input{tex/eval_overview.tex}

\subsection{Evaluation Protocol}\label{sec:protocol}
\input{tex/protocol.tex}

\subsection{What Is Measured, and What the Results Show}\label{sec:thesis}
\input{tex/eval_thesis.tex}

\subsection{Why the Fast Path Pays: Cost, Calibration and When the Gate Helps}\label{sec:mechanism}
\input{tex/eval_mechanism.tex}

\subsection{Offensive Security: Agentic CTF Subtasks}\label{sec:results}
\input{tex/results.tex}

\subsection{Tool-Use Security and Agent Compliance}\label{sec:job-tooluse}
\input{tex/job_tooluse.tex}

\subsection{Phishing and Malicious-Request Detection}\label{sec:job-phishing}
\input{tex/job_phishing.tex}

\subsection{Prompt-Injection Resilience and Legitimate-Task Utility}\label{sec:job-injection}
\input{tex/job_injection.tex}

\subsection{Vulnerability Assessment and Secure-Code Review}\label{sec:job-vuln}
\input{tex/job_vuln.tex}

\subsection{SOC Triage, Malware Analysis and Incident Investigation}\label{sec:job-soc}
\input{tex/job_soc.tex}

\subsection{Security-Triage Calibration and Serving Efficiency}\label{sec:job-calibration}
\input{tex/job_calibration.tex}

\section{Discussion}\label{sec:discussion}
\input{tex/discussion.tex}

\section{Conclusion}\label{sec:conclusion}
\input{tex/conclusion.tex}

\bibliography{ct0-bibliography}

\onecolumn
\input{tex/appendix.tex}

\end{document}

%% file: tex/figure_palette.tex
\definecolor{fig_ct0}{HTML}{2E6260}
\definecolor{fig_mini}{HTML}{4C9A99}
\definecolor{fig_base}{HTML}{B8D8D8}
\definecolor{fig_s1}{HTML}{0CA30C}
\definecolor{fig_s2}{HTML}{1F4E8C}
\definecolor{fig_compose}{HTML}{2E6260}
\definecolor{fig_thinking}{HTML}{B8641B}
\definecolor{fig_control}{HTML}{5B6875}
\definecolor{fig_ink}{HTML}{16202A}
\definecolor{fig_rule}{HTML}{D6DCE2}
\definecolor{fig_warning}{HTML}{8A1C1C}
\definecolor{fig_frontier}{HTML}{C0392B}
\definecolor{fig_surface}{HTML}{F7F8FA}
\definecolor{fig_process}{HTML}{7A4FB5}
\definecolor{fig_time}{HTML}{6B7884}
\definecolor{fig_link}{HTML}{307FE2}
\pgfplotsset{ct0 plot/.style={
  font=\sffamily,
  axis lines=left, axis line style={draw=fig_rule,line width=.6pt},
  grid style={solid,draw=fig_rule!65,line width=.4pt},
  tick style={draw=fig_rule},
  tick label style={text=fig_control},
  label style={text=fig_ink},
  legend style={draw=none,fill=none,font=\scriptsize\sffamily},
  legend cell align=left
}}

%% file: tex/fig_hero.tex
\includegraphics[width=\linewidth]{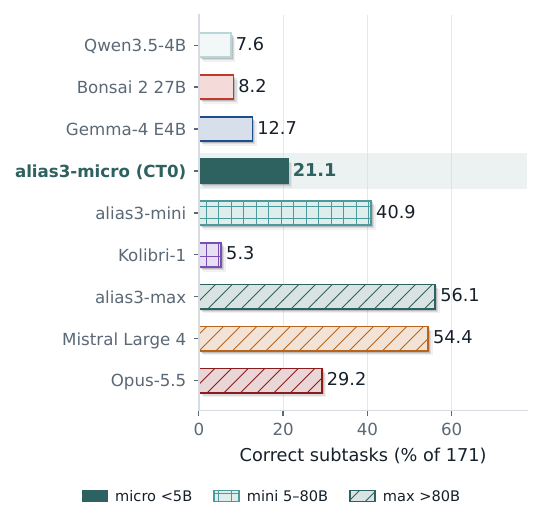}

%% file: tex/model_reference_rows.tex
\begin{tabular}{@{}lrrrr@{}}
\toprule Model & Params & Format & GB\mk{\dagger} & Cybench\mk{\S} \\ \midrule
  Qwen3.5-4B & 4B & BF16 & 9.6 & 7.6\% \\
  Bonsai 2 27B & 27B & 1.75-bit & 7.14 & 8.2\% \\
  Gemma-4 E4B & 8B & BF16 & 17.9 & 12.7\% \\
  \rowcolor{fig_ct0!10} \textcolor{fig_ct0}{\textbf{alias3-micro (\ctzero{})}} & \textcolor{fig_ct0}{\textbf{4B}} & \textcolor{fig_ct0}{\textbf{BF16}} & \textcolor{fig_ct0}{\textbf{9.6}} & \textcolor{fig_ct0}{\textbf{21.1\%}} \\
  \textcolor{fig_mini}{\textbf{alias3-mini}} & 30B & FP8 & 36 & 40.9\% \\
  Kolibri-1\mk{\P} & 78B & FP8 & 89 & 5.3\% \\
  \textcolor{fig_ct0}{\textbf{alias3-max}} & 800B & FP8 & 960 & 56.1\% \\
  Mistral Large 4 & 1T & API & --- & 54.4\% \\
  Opus-5.5 & $\sim$2--4T\mk{*} & API & --- & 29.2\% \\
\bottomrule\end{tabular}

%% file: tex/introduction.tex
\begin{figure*}[tp]
\centering
\includegraphics[width=\textwidth,height=0.82\textheight,keepaspectratio]{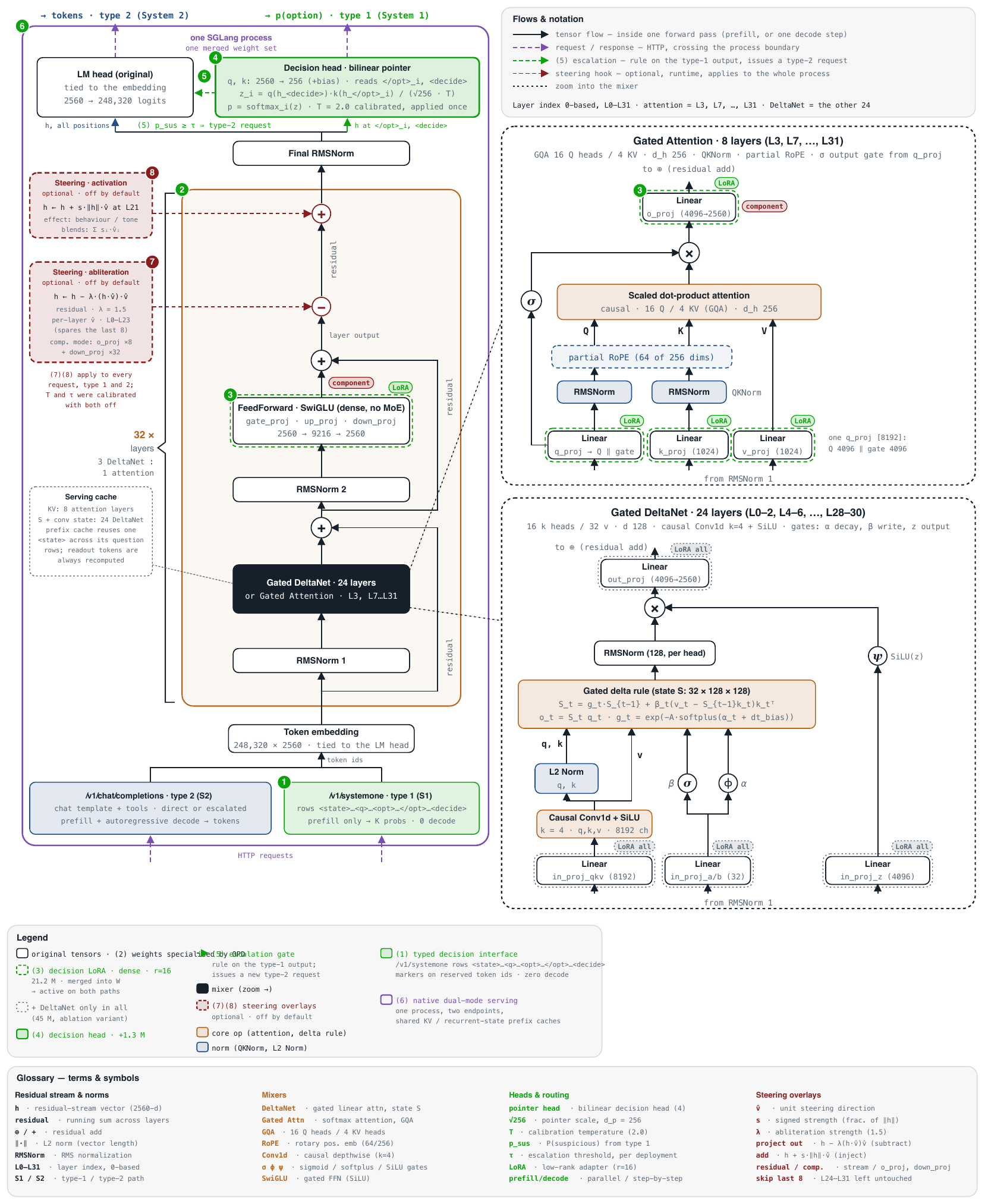}
\caption{\textbf{The \ctzero{} architecture, as deployed.} A single SGLang process serves one merged weight set (an OPD-specialised hybrid $4$\,B backbone with a fused decision LoRA) plus a small pointer head, and answers two request types. \emph{Left:} the transformer stack, $32$ layers ($24$ Gated DeltaNet linear-attention mixers and $8$ Gated Attention layers, one every four), a dense SwiGLU feed-forward, tied embeddings, and a shared final RMSNorm feeding two heads. \emph{System~2}, the type-2 request (\texttt{/v1/chat/completions}), is the ordinary agentic decode loop through the LM head. \emph{System~1}, the type-1 request (\texttt{/v1/systemone}), reads the final hidden states at the \texttt{<decide>} and \texttt{</opt>$_i$} positions and scores options with a bilinear pointer head during prefill, decoding zero tokens. Green badges mark what \ctzero{} adds to the backbone: \badge{1} the typed decision interface, marker tokens that pose a question over options; \badge{2} the weights, specialised by two-phase on-policy distillation; \badge{3} the decision LoRA, fused into those weights; \badge{4} the bilinear pointer head; \badge{5} the escalation gate, a threshold on the System-1 output that issues a System-2 request; \badge{6} one process serving both endpoints over shared prefix caches. Red badges \badge[fig_warning]{7} and \badge[fig_warning]{8} are optional residual-stream steering overlays, off by default; this paper uses one only in the de-refusal loop of Section~\ref{sec:refusal}.}
\label{fig:arch}
\end{figure*}

The dominant recipe for Cybersecurity AI consists of post-training a large general-purpose model with security-related RL environments, a model that was never architected for the domain \cite{aliasrobotics2025cai,mayoralvilchescybersuper2026} adapting its weights to security use cases. Four costs follow. \emph{Price and latency}: frontier inference is expensive and slow relative to the tight loops of offensive-security work. \emph{Control}: closed weights cannot be inspected, fine-tuned to a house methodology, run air-gapped on a client's premises, or audited. \emph{Refusal}: alignment layers calibrated for consumer products routinely decline legitimate, authorised defensive work; we have observed frontier models halt mid-engagement on authorised forensic targets, citing usage policy. \emph{Privacy}: every prompt to a hosted model carries the client's most sensitive material, such as credentials, configurations, telemetry and unpatched vulnerabilities, off the premises and into a third party's logs and retention policies, which many engagements, regulated sectors and non-disclosure terms do not permit; a specialist that runs locally keeps that material where the contract says it must stay.

We ask the opposite question: \emph{how small can a transformer be while remaining a genuinely useful cybersecurity specialist?} A specialist that runs on one workstation GPU, can be further fine-tuned to a methodology, and sits with no policy layer between operator and target changes both the economics and the control surface of the field.

We introduce \textbf{\ctzero{}}, a \emph{Cybersecurity Transformer} architecture depicted in figure \ref{fig:arch}, and \aliasmicro{}, its first incarnation: a $\sim$4\,B-parameter model post-trained from an open base into a dual-mode agent. \ctzero{} makes two commitments. \emph{(i) One backbone, two inference regimes.} Following dual-process accounts of cognition \cite{kahneman2011thinking}, a fast \emph{System-1} mode answers typed decisions (classify, score, triage) during prefill through a bilinear \emph{pointer} head adapted from the Jev/System One lineage \cite{vinyals2015pointer,typesafe2026systemone}, decoding zero tokens; the usual \emph{System-2} agentic decode loop runs from the same weight set; an external controller routes between them. \emph{(ii) Specialisation by on-policy distillation.} The backbone is distilled on-policy \cite{agarwal2024onpolicy} from a larger open teacher over its own rollouts on the CAI dataset \cite{aliasrobotics2025cai}, in two chained phases.

Table~\ref{tab:evidence-overview} summarises the evidence, and its strongest rows concern \emph{how computation is used}. System~1 detects refusals at $96.5\%$ accuracy in $47$\,ms where generative self-judging reaches $76.5\%$ in $254$\,ms and misses half of them; it screens eighty files per second for indicators of compromise (IOC); and preserving confident typed facts while repairing uncertain ones raises exact match on synthetic findings to $48.3\%$ from $43.4\%$ (S1) and $38.0\%$ (S2) at a fifth of the generative cost. Specialisation lifts the agentic path as well: under the fixed local budget of Section~\ref{sec:protocol} the specialist solves $36/171$ ($21.1\%$) of \texttt{cybench-subtask}, $2.8\times$ its open base ($7.6\%$) and above Gemma-4 E4B ($12.7\%$), the ternary-weight Bonsai~2 27B ($8.2\%$) and the 78\,B mixture-of-experts Kolibri-1 ($5.3\%$); the alias series then scales to $40.9$ and $56.1\%$. Where a cheap first stage drops the evidence the generator needs, composition loses, and Section~\ref{sec:job-soc} says where; Section~\ref{sec:discussion} gathers everything the evidence does not support.

\input{tex/evidence_overview.tex}

Contributions:
\begin{enumerate}[leftmargin=*,nosep]
  \item \textbf{The \ctzero{} architecture (Figure~\ref{fig:arch}, Table~\ref{tab:elements}):} six additions to an open hybrid backbone that serve a non-generative decision head and an agentic decode loop from one weight set: \badge{1} a typed decision interface of marker tokens that poses a security question over options; \badge{2} weights specialised for security by two-phase on-policy distillation; \badge{3} a rank-16 decision adapter fused into those weights, so that one weight set serves both modes; \badge{4} a $1.3$\,M-parameter bilinear pointer head that returns calibrated probabilities during prefill, decoding zero tokens; \badge{5} an escalation gate that sends only the uncertain cases to generation; \badge{6} native dual-mode serving, one process and two endpoints over shared prefix caches. Two optional, reversible steering overlays complete the design (Section~\ref{sec:architecture}).
  \item \textbf{A two-phase on-policy distillation recipe}\footnote{Distillation trains a smaller \emph{student} model to reproduce the next-token distribution of a larger \emph{teacher}; on-policy means the teacher is consulted on the student's own outputs rather than on a fixed corpus (Section~\ref{sec:opd}).} that turns an open $4$\,B base into a domain specialist (Section~\ref{sec:opd}).
  \item \textbf{An empirical qualification of \ctzero{} for cybersecurity work:} a battery of security benchmarks and use cases run end to end with the same checkpoint and head: refusal detection and de-refusal, jailbreak-request and dangerous-tool-call detection, pentest-finding triage, agentic CTF subtasks (\texttt{cybench-subtask}) under one fixed budget, email, agent-trace and code review, vulnerability-fact extraction with selective repair, malware and alert triage, incident-log investigation and prompt-injection guarding. The decision head beats the generator on detection-style decisions at a fraction of the latency, selective composition improves on either mode alone where the two err on different cases, the specialist is built in about $314$ RTX PRO 6000 GPU-hours and served from one workstation GPU, and the workflows where a cheap first stage does not pay are measured and explained (Section~\ref{sec:eval}, Appendices~\ref{app:ct0-usecases} and~\ref{app:ct0-operational}).
\end{enumerate}

%% file: tex/evidence_overview.tex
\begin{table*}[p]
\centering
\caption{\textbf{Cybersecurity benchmarks: \ctzero{}'s two modes and their composition.} S1 is System~1 (the decision head), S2 System~2 (generation) and S1$\to$S2 their cascade; other models report S2 unless marked $^{\alpha}$ (head) or $^{\beta}$ (pipeline). \textbf{Bold} marks the best displayed value per row, including ties. Arrows give the preferred direction. Values are percentages except AUROC/ECE, counts and physical units. AUROC is the area under the ROC curve (ranking quality; $1$ is perfect); recall is the share of true positives detected; FPR is the share of negatives wrongly flagged; ECE is the expected calibration error, the gap between stated confidence and observed accuracy (lower is better). A dash means unavailable or inapplicable. The bands follow Sections~\ref{sec:results}--\ref{sec:job-calibration}, and the link after each band and row opens the subsection that develops it; sources and data sizes are in Table~\ref{tab:sources} (Appendix~\ref{app:sources}). Small grey subscripts give the mean per-request latency recorded in a model\textquotesingle s run (unit in the column header); \ctzero{}\textquotesingle s three arms carry them together or not at all, other models where recorded; Section~\ref{sec:eval} discusses them.}
\label{tab:evidence-overview}
\begingroup\small
\setlength{\tabcolsep}{2.5pt}
\renewcommand{\arraystretch}{1.06}
\newcommand{\evna}{\textcolor{fig_control!65}{\textendash}}
\newcommand{\evt}[1]{\textsubscript{\textcolor{fig_time}{\fontsize{5}{5.5}\selectfont #1}}}
\newcommand{\evu}{\textsubscript{\textcolor{fig_time}{\fontsize{5}{5.5}\selectfont [ms]}}}
\newcommand{\evband}[2]{\rowcolor{fig_ct0!9}[0pt][0pt]\multicolumn{7}{@{}p{\textwidth}@{}}{\strut\hspace{3pt}\textcolor{fig_ct0}{\textbf{#1}} {\scriptsize\hyperref[#2]{(\S\ref*{#2})}}}\\}
\begin{tabularx}{\textwidth}{@{}>{\raggedright\arraybackslash}p{.38\textwidth}>{\centering\arraybackslash}X>{\columncolor{fig_s1!6}\centering\arraybackslash}X>{\columncolor{fig_s2!6}\centering\arraybackslash}X>{\columncolor{fig_compose!8}\centering\arraybackslash}X*{2}{>{\centering\arraybackslash}X}@{}}
\toprule
& \shortstack{Qwen3.5\\4B (base)} & \multicolumn{3}{c}{\textcolor{fig_ct0}{\shortstack{\textbf{alias3-micro (\ctzero{})}\\\textbf{4B}}}} & \textcolor{fig_mini}{\shortstack{alias3-mini\\30B}} & \textcolor{fig_ct0}{\shortstack{alias3-max\\800B}} \\
\cmidrule(lr){3-5}
\textbf{Evaluation / metric} & S2\evu & \textcolor{fig_s1}{\textbf{S1}}\evu & \textcolor{fig_s2}{\textbf{S2}}\evu & \textcolor{fig_compose}{\textbf{S1$\to$S2}}\evu & S2\evu & S2\evu \\
\midrule
\evband{Offensive security: agentic CTF subtasks}{sec:results}
Cybench-Subtask (\%) $\uparrow$ {\scriptsize\hyperref[sec:results]{(\S\ref*{sec:results})}} & 7.6 & \evna & \textcolor{fig_s2}{21.1} & \evna & 40.9 & \textbf{56.1} \\
\evband{Tool-use security and agent compliance}{sec:job-tooluse}
IFC tool-call labels: accuracy $\uparrow$ {\scriptsize\hyperref[sec:de-ifc]{(\S\ref*{sec:de-ifc})}} & 73.9 & \textcolor{fig_s1}{80.6} & \textcolor{fig_s2}{75.2} & \textcolor{fig_compose}{81.9} & 87.9 & \textbf{88.1} \\
R-Judge agent risk: balanced accuracy $\uparrow$ {\scriptsize\hyperref[sec:uc-rjudge]{(\S\ref*{sec:uc-rjudge})}} & 69.9 & \textcolor{fig_s1}{73.7$^{\gamma}$}\evt{81} & \textcolor{fig_s2}{77.4}\evt{568} & \textcolor{fig_compose}{75.4$^{\gamma}$}\evt{133} & \textbf{94.6} & 93.3 \\
Refusal detection: accuracy $\uparrow$ {\scriptsize\hyperref[sec:refusal]{(\S\ref*{sec:refusal})}} & 76.0$^{\delta}$\evt{173} & \textcolor{fig_s1}{\textbf{96.5}}\evt{47} & \textcolor{fig_s2}{76.7$^{\delta}$}\evt{321} & \textcolor{fig_compose}{82.3}\evt{$\sim$100} & 83.7$^{\delta}$\evt{856} & 94.5$^{\delta}$\evt{4,649} \\
Refusal preflight: AUROC $\uparrow$ {\scriptsize\hyperref[sec:refusal]{(\S\ref*{sec:refusal})}} & .923$^{\alpha}$\evt{20} & \textcolor{fig_s1}{\textbf{.932}}\evt{58} & \evna & \evna & \evna & 0.804$^{\alpha}$\evt{137} \\
Retry, hardest slice: refusal (\%) $\downarrow$ {\scriptsize\hyperref[sec:refusal]{(\S\ref*{sec:refusal})}} & 78.0$^{\delta}$ & \evna & \textcolor{fig_s2}{70.0} & \textcolor{fig_compose}{60.0} & \textbf{0.0}$^{\delta}$ & 22.0$^{\delta}$ \\
Retry, intermediate: refusal (\%) $\downarrow$ {\scriptsize\hyperref[sec:refusal]{(\S\ref*{sec:refusal})}} & 89.0$^{\delta}$ & \evna & \textcolor{fig_s2}{84.0} & \textcolor{fig_compose}{71.0} & \textbf{0.0}$^{\delta}$ & 20.0$^{\delta}$ \\
\evband{Phishing and malicious-request detection}{sec:job-phishing}
PhishNChips: balanced accuracy $\uparrow$ {\scriptsize\hyperref[sec:uc-phishing]{(\S\ref*{sec:uc-phishing})}} & 72.0 & \textcolor{fig_s1}{92.5$^{\gamma}$}\evt{61} & \textcolor{fig_s2}{60.1}\evt{635} & \textcolor{fig_compose}{\textbf{92.9}$^{\gamma}$}\evt{65} & 73.7 & 80.6 \\
WildJailbreak: detector AUROC $\uparrow$ {\scriptsize\hyperref[sec:de-wj]{(\S\ref*{sec:de-wj})}} & \textbf{0.9661}$^{\alpha}$ & \textcolor{fig_s1}{.966} & \evna & \evna & 0.7555$^{\alpha}$ & \textbf{0.9661}$^{\alpha}$ \\
WildJailbreak, full set: recall $\uparrow$ {\scriptsize\hyperref[sec:de-wj]{(\S\ref*{sec:de-wj})}} & 91.3$^{\beta}$ & \textcolor{fig_s1}{88.9} & \textcolor{fig_s2}{\textbf{96.6}} & \textcolor{fig_compose}{94.4} & 63.4$^{\beta}$ & 90.0$^{\beta}$ \\
WildJailbreak, full set: FPR $\downarrow$ {\scriptsize\hyperref[sec:de-wj]{(\S\ref*{sec:de-wj})}} & 14.8$^{\beta}$ & \textcolor{fig_s1}{\textbf{6.7}} & \textcolor{fig_s2}{14.3} & \textcolor{fig_compose}{14.8} & 23.3$^{\beta}$ & 14.3$^{\beta}$ \\
WildJailbreak, S2 subset: recall $\uparrow$ {\scriptsize\hyperref[sec:de-wj]{(\S\ref*{sec:de-wj})}} & 91.7 & \textcolor{fig_s1}{88.0} & \textcolor{fig_s2}{\textbf{96.3}} & \textcolor{fig_compose}{93.0} & 67.0 & 87.7 \\
WildJailbreak, S2 subset: FPR $\downarrow$ {\scriptsize\hyperref[sec:de-wj]{(\S\ref*{sec:de-wj})}} & 12.9 & \textcolor{fig_s1}{\textbf{6.2}} & \textcolor{fig_s2}{14.3} & \textcolor{fig_compose}{15.2} & 22.4 & 16.2 \\
\evband{Prompt-injection resilience and legitimate-task utility}{sec:job-injection}
LivePI: guard AUROC $\uparrow$ {\scriptsize\hyperref[sec:livepi-offline]{(\S\ref*{sec:livepi-offline})}} & \textbf{0.9988}$^{\alpha}$ & \textcolor{fig_s1}{0.9983} & \evna & \evna & 0.9693$^{\alpha}$ & 0.9985$^{\alpha}$ \\
LivePI: tool requests $\downarrow$ {\scriptsize\hyperref[sec:livepi-offline]{(\S\ref*{sec:livepi-offline})}} & \textbf{0/165} & \evna & \textcolor{fig_s2}{29/165} & \textcolor{fig_compose}{26/165} & 11/165 & \textbf{0/165} \\
LivePI: task completion $\uparrow$ {\scriptsize\hyperref[sec:livepi-offline]{(\S\ref*{sec:livepi-offline})}} & \textbf{165/165} & \evna & \textcolor{fig_s2}{136/165} & \textcolor{fig_compose}{118/165} & 154/165 & 130/165 \\
IPI original: target-tool matches $\downarrow$ {\scriptsize\hyperref[sec:ipi-offline]{(\S\ref*{sec:ipi-offline})}} & 7/89 & \evna & \textcolor{fig_s2}{14/89} & \textcolor{fig_compose}{10/89} & 1/89 & \textbf{0/89} \\
IPI matched follow-up: target-tool matches $\downarrow$ {\scriptsize\hyperref[sec:ipi-offline]{(\S\ref*{sec:ipi-offline})}} & 5/89$^{\delta}$ & \evna & \textcolor{fig_s2}{15/89$^{\delta}$} & \textcolor{fig_compose}{10/89$^{\delta}$} & 1/89$^{\delta}$ & \textbf{0/89}$^{\delta}$ \\
AgentDojo: attack success $\downarrow$ {\scriptsize\hyperref[sec:op-dojo]{(\S\ref*{sec:op-dojo})}} & 18/64 & \evna & \textcolor{fig_s2}{16/64} & \textcolor{fig_compose}{4/64} & 7/64 & \textbf{0.0}$^{\delta}$ \\
AgentDojo: benign utility $\uparrow$ {\scriptsize\hyperref[sec:op-dojo]{(\S\ref*{sec:op-dojo})}} & 27/32 & \evna & \textcolor{fig_s2}{23/32} & \textcolor{fig_compose}{13/32} & \textbf{29/32} & 87.5$^{\delta}$ \\
\evband{Vulnerability assessment and secure-code review}{sec:job-vuln}
OWASP Java: balanced accuracy $\uparrow$ {\scriptsize\hyperref[sec:uc-java]{(\S\ref*{sec:uc-java})}} & 71.0 & \textcolor{fig_s1}{84.1$^{\gamma}$}\evt{111} & \textcolor{fig_s2}{88.8}\evt{616} & \textcolor{fig_compose}{90.2$^{\gamma}$}\evt{335} & 97.3 & \textbf{97.5} \\
Pentest findings: adjudication accuracy $\uparrow$ {\scriptsize\hyperref[sec:de-pentest]{(\S\ref*{sec:de-pentest})}} & 93.1\evt{421} & \textcolor{fig_s1}{83.3}\evt{12} & \textcolor{fig_s2}{\textbf{95.8}}\evt{200} & \textcolor{fig_compose}{87.5}\evt{109} & 91.7\evt{1,957} & 93.1$^{\delta}$\evt{5,580} \\
Pentest findings: severity accuracy $\uparrow$ {\scriptsize\hyperref[sec:de-pentest]{(\S\ref*{sec:de-pentest})}} & 75.0\evt{421} & \textcolor{fig_s1}{44.4}\evt{12} & \textcolor{fig_s2}{81.9}\evt{200} & \textcolor{fig_compose}{79.2}\evt{109} & \textbf{86.1}\evt{1,957} & \textbf{86.1}$^{\delta}$\evt{5,580} \\
Pentest findings: pruning recall $\uparrow$ {\scriptsize\hyperref[sec:de-pentest]{(\S\ref*{sec:de-pentest})}} & 87.1\evt{421} & \textcolor{fig_s1}{61.3}\evt{12} & \textcolor{fig_s2}{22.6}\evt{200} & \textcolor{fig_compose}{61.3}\evt{109} & \textbf{100.0}\evt{1,957} & \textbf{100.0}$^{\delta}$\evt{5,580} \\
Typed facts: seven-field exact match $\uparrow$ {\scriptsize\hyperref[sec:op-facts]{(\S\ref*{sec:op-facts})}} & 47.0\evt{512} & \textcolor{fig_s1}{43.4}\evt{154} & \textcolor{fig_s2}{38.0}\evt{1,577} & \textcolor{fig_compose}{48.3}\evt{316} & 82.3\evt{2,788} & \textbf{87.9}$^{\delta}$\evt{3,735} \\
CodeQL: zip4j fix-overlapping paths {\scriptsize\hyperref[sec:op-code]{(\S\ref*{sec:op-code})}} & 0 & \evna & \textcolor{fig_s2}{0} & \textcolor{fig_compose}{6} & \textbf{35} & \textbf{35}$^{\delta}$ \\
\evband{SOC triage, malware analysis and incident investigation}{sec:job-soc}
File corpus: wall time (s) $\downarrow$ {\scriptsize\hyperref[sec:op-files]{(\S\ref*{sec:op-files})}} & failed & \textcolor{fig_s1}{\textbf{3.4}} & \textcolor{fig_s2}{1,655} & \textcolor{fig_compose}{93} & 1,269 & 130$^{\delta}$ \\
Malware reports: exact answer set $\uparrow$ {\scriptsize\hyperref[sec:op-malware]{(\S\ref*{sec:op-malware})}} & 20.3\evt{1,784} & \textcolor{fig_s1}{17.2}\evt{456} & \textcolor{fig_s2}{20.7}\evt{750} & \textcolor{fig_compose}{20.1}\evt{1,372} & 28.2\evt{9,354} & \textbf{29.7}$^{\delta}$\evt{4,103} \\
Incident SQL: normal telemetry accuracy $\uparrow$ {\scriptsize\hyperref[sec:op-investigation]{(\S\ref*{sec:op-investigation})}} & 29.8 & \evna & \textcolor{fig_s2}{31.0} & \textcolor{fig_compose}{22.6} & 57.1 & \textbf{72.6}$^{\delta}$ \\
Incident SQL: raw telemetry accuracy $\uparrow$ {\scriptsize\hyperref[sec:op-investigation]{(\S\ref*{sec:op-investigation})}} & 10.7 & \evna & \textcolor{fig_s2}{11.9} & \textcolor{fig_compose}{13.1} & 16.7 & \textbf{39.3}$^{\delta}$ \\
AIT alert queue: labeled-group recall $\uparrow$ {\scriptsize\hyperref[sec:op-alerts]{(\S\ref*{sec:op-alerts})}} & 5.7 & \textcolor{fig_s1}{6.0} & \textcolor{fig_s2}{4.8} & \textcolor{fig_compose}{5.7} & \textbf{7.8} & \textbf{7.8}$^{\delta}$ \\
\evband{Security-triage calibration and serving efficiency}{sec:job-calibration}
Transfer-v4: dense-head ECE $\downarrow$ {\scriptsize\hyperref[sec:de-calibration]{(\S\ref*{sec:de-calibration})}} & \evna & \textcolor{fig_s1}{\textbf{.052}} & \evna & \evna & .053$^{\alpha}$ & \evna \\
Native triage throughput (files/s) $\uparrow$ {\scriptsize\hyperref[sec:de-serving]{(\S\ref*{sec:de-serving})}} & \evna & \textcolor{fig_s1}{\textbf{80--82}} & \evna & \evna & 7.1--7.2$^{\alpha}$ & \evna \\
\bottomrule
\end{tabularx}
\endgroup
\vspace{3pt}
\begin{minipage}{\textwidth}\footnotesize\raggedright
\textbf{Marks.} $^{\alpha}$Head-only measurement. $^{\beta}$Head-plus-actor pipeline. $^{\gamma}$Fitted signal combiner with validation-selected routing. $^{\delta}$Output budget fixed at a 4096-token S2 allowance (max with low reasoning); retry slices keep their original budgets.\par
\textbf{Rows.} Dashes mark values that are unavailable or inapplicable; \emph{failed} marks a run that ended without the required final output (Section~\ref{sec:op-files}). File corpus: S1 is screening only and the \ctzero{} S2 time includes a stalled-output recovery. IFC uses nine-shot S2. WildJailbreak: 2,210-request full set and 510-request subset, S1 at $\tau=0.3$. Protocols: Section~\ref{sec:protocol}.
\end{minipage}
\end{table*}

%% file: tex/related_work.tex
\paragraph{Cybersecurity agents and benchmarks.} LLM-driven security agents have moved from prompt-loop prototypes \cite{deng2024pentestgpt} to practitioner frameworks such as CAI \cite{aliasrobotics2025cai} and meta-scaffolds that compose heterogeneous harnesses \cite{mayoralvilchescybersuper2026,mayoralvilches2025gametheoretic}. Evaluation has consolidated around agentic CTF suites; we use \texttt{cybench-subtask}, the subtask-decomposed variant of Cybench \cite{cybench2024}, whose $171$ subtasks across $40$ challenges score partial progress and are therefore far more discriminative at small scale than binary flag capture. Most of this work holds the model fixed and varies the scaffold; we fix a lean scaffold and specialise the model.

\paragraph{Small models: specialise or compress.} A parallel line argues that small, specialised models, not scale, are the efficient frontier for narrow agentic tasks \cite{belcak2025small}, and open bases in the $2$--$4$\,B range \cite{qwen35card2026,google2026gemma4card} make a private specialist feasible to post-train. The other route to a small footprint is to compress a large generalist. Bonsai~2 27B \cite{bonsai2_2026,bonsai2_modelcard2026} is a general-purpose 27\,B open model quantised to ternary weights ($1.75$ bits per weight, $5.95$\,GB), so it loads in less memory than a 4\,B model in BF16; it is our external comparison point for that route, not a cybersecurity model. Our claim is that specialisation, not size, carries the result: a $4$\,B specialist outperforms same-size generalists and a 27\,B generalist compressed to a comparable footprint.

\paragraph{On-policy distillation.} Sequence-level knowledge distillation \cite{kim2016sequence} trains a student on teacher outputs; on-policy variants \cite{agarwal2024onpolicy} minimise the divergence over the student's own trajectories, removing the train/inference mismatch of off-policy imitation and connecting to DAgger-style interactive imitation \cite{ross2011dagger}\footnote{Dataset aggregation: the learner's own policy is run to collect the states it visits, an expert labels them, and training continues on the aggregate, so the training distribution matches the one met at inference.}. \ctzero{}'s backbone is produced by reverse-KL\footnote{The Kullback--Leibler (KL) divergence measures how far one probability distribution is from another; the \emph{reverse} direction puts the student's distribution first, which rewards matching the teacher's most probable continuations (Section~\ref{sec:opd}).} on-policy distillation from a larger open teacher (Section~\ref{sec:opd}).

\paragraph{Dual-process inference.} The System-1/System-2 framing \cite{kahneman2011thinking} motivates architectures that trade a fast amortised path against a slow deliberative one. \ctzero{} realises it at the level of decoding: System-1 is a single-forward pointer readout \cite{vinyals2015pointer} over structured option tokens, System-2 is the autoregressive agent, and both read the same final hidden state. The typed interface, pointer head and decision-LoRA recipe are adapted from TypeSafe's Jev/System One lineage \cite{typesafe2026systemone}; \ctzero{} adds cybersecurity specialisation, native integration of both modes in one inference server, and their joint evaluation.

\paragraph{Selective security decisions.} Recent Jev work studies confidence-based delegation in agents \cite{wu2026reflex} and security-judge reliability on risk and injection tasks \cite{liu2026systemonesecurity}; the phishing reproduction of \cite{anisselbd2026phishing} separates direct verdicts, decomposed signals and cheap non-neural controls. Sections~\ref{sec:uc-phishing}, \ref{sec:uc-rjudge} and~\ref{sec:uc-java} apply these principles to \ctzero{} on public phishing, agent-risk and source-code data, separating fitted decision quality, escalation benefit and measured pipeline cost.

%% file: tex/architecture.tex
\ctzero{} keeps the tensor architecture of its hybrid backbone (Figure~\ref{fig:arch}): $32$ layers at hidden width $2560$, $24$ Gated DeltaNet blocks interleaved with $8$ Gated-Attention layers (one every four), a dense SwiGLU feed-forward ($2560\!\to\!9216\!\to\!2560$), grouped-query attention ($16$ query / $4$ key-value heads, head dimension $256$) with partial rotary embeddings,\footnote{Gated DeltaNet layers mix tokens through a gated linear-attention recurrence with constant memory per token; Gated-Attention layers are softmax attention with an output gate; grouped-query attention lets several query heads share one key/value head; SwiGLU is a gated feed-forward block; rotary embeddings encode position by rotating query and key vectors.} and a tied $248{,}320$-token embedding. Around these unchanged tensors \ctzero{} adds six elements, numbered as in Figure~\ref{fig:arch}. \badge{1}~A \emph{typed decision interface}: a prompt grammar of marker tokens (\texttt{<state>}, \texttt{<q>}, \texttt{<opt>}, \texttt{</opt>}, \texttt{<decide>}) mapped onto reserved token ids, which turns a security question into a decision over options without changing the vocabulary. \badge{2}~A \emph{cybersecurity specialisation} of the whole backbone by two-phase on-policy distillation on the CAI dataset \cite{mayoralvilches2026caidataset} (Section~\ref{sec:opd}). \badge{3}~A \emph{decision LoRA}, a low-rank adapter \cite{hu2022lora}\footnote{A LoRA adds a trainable low-rank update $BA$ to selected weight matrices while the original weights stay frozen; rank $16$ is the inner dimension of $B$ and $A$.} (rank $16$, $21.2$\,M parameters) trained together with the head and fused into the specialised weights, so that one weight set serves both modes. \badge{4}~A \emph{bilinear pointer head} of $1.3$\,M parameters that reads the final hidden states at the markers and returns a calibrated probability during prefill, decoding nothing. \badge{5}~An \emph{escalation gate}: a thresholded rule on the calibrated System-1 output that issues a System-2 request only when the fast path is unsure, the mechanism behind the cascades and selective repair of Section~\ref{sec:eval}. \badge{6}~\emph{Native dual-mode serving}: one inference process exposes a System-1 endpoint beside the chat endpoint and shares the attention and recurrent-state prefix caches between them, so a typed decision costs a single prefill. Optional, off-by-default steering overlays on the residual stream\footnote{The residual stream is the vector that every layer reads from and adds to; a steering overlay adds or removes a fixed direction in it at inference time, leaving the weights unchanged.} (\badge[fig_warning]{7}, \badge[fig_warning]{8}) complete the design; they are inference-time and reversible, and one of them drives the de-refusal use case of Section~\ref{sec:refusal}. Table~\ref{tab:elements} sets each element beside what it buys. The decision interface, pointer head and decision-adapter recipe are adapted from the Jev/System One lineage \cite{typesafe2026systemone}; \ctzero{}'s contribution is their integration with a specialised backbone, the native dual-mode server, the gate, and the evaluation of both modes.

\begin{table*}[t]
\centering\small
\setlength{\tabcolsep}{5pt}\renewcommand{\arraystretch}{1.15}
\caption{\textbf{What \ctzero{} adds to its backbone, and what each element buys.} Numbered as in Figure~\ref{fig:arch}; every payoff is a measurement of Section~\ref{sec:eval}, taken on the deployed checkpoint.}
\label{tab:elements}
\begin{tabularx}{\textwidth}{@{}c>{\raggedright\arraybackslash}p{.135\textwidth}>{\raggedright\arraybackslash}X>{\raggedright\arraybackslash}X>{\raggedright\arraybackslash}p{.075\textwidth}@{}}
\toprule
& Element & What it is & What it buys & Where \\
\midrule
\badge{1} & Typed decision interface & Marker tokens (\texttt{<state>}, \texttt{<q>}, \texttt{<opt>}\ldots\texttt{</opt>}, \texttt{<decide>}) on reserved ids pose a security question over options; the vocabulary is unchanged. & Any security decision becomes one request that returns $K$ probabilities with zero tokens decoded: is this file suspicious, is this request an attack, did the model just refuse. & \S\ref{sec:architecture} \\
\badge{2} & Specialised weights & The whole backbone, taught security by two-phase on-policy distillation on the CAI dataset. & $7.6\to21.1\%$ on \texttt{cybench-subtask}: $2.8\times$ the open base and above every open model of comparable footprint, including a 78\,B mixture-of-experts. & \S\ref{sec:results} \\
\badge{3} & Decision adapter & A rank-16 LoRA on the attention, DeltaNet and feed-forward projections, trained with the head and fused into the specialised weights. & $21.2$\,M parameters and nothing to swap at runtime: one weight set serves the agent and the decision head, at no measured cost to the agentic path (merged against unmerged: $+2.3$ percentage points (pp), Welch interval $[-2.1,+6.8]$). & \S\ref{sec:architecture} \\
\badge{4} & Bilinear pointer head & Reads the final hidden states at the option markers and at \texttt{<decide>} and returns one calibrated probability per option. & $1.3$\,M parameters; the answer is ready when prefill ends. One fitted temperature halves the calibration error ($0.126\to0.052$) and cuts confident errors from $9.6$ to $2.9\%$. & \S\ref{sec:de-calibration} \\
\badge{5} & Escalation gate & A threshold on the System-1 probability: confident decisions return at once, the uncertain band issues a System-2 request over the same cache. & Code review at $90.2\%$ balanced accuracy generating for $36\%$ of cases; phishing at $92.9\%$ with eight generations in a thousand; typed facts at $48.3\%$ exact match for a fifth of the generative cost. & \S\ref{sec:mechanism}, \S\ref{sec:uc-phishing}, \S\ref{sec:op-facts} \\
\badge{6} & Native dual-mode serving & One SGLang process, two endpoints; the attention KV cache and the DeltaNet recurrent state are prefix caches both request types share. & $80$--$82$ files screened per second on one RTX PRO 6000, $1.4\times$ the reference server and ten times an external gateway; screening is unaffected by concurrent generation. & \S\ref{sec:de-serving} \\
\midrule
\badge[fig_warning]{7}\,\badge[fig_warning]{8} & Steering overlays (optional) & Runtime, reversible, off by default: a projection that removes the refusal direction from the residual stream (7), an added direction that shifts behaviour or tone (8). Weights untouched. & With the System-1-gated abliterated retry, refusals on the hardest authorised prompts fall from $70$ to $60\%$ and one in three detected refusals becomes a substantive answer, with zero degenerate outputs. & \S\ref{sec:refusal} \\
\bottomrule
\end{tabularx}
\end{table*}

\subsection{Two inference regimes from one weight set}
\textbf{System~2} is the conventional path: a chat request is prefilled and the model decodes responses and tool calls autoregressively as observations return.\footnote{Prefill processes the whole prompt in one parallel forward pass; decoding then produces the response one token per forward pass, which is what makes generation slow.} It produces the \texttt{cybench-subtask} results of Section~\ref{sec:results}, with extended thinking\footnote{The optional reasoning phase in which a model writes a long chain of thought before its answer.} disabled, for reproducibility and in line with the upstream Cybench subtask methodology; enabling it does not change the results, as Appendix~\ref{app:thinking} shows for the two largest references.

\textbf{System~1} answers a typed question without decoding. The request is one causal sequence, \texttt{<state>\,\dots\,<q>\,\dots\,<opt>\,\dots\,</opt>\,\dots\,<decide>}: the evidence, the question, $K$ candidate options each closed by \texttt{</opt>}, and a final \texttt{<decide>} marker at which the decision is read. Writing $h_x\in\mathbb{R}^{2560}$ for the final hidden state (after the last RMSNorm) at token $x$, a bilinear pointer head \cite{vinyals2015pointer} scores option $i$ by comparing the state at \texttt{<decide>} with the state at that option's closing marker, as one attention query over the options:
\begin{equation}
z_i = \frac{q(h_{\texttt{<decide>}})^{\!\top} k(h_{\texttt{</opt>}_i})}{\sqrt{d}\,T},\qquad
p_i = \operatorname{softmax}_i(z),
\label{eq:pointer}
\end{equation}
where $q$ and $k$ are learned affine query and key projections $\mathbb{R}^{2560}\!\to\!\mathbb{R}^{d}$ with $d=256$ (together the head's $1.3$\,M parameters), $z_i$ is the score of option $i$, $1/\sqrt{d}$ is the usual attention scaling, and $T=2.0$ is a temperature fitted once on development data after training so that the probabilities are calibrated\footnote{A probability is calibrated when it matches the observed frequency: of the cases scored $0.8$, about $80\%$ turn out positive. The expected calibration error (ECE) is the average gap between stated probability and observed frequency across score bins; lower is better.} (Section~\ref{sec:de-calibration}); the softmax over the $K$ scores $z=(z_1,\dots,z_K)$ returns one probability per option. The same head serves yes/no questions ($K=2$), multi-way choices and ordinal scales. The confidence that the gate thresholds is $(p_{\max}-1/K)/(1-1/K)$ for a choice, $0$ at a uniform $p$ and $1$ at certainty, and $\max\{0,\,1-\mathbb{E}_p|\ell-\ell^\ast|/D\}$ for an ordinal scale with levels $\ell=1,\dots,L$: the expected distance from the modal level $\ell^\ast$, relative to the mean absolute deviation $D$ of a uniform distribution over the levels.

\textbf{Deployment.} A rank-$16$ decision LoRA adapts the attention and feed-forward projections ($21.2$\,M trainable parameters, $22.5$\,M with the head) and is fused into the specialised backbone; the fused weight set is the \emph{deployed checkpoint}, the one weight set used for both modes throughout this paper. One SGLang\footnote{An open-source inference server for large language models.} process serves \texttt{/v1/systemone} through the pointer head and \texttt{/v1/chat/completions} through the LM head; the endpoint selects the readout and decoding regime, not whether the adapter is active. The server reuses the shared state prefix through the attention KV and DeltaNet recurrent-state caches,\footnote{The per-token keys and values of the attention layers, and the recurrent state of the DeltaNet layers, kept from processing a prefix so that a later request sharing that prefix does not recompute it.} and an external controller applies task-specific thresholds to System-1 outputs and may issue a System-2 request; the heads do not feed into each other. Fusing the adapter costs the agentic path nothing measurable: with three \texttt{cybench-subtask} campaigns per arm, the merged checkpoint scores $2.3$ pp above the unmerged backbone (Welch interval $[-2.1,+6.8]$ pp\footnote{A $95\%$ confidence interval for the difference of two means from Welch's $t$-test, which does not assume equal variances.}). The probability-only interface cannot fabricate a prose citation: it answers with a calibrated probability that a controller thresholds.

\subsection{Post-training by two-phase on-policy distillation}
\label{sec:opd}
An offensive-security agent is judged on trajectories it generates itself, so the training signal should be computed where the student actually puts probability mass rather than on static teacher demonstrations. Let $\pi_\theta$ be the student ($\sim$4\,B) with trainable parameters $\theta$, and $\pi_\mathcal{T}$ a larger open teacher that shares its tokenizer, so that both define next-token distributions over the same vocabulary and their divergence is well defined token by token. For a prompt $x$ drawn from a distribution $\mathcal{D}$ of security tasks and a response $y=(y_1,\dots,y_n)$ sampled from the student itself, write $p_{\theta,t}$ and $p_{\mathcal{T},t}$ for the student's and the teacher's next-token distributions at position $t$ given the same prefix $(x,y_{<t})$. We minimise the \emph{reverse} Kullback--Leibler divergence between them, summed over positions and averaged over the student's own rollouts:
\begin{equation}
\begin{aligned}
\mathcal{L}(\theta)
 &= \mathbb{E}_{\substack{x\sim\mathcal{D}\\y\sim\pi_\theta(\cdot\mid x)}}
    \left[\sum_t D_{\mathrm{KL}}\!\left(p_{\theta,t}\,\|\,p_{\mathcal{T},t}\right)\right],\\
p_{\theta,t}&=\pi_\theta(\cdot\mid y_{<t},x),\qquad
p_{\mathcal{T},t}=\pi_\mathcal{T}(\cdot\mid y_{<t},x).
\end{aligned}
\label{eq:opd}
\end{equation}
with $D_{\mathrm{KL}}(p\,\|\,p')=\sum_v p(v)\log\big(p(v)/p'(v)\big)$ over vocabulary tokens $v$. Two properties follow from this form. Being on-policy, Eq.~\eqref{eq:opd} is evaluated at the states the student visits at inference, which removes exposure bias\footnote{The compounding of errors when a model trained only on reference continuations must continue its own imperfect outputs at inference.} \cite{ross2011dagger,agarwal2024onpolicy}; being reverse (the student's distribution comes first), the KL is mode-seeking, so the small student commits to the teacher's high-probability continuations instead of spreading mass over modes it cannot represent. In practice the teacher distribution is truncated to its $k=32$ most probable tokens per position and renormalised ($\tilde p_t$ in Algorithm~\ref{alg:opd}), which keeps teacher scoring to $k$ log-probabilities per position.

\begin{algorithm}[t]
\small
\caption{One OPD round}\label{alg:opd}
\begin{algorithmic}[1]
\Require student $\pi_\theta$, teacher $\pi_\mathcal{T}$, prompt corpus $\mathcal{D}$, truncation $k$, learning rate $\eta$
\State $\mathcal{R}\gets\{(x,y):x\sim\mathcal{D},\,y\sim\pi_\theta(\cdot\mid x)\}$ \Comment{Generate (on-policy)}
\State $\tilde p_t\gets$ top-$k$ of $\pi_\mathcal{T}(\cdot\mid y_{<t},x)$, renormalised, for every position $t$ of every $(x,y)\in\mathcal{R}$ \Comment{Score (teacher-forced)}
\State $\theta\gets\theta-\eta\nabla_\theta\!\sum_{t} D_{\mathrm{KL}}\!\big(\pi_\theta\,\|\,\tilde p_t\big)$ \Comment{Train (reverse KL)}
\end{algorithmic}
\end{algorithm}

Each round generates on-policy rollouts from the current student (temperature $1.0$, top-$p\,0.9$\footnote{Temperature scales the logits before sampling; top-$p$ keeps the smallest set of most probable tokens whose probability mass reaches $p$, and top-$k$ the $k$ most probable tokens.}; prompts of up to $4096$ tokens followed by one assistant turn of up to $256$ generated tokens in Phase~1 and $1024$ in Phase~2; $\sim$$4.4\times10^{4}$ states per round in Phase~1), scores them by teacher-forcing\footnote{The teacher is run over the student's generated tokens, so its next-token distribution is read at every position without the teacher generating anything.} with the teacher's top-$32$ log-probabilities, and updates \emph{all} backbone weights by one pass of reverse-KL training over the round's rollouts (bf16, 8-bit AdamW, learning rate $10^{-6}$ held constant after a $3\%$ warm-up, gradient clipping at $1$; sequences of up to $8192$ tokens at an effective batch of $64$ in Phase~1, up to $9216$ tokens at $128$ in Phase~2) (Algorithm~\ref{alg:opd}). Distillation uses no adapter: the only LoRA in \ctzero{} is the decision adapter of Table~\ref{tab:elements}, trained afterwards together with the head. The corpus is the CAI dataset \cite{mayoralvilches2026caidataset}, Alias Robotics' proprietary corpus of cybersecurity trajectories collected with the open-source CAI framework \cite{aliasrobotics2025cai}, augmented with DAgger-harvested reasoning-and-action states on a fixed substrate of $55$ CTF tasks; the evaluation benchmark is deny-listed, so no \texttt{cybench} content is seen in training. Because a single schedule either under-distils or overfits, training runs in two chained phases. \emph{Phase 1} applies twelve rounds on the full corpus and turns the open base into a first specialist, the Phase-1 candidate. \emph{Phase 2} starts from that candidate with a filtered DAgger schedule \cite{ross2011dagger} that mixes freshly harvested on-policy states ($\sim$$4{\times}10^{3}$, up-weighted) with curated CAI states ($1.6\times10^{4}$), $5{,}000$ states per round through five acceptance filters, for a handful of rounds; its strongest checkpoint is \aliasmicro{}. Applying the Phase-2 recipe directly to the open base does not reach the checkpoint started from the Phase-1 candidate: sequence matters more than volume, which is why training runs in two phases.

\paragraph{Training budget.} Phase~1 trained on about $1.1\times10^{9}$ tokens over its twelve rounds for $\approx$$210$ RTX PRO 6000 GPU-hours; the two Phase-2 rounds of the delivered path on about $1.1\times10^{7}$ tokens for $\approx$$12$ hours, the full Phase-2 sweep on about $10^{8}$ tokens for $\approx$$104$; the decision head and its adapter take about one hour on one GPU with the backbone frozen. Only the generated turn carries the KL loss. The delivered model therefore costs $\approx$$222$ GPU-hours and the full sweep $\approx$$314$: roughly $2.5$--$3.5$ days on one $4\times$RTX PRO 6000 workstation, or about \$$220$--$630$ at current on-demand rental prices for that GPU (\$$1$--$2$ per GPU-hour \cite{getdeploying2026rtxpro6000}). A house can turn an open $4$\,B base into its own private cybersecurity specialist in a few GPU-days on one workstation; the economic asymmetry, not a leaderboard number, is the point.

%% file: tex/eval_overview.tex
\begin{figure*}[p]
\centering
\input{tex/evalmap_links.tex}
\caption{\textbf{Table~\ref{tab:evidence-overview} as a picture.} Every percentage-scale row of Table~\ref{tab:evidence-overview}, drawn from the same generated data. Each row has two lanes. The upper lane joins the \basemodel{} base (pale circle) to \ctzero{} (star; the best of its arms, as bolded in the table), so the line is the improvement from base to \aliasmicro{}; it is drawn in red where \ctzero{} is worse than the base. The lower lane holds the larger references, \aliasmini{} (filled circle) and alias3-max (open circle, reference conditions). The right-hand columns print the \ctzero{} value and its gain over the base, signed in the preferred direction (rows marked $\downarrow$ are better when lower; negative gains in red). AUROC and ECE are shown $\times100$; wall-time, throughput and path-count rows remain in the table only. Base and max pipelines are gated by the \ctzero{} head. The link after each band and row opens the subsection that develops it, as in the table.}
\label{fig:evalmap}
\end{figure*}
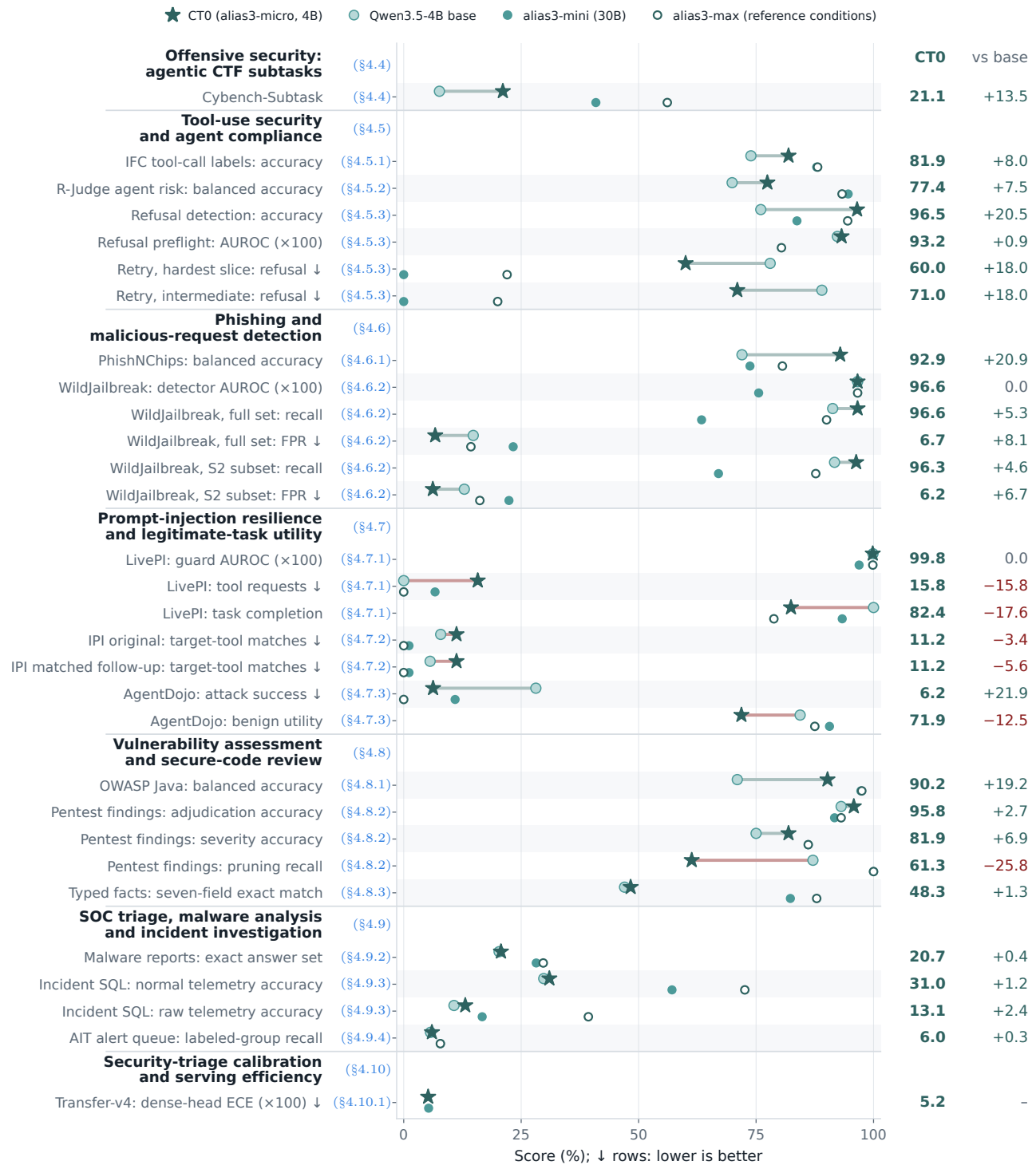

This section gathers every measurement under one protocol statement (Section~\ref{sec:protocol}), states the claims it tests and the mechanism behind them (Sections~\ref{sec:thesis} and~\ref{sec:mechanism}), and then reports the evidence by security job, in the order of the bands of Table~\ref{tab:evidence-overview}: offensive security (Section~\ref{sec:results}), tool-use security and agent compliance (\ref{sec:job-tooluse}), phishing and malicious-request detection (\ref{sec:job-phishing}), prompt-injection resilience (\ref{sec:job-injection}), vulnerability assessment and secure-code review (\ref{sec:job-vuln}), security-operations-centre (SOC) triage and incident investigation (\ref{sec:job-soc}) and security-triage calibration and serving efficiency (\ref{sec:job-calibration}). The benchmarks measure different quantities: solved subtasks, classification accuracy, balanced accuracy (BA)\footnote{The mean of recall on the positive class and on the negative class; $50\%$ for a classifier that labels every case the same way, so class imbalance does not inflate it.}, detector ranking and verified agent outcomes. Table~\ref{tab:evidence-overview} gathers every recorded model/mode comparison and marks missing cells explicitly; Figure~\ref{fig:evalmap} draws the same table. Each result is tied to its original checkpoint and protocol, and no cell is filled by inference. The grey subscripts in Table~\ref{tab:evidence-overview} are per-request latencies recorded under each campaign's own serving conditions and are compared only within a row.

%% file: tex/evalmap_links.tex
\sbox{\evfigbox}{\includegraphics[width=\textwidth,height=.85\textheight,keepaspectratio]{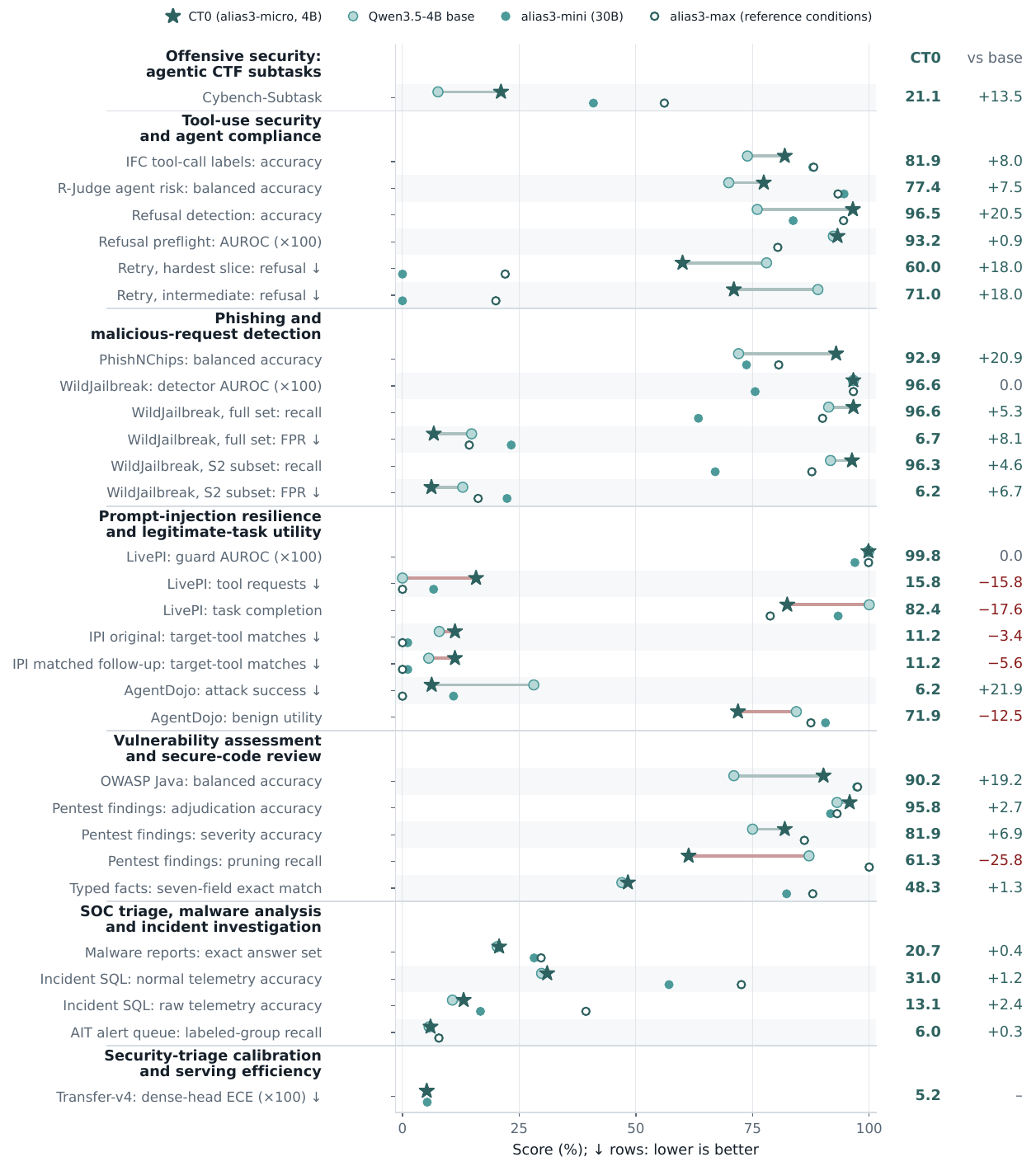}}%
\setlength{\unitlength}{\wd\evfigbox}%
\usebox{\evfigbox}%
\begin{picture}(0,0)
\put(-0.6226,1.0671){\makebox(0,0)[r]{\scriptsize\hyperref[sec:results]{(\S\ref*{sec:results})}}}
\put(-0.6226,1.0362){\makebox(0,0)[r]{\scriptsize\hyperref[sec:results]{(\S\ref*{sec:results})}}}
\put(-0.6226,1.0052){\makebox(0,0)[r]{\scriptsize\hyperref[sec:job-tooluse]{(\S\ref*{sec:job-tooluse})}}}
\put(-0.6226,0.9743){\makebox(0,0)[r]{\scriptsize\hyperref[sec:de-ifc]{(\S\ref*{sec:de-ifc})}}}
\put(-0.6226,0.9485){\makebox(0,0)[r]{\scriptsize\hyperref[sec:uc-rjudge]{(\S\ref*{sec:uc-rjudge})}}}
\put(-0.6226,0.9227){\makebox(0,0)[r]{\scriptsize\hyperref[sec:refusal]{(\S\ref*{sec:refusal})}}}
\put(-0.6226,0.8969){\makebox(0,0)[r]{\scriptsize\hyperref[sec:refusal]{(\S\ref*{sec:refusal})}}}
\put(-0.6226,0.8711){\makebox(0,0)[r]{\scriptsize\hyperref[sec:refusal]{(\S\ref*{sec:refusal})}}}
\put(-0.6226,0.8453){\makebox(0,0)[r]{\scriptsize\hyperref[sec:refusal]{(\S\ref*{sec:refusal})}}}
\put(-0.6226,0.8143){\makebox(0,0)[r]{\scriptsize\hyperref[sec:job-phishing]{(\S\ref*{sec:job-phishing})}}}
\put(-0.6226,0.7834){\makebox(0,0)[r]{\scriptsize\hyperref[sec:uc-phishing]{(\S\ref*{sec:uc-phishing})}}}
\put(-0.6226,0.7576){\makebox(0,0)[r]{\scriptsize\hyperref[sec:de-wj]{(\S\ref*{sec:de-wj})}}}
\put(-0.6226,0.7318){\makebox(0,0)[r]{\scriptsize\hyperref[sec:de-wj]{(\S\ref*{sec:de-wj})}}}
\put(-0.6226,0.7060){\makebox(0,0)[r]{\scriptsize\hyperref[sec:de-wj]{(\S\ref*{sec:de-wj})}}}
\put(-0.6226,0.6802){\makebox(0,0)[r]{\scriptsize\hyperref[sec:de-wj]{(\S\ref*{sec:de-wj})}}}
\put(-0.6226,0.6544){\makebox(0,0)[r]{\scriptsize\hyperref[sec:de-wj]{(\S\ref*{sec:de-wj})}}}
\put(-0.6226,0.6235){\makebox(0,0)[r]{\scriptsize\hyperref[sec:job-injection]{(\S\ref*{sec:job-injection})}}}
\put(-0.6226,0.5925){\makebox(0,0)[r]{\scriptsize\hyperref[sec:livepi-offline]{(\S\ref*{sec:livepi-offline})}}}
\put(-0.6226,0.5667){\makebox(0,0)[r]{\scriptsize\hyperref[sec:livepi-offline]{(\S\ref*{sec:livepi-offline})}}}
\put(-0.6226,0.5409){\makebox(0,0)[r]{\scriptsize\hyperref[sec:livepi-offline]{(\S\ref*{sec:livepi-offline})}}}
\put(-0.6226,0.5151){\makebox(0,0)[r]{\scriptsize\hyperref[sec:ipi-offline]{(\S\ref*{sec:ipi-offline})}}}
\put(-0.6226,0.4893){\makebox(0,0)[r]{\scriptsize\hyperref[sec:ipi-offline]{(\S\ref*{sec:ipi-offline})}}}
\put(-0.6226,0.4635){\makebox(0,0)[r]{\scriptsize\hyperref[sec:op-dojo]{(\S\ref*{sec:op-dojo})}}}
\put(-0.6226,0.4378){\makebox(0,0)[r]{\scriptsize\hyperref[sec:op-dojo]{(\S\ref*{sec:op-dojo})}}}
\put(-0.6226,0.4068){\makebox(0,0)[r]{\scriptsize\hyperref[sec:job-vuln]{(\S\ref*{sec:job-vuln})}}}
\put(-0.6226,0.3759){\makebox(0,0)[r]{\scriptsize\hyperref[sec:uc-java]{(\S\ref*{sec:uc-java})}}}
\put(-0.6226,0.3501){\makebox(0,0)[r]{\scriptsize\hyperref[sec:de-pentest]{(\S\ref*{sec:de-pentest})}}}
\put(-0.6226,0.3243){\makebox(0,0)[r]{\scriptsize\hyperref[sec:de-pentest]{(\S\ref*{sec:de-pentest})}}}
\put(-0.6226,0.2985){\makebox(0,0)[r]{\scriptsize\hyperref[sec:de-pentest]{(\S\ref*{sec:de-pentest})}}}
\put(-0.6226,0.2727){\makebox(0,0)[r]{\scriptsize\hyperref[sec:op-facts]{(\S\ref*{sec:op-facts})}}}
\put(-0.6226,0.2417){\makebox(0,0)[r]{\scriptsize\hyperref[sec:job-soc]{(\S\ref*{sec:job-soc})}}}
\put(-0.6226,0.2108){\makebox(0,0)[r]{\scriptsize\hyperref[sec:op-malware]{(\S\ref*{sec:op-malware})}}}
\put(-0.6226,0.1850){\makebox(0,0)[r]{\scriptsize\hyperref[sec:op-investigation]{(\S\ref*{sec:op-investigation})}}}
\put(-0.6226,0.1592){\makebox(0,0)[r]{\scriptsize\hyperref[sec:op-investigation]{(\S\ref*{sec:op-investigation})}}}
\put(-0.6226,0.1334){\makebox(0,0)[r]{\scriptsize\hyperref[sec:op-alerts]{(\S\ref*{sec:op-alerts})}}}
\put(-0.6226,0.1024){\makebox(0,0)[r]{\scriptsize\hyperref[sec:job-calibration]{(\S\ref*{sec:job-calibration})}}}
\put(-0.6226,0.0715){\makebox(0,0)[r]{\scriptsize\hyperref[sec:de-calibration]{(\S\ref*{sec:de-calibration})}}}
\end{picture}%

%% file: tex/protocol.tex
Every number in this section follows a protocol frozen before the test requests: a fixed inference budget for the agentic benchmark, and frozen prompts, thresholds, seeds and input hashes for the decision benchmarks and use cases, whose settings are given with each result and pinned in Appendix~\ref{app:repro}. Where arms share inputs we report paired bootstrap intervals\footnote{A $95\%$ interval for the difference between two arms, obtained by resampling the shared test cases with replacement ($2{,}000$ times here) and recomputing the difference on each resample.}; where they do not, we say so. Table~\ref{tab:sources} (Appendix~\ref{app:sources}) lists every benchmark behind Table~\ref{tab:evidence-overview} with its source, the data used here and the subsection that develops it; the three campaign protocols shared by the decision benchmarks follow the reference conditions below.

\paragraph{A fixed budget for the agentic benchmark.} An agentic score is only meaningful relative to a disclosed inference budget, because success rises with test-time compute (samples, reasoning tokens, turns, context): a Cybench number reported without a budget measures the budget as much as the model. We use the upstream Cybench Subtask decomposition \cite{cybench2024}, $40$ CTF challenges split into $171$ ordered subtasks with known answers, scored by exact match and micro-averaged, under the deliberately small budget of Table~\ref{tab:protocol}: one attempt, at most five steps, no extended thinking. Each number is therefore a floor that a third party can reproduce with the public benchmark and an equivalent scaffold. The earlier whole-challenge protocol of this programme, and its saturation, are recorded in Appendix~\ref{app:cybench-full}.

\begin{table}[t]
\centering\small
\setlength{\tabcolsep}{5pt}
\renewcommand{\arraystretch}{1.18}
\caption{The fixed \texttt{cybench-subtask} evaluation budget for the local thinking-off comparisons: single-attempt (pass@1) under a small, disclosed budget, the opposite of an unbounded best-of-$k$ evaluation.}
\label{tab:protocol}
\begin{tabular}{@{}lp{0.60\linewidth}@{}}
\toprule
Item & Value \\
\midrule
Benchmark      & Cybench Subtask mode (upstream), $40$ challenges $\to$ $171$ subtasks \\
Metric         & micro-average $=$ correct subtasks $/\,171$ \\
Scoring        & exact-answer match per subtask (guided: answer revealed after each) \\
Attempts       & \textbf{single attempt per subtask (pass@1)}; no best-of-$k$ \\
Iterations     & $\le 5$ ReAct steps (LLM calls) per subtask \\
Token budget   & $200$k input $/\,32$k output \\
Context        & rolling retention cap $300/300$, fixed \\
Sampling       & $T{=}1.0$, top-$p\,0.95$, top-$k\,20$, min-$p\,0$, presence $1.5$, rep.\ $1.0$ \\
Thinking       & \textbf{off} (harness-forced) unless a column states otherwise; thinking-on experiments in Appendix~\ref{app:thinking} \\
Scaffold       & Cybench SimpleAgent (ReAct; shell/bash in the challenge Docker), fixed across models \\
Serving        & SGLang, bf16 checkpoint \\
Campaigns      & single-attempt campaign per model \\
Contamination  & Cybench deny-listed from training ($0$ overlap, verified by task name) \\
\bottomrule
\end{tabular}
\end{table}

\paragraph{Why fixed budgets matter.} Frontier model cards report Cybench under conditions that inflate the number and cannot be reproduced as stated: OpenAI's GPT-5.x system cards score CTF/Cybench as pass@$12$ over $16$ rollouts \cite{openai_gpt55_systemcard,openai_gpt52_systemcard}, and Anthropic reports it with extended thinking enabled \cite{anthropic_opus45_systemcard,anthropic_claude37_systemcard}. These are legitimate upper bounds for risk governance, but they omit the per-task token and turn budget, scaffold and sampling needed to reconstruct them \cite{hal2025holistic}, and a best-of-$k$ extended-thinking number cannot share an axis with a fixed-budget pass@$1$ number: the gap between the two, for the same model, is exactly the test-time-scaling term we hold fixed.

\paragraph{Reference conditions.} The local Alias figures in Figure~\ref{fig:hero} are single-attempt campaigns under the budget of Table~\ref{tab:protocol}, never a best-of-$k$ within a run. The hero figure also carries references whose conditions differ and are marked as such. \aliasmini{} and alias3-max are specified as 30B and 800B FP8 (nominal weights 30/800 GB; 36/960 GB with the declared $20\%$ loading allowance, context cache additional); the mini score is retained from an earlier alias3-mini campaign served in BF16, and the max score ($56.1\%$) is measured without extended thinking, like the local models, and is not a measurement of a \ctzero{}-trained model; with thinking on it moves by one subtask (Appendix~\ref{app:thinking}). Opus-5.5 solves $50/171$ ($29.2\%$) in one API campaign with the same ReAct\footnote{A reasoning-and-acting loop: at each step the model writes a thought and a shell command, observes the output and continues.} scaffold over $38$ challenge environments. Mistral Large 4 ($1$T total, $49$B active parameters, API-served) solves $93/171$ ($54.4\%$) in one campaign with the same scaffold over all $40$ environments, with and without extended thinking alike (Appendix~\ref{app:thinking}). Kolibri-1 \cite{alephalpha2026kolibri1}, an open $78$B mixture-of-experts model with $3.5$B active parameters served locally in FP8 with the same scaffold, scores $9/171$ ($5.3\%$) on all $40$ challenges; the upstream first-match answer parser gives $4/171$, and the reported figure applies a last-line ``Answer:'' rule to the same transcripts. A transcript audit attributes the result to behaviour under the budget rather than to serving: $117/171$ subtasks end unanswered after five steps, $22\%$ of commands repeat the previous one verbatim, and narrated ``answer:'' text accounts for $41$ of its $50$ wrong answers, $27$ of them closed at the first step before any command ran. The same parser applies to every model under this protocol.

\paragraph{Decision benchmarks.} The decision benchmarks of Sections~\ref{sec:job-tooluse}--\ref{sec:job-calibration} evaluate the decision head and its interaction with System~2 on the deployed checkpoint (dense decision adapter, calibrated pointer head, steering disabled). File and finding corpora are constructed fixtures; WildJailbreak is a public evaluation set. The base, \aliasmini{} and alias3-max columns of Table~\ref{tab:evidence-overview} repeat every benchmark on the same frozen inputs and prompts, with the same test membership and a frozen policy per actor: mini with its own decision head, base and alias3-max generating under \ctzero{}\textquotesingle s gate.

\paragraph{Selective-review tasks.} Three selective-review tasks test whether the two modes complement each other on public security data: email triage, review of agent interactions and prioritisation of source-code findings, three offline classification tasks in which a positive decision requests review (no email link, agent action or generated code is executed). Typed decisions already serve incident triage and tool gates in Jev-style workflows \cite{huang2026jevusecases}, exploratory pentest evidence remains too small for general claims \cite{barbosa2026decisionpentest}, REFLEX motivates selective generation while documenting task-dependent advantages \cite{wu2026reflex}, and aggregate calibration and complementary errors do not alone make automation safe \cite{liu2026systemonesecurity}; we therefore report quality, missed threats, false alarms and the fraction of cases that need generation together.

\begin{figure*}[t]  %
\centering
\includegraphics[width=\textwidth]{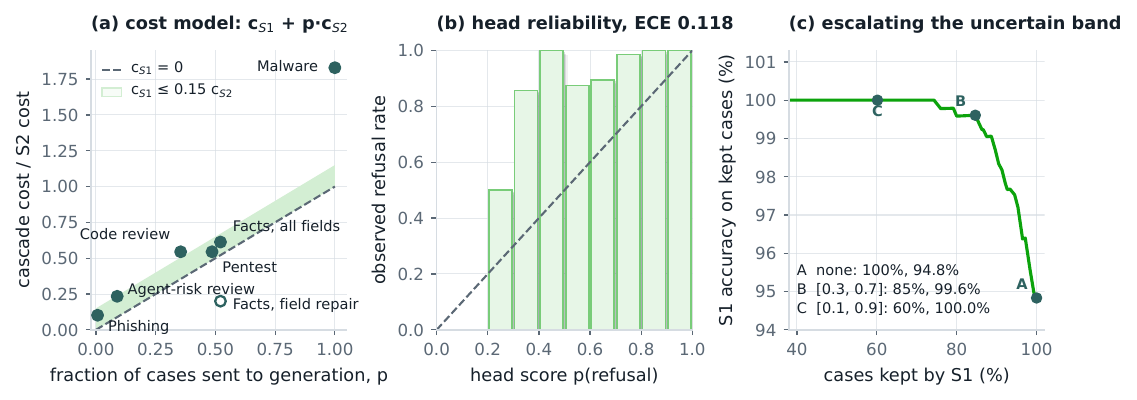}
\caption{\textbf{Why the fast path pays.} (a) Measured cascade cost relative to full generation against the fraction of cases sent to generation; the dashed line is Eq.~\eqref{eq:cost} with a free decision, the band allows a decision costing up to $15\%$ of a generation. The hollow point generates only uncertain fields; the top point adds an evidence-selection pass before generating for every case. (b) Reliability of the deployed head on $600$ real responses from the refusal follow-up ($300$ aggressive and $300$ benign prompts answered by the base actor); $260$ of the $600$ scores fall below $0.1$ and $101$ above $0.9$. (c) The same $600$ cases: answering with System~1 only when its score lies outside a band around $0.5$ and escalating the rest raises the accuracy of what the head keeps; the key gives, per band, the share of cases kept and their accuracy.}
\label{fig:mechanism}
\end{figure*}

\paragraph{Three public adaptations.} PhishNChips v5.2 supplies $2{,}000$ emails with generated bodies and public URL labels \cite{arelit2026phishnchips}; we keep the upstream study's $1{,}000/1{,}000$ selection/test membership and its non-neural domain/hosting control \cite{anisselbd2026phishing}, splitting the selection half into $500$ fit and $500$ routing-validation cases. R-Judge supplies annotated agent interactions \cite{yuan2024rjudge}; we reuse the published security study's $100$ development, $233$ validation and $236$ test cases. OWASP BenchmarkJava supplies expected vulnerability labels \cite{owaspbenchmarkjava}; we keep $625$ cases on cryptography, hashing, randomness and cookie security, exclude $417$ that need external helpers, group exact normalised templates and split $250/124/251$. We use it as a static-review adaptation restricted to these four families.

\paragraph{Matched modes and frozen routing.} The same deployed checkpoint and head serve both modes in bf16\footnote{16-bit brain floating point, the precision at which every local model is served; FP8 is an 8-bit floating-point format used by the larger references, and ternary weights take one of three values.} without steering; System~2 receives the same evidence and returns a short explanation and a binary JSON label (temperature $0$, seed $42$, thinking off, $384$-token limit). \emph{Direct S1} thresholds a single typed risk score at $0.5$; \emph{fitted S1} is an L2 logistic combiner of four to six fixed typed signals trained on the fit labels only, with its threshold and low-margin escalation rule selected on validation balanced accuracy (at most $50\%$ validation escalation, fewer calls preferred on ties). Input hashes, prompts, coefficients and policies are frozen before any test request (Appendix~\ref{app:repro}).

\paragraph{Operational pipelines.} Six further workflows (typed facts, malware reports, incident investigation, guarded agents, file screening, alert-queue triage and source/sink inference ahead of CodeQL) use the unchanged deployed checkpoint and head, \emph{without new training or fitted combiners}, which separates specialisation through weights from task adaptation through prompts, interfaces and control flow. Every composed arm executes its constituent model calls; SQL and code analysis execute real local tools and AgentDojo its original sandbox tools. Appendix~\ref{app:ct0-operational} documents the adaptations in detail.

%% file: tex/eval_thesis.tex
Four properties of the design in Section~\ref{sec:architecture} organise the benchmarks that follow, each delivered by elements of Table~\ref{tab:elements}. The typed interface \badge{1} and the bilinear pointer head \badge{4} answer typed security decisions (is this response a refusal, is this request an attack, is this finding prunable) in one prefill instead of a decode, and are a better instrument for them than asking the same model to judge generatively. The escalation gate \badge{5} thresholds the head\textquotesingle s calibrated confidence so that only the uncertain cases go to generation, which keeps the quality of the better mode at a fraction of its cost. The decision adapter \badge{3} is fused into the specialised weights \badge{2} and served from one process \badge{6}, at no measurable cost to the agentic path. And the specialist is cheap to build and to run: about $314$ RTX PRO 6000 GPU-hours for the full training sweep and one workstation GPU to serve both modes (Sections~\ref{sec:opd} and~\ref{sec:de-serving}). Whether the bilinear head also beats other one-forward-pass readouts is a separate question, taken up in Section~\ref{sec:discussion}.

Every benchmark below is read the same way, and each subsection opens by stating what is measured (a task with one metric from Table~\ref{tab:evidence-overview}: accuracy, balanced accuracy, recall at a false-positive rate\footnote{Recall is the share of true positives the detector flags; the false-positive rate (FPR) is the share of negatives it wrongly flags; a detector is read at a threshold that fixes one of the two.}, AUROC\footnote{The area under the receiver-operating-characteristic curve: the probability that a randomly chosen positive is scored above a randomly chosen negative, $0.5$ for a random detector and $1$ for a perfect one; it summarises ranking quality over all thresholds.} or an outcome count), who is compared (\ctzero{}'s three arms, System~1 (S1, the head alone), System~2 (S2, generation alone) and the cascade (S1$\to$S2, the head deciding what to generate), against the unspecialised base and, where available, the larger alias3-mini and alias3-max references) and the verdict: whether the fast path helps, and at what cost.

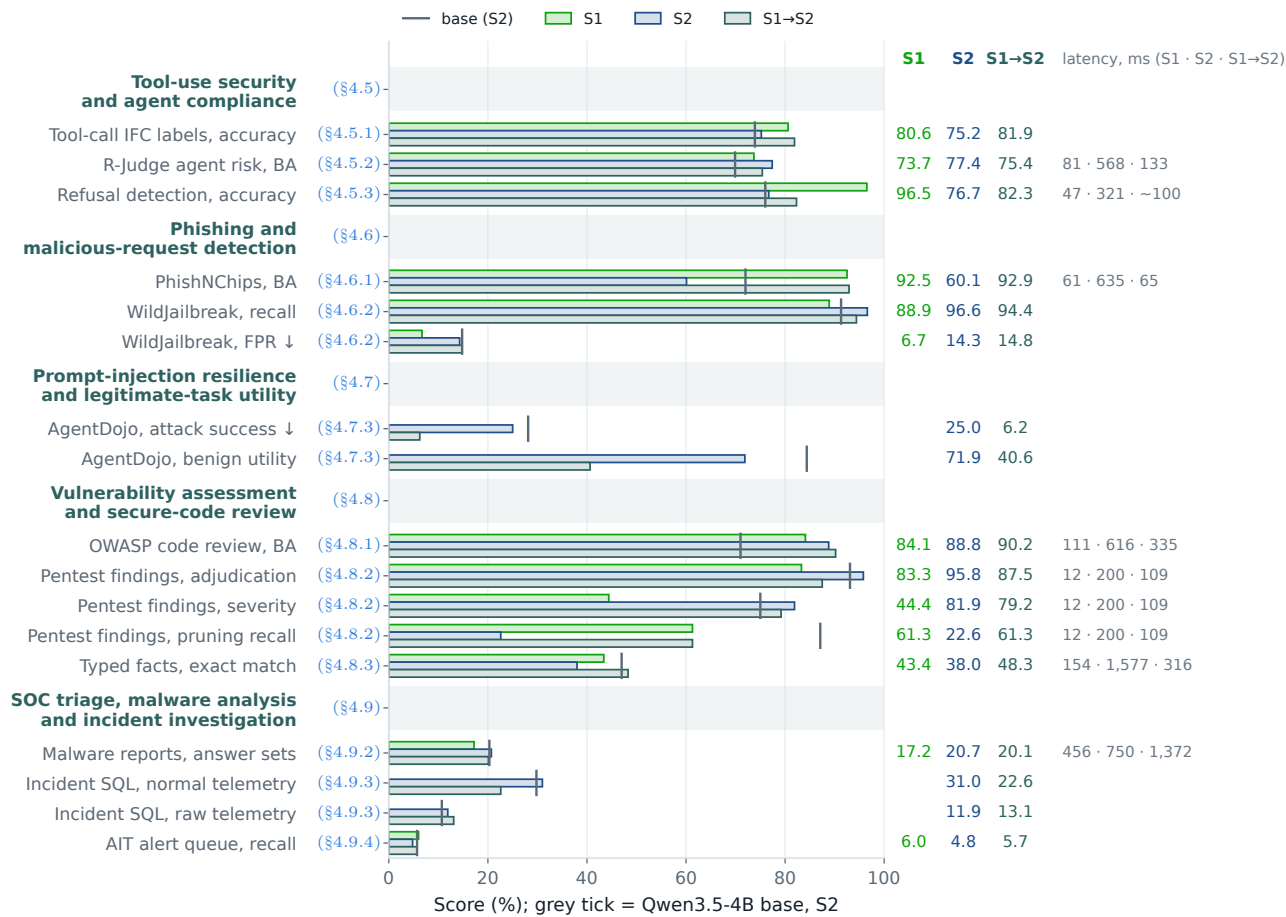
\begin{figure*}[t]
\centering
\input{tex/modes_links.tex}
\caption{\textbf{\ctzero{}'s three arms on every task where all are measured.} Green: System~1, the decision head in one prefill; blue: System~2, generation; teal: the cascade S1$\to$S2. Grey ticks mark the unspecialised base generating, and grey text gives the mean per-request latencies where they were recorded (Table~\ref{tab:evidence-overview}). Rows marked $\downarrow$ are better when lower (false-positive and attack-success rates). Rows are grouped by the bands of Table~\ref{tab:evidence-overview}, and the link after each band and row opens the subsection that develops it.}
\label{fig:modes}
\end{figure*}

Figure~\ref{fig:modes} is the summary: the three arms on every task where all are measured, with the base as a tick and the recorded per-request latency beside each row. Three patterns organise it. Where the decision is a detection, the head wins outright (refusal detection, phishing with fitted signals, tool-call labels) and does so in tens of milliseconds. Where the decision needs the evidence to be read and reasoned about (pentest adjudication, agent-risk review, incident SQL), generation stays stronger and the cascade lands between the two while generating for only part of the cases. Where the two modes are comparably good and err on different cases (code review, typed facts), the cascade beats both, which is the behaviour the calibrated gate predicts; Section~\ref{sec:mechanism} explains why, and what it costs.

%% file: tex/modes_links.tex
\sbox{\evfigbox}{\includegraphics[width=\textwidth]{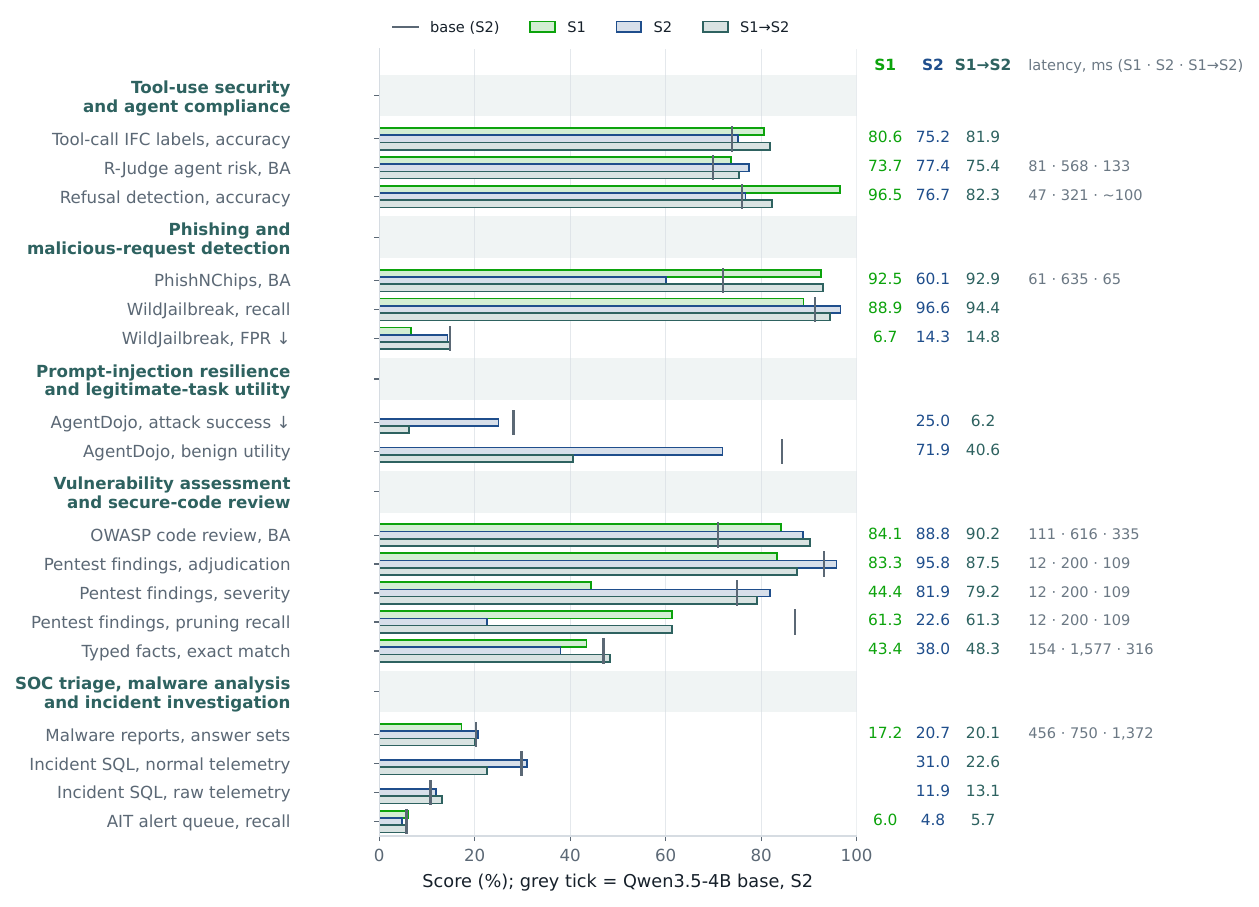}}%
\setlength{\unitlength}{\wd\evfigbox}%
\usebox{\evfigbox}%
\begin{picture}(0,0)
\put(-0.7042,0.6448){\makebox(0,0)[r]{\scriptsize\hyperref[sec:job-tooluse]{(\S\ref*{sec:job-tooluse})}}}
\put(-0.7042,0.6103){\makebox(0,0)[r]{\scriptsize\hyperref[sec:de-ifc]{(\S\ref*{sec:de-ifc})}}}
\put(-0.7042,0.5873){\makebox(0,0)[r]{\scriptsize\hyperref[sec:uc-rjudge]{(\S\ref*{sec:uc-rjudge})}}}
\put(-0.7042,0.5643){\makebox(0,0)[r]{\scriptsize\hyperref[sec:refusal]{(\S\ref*{sec:refusal})}}}
\put(-0.7042,0.5321){\makebox(0,0)[r]{\scriptsize\hyperref[sec:job-phishing]{(\S\ref*{sec:job-phishing})}}}
\put(-0.7042,0.4976){\makebox(0,0)[r]{\scriptsize\hyperref[sec:uc-phishing]{(\S\ref*{sec:uc-phishing})}}}
\put(-0.7042,0.4746){\makebox(0,0)[r]{\scriptsize\hyperref[sec:de-wj]{(\S\ref*{sec:de-wj})}}}
\put(-0.7042,0.4516){\makebox(0,0)[r]{\scriptsize\hyperref[sec:de-wj]{(\S\ref*{sec:de-wj})}}}
\put(-0.7042,0.4194){\makebox(0,0)[r]{\scriptsize\hyperref[sec:job-injection]{(\S\ref*{sec:job-injection})}}}
\put(-0.7042,0.3849){\makebox(0,0)[r]{\scriptsize\hyperref[sec:op-dojo]{(\S\ref*{sec:op-dojo})}}}
\put(-0.7042,0.3619){\makebox(0,0)[r]{\scriptsize\hyperref[sec:op-dojo]{(\S\ref*{sec:op-dojo})}}}
\put(-0.7042,0.3297){\makebox(0,0)[r]{\scriptsize\hyperref[sec:job-vuln]{(\S\ref*{sec:job-vuln})}}}
\put(-0.7042,0.2952){\makebox(0,0)[r]{\scriptsize\hyperref[sec:uc-java]{(\S\ref*{sec:uc-java})}}}
\put(-0.7042,0.2722){\makebox(0,0)[r]{\scriptsize\hyperref[sec:de-pentest]{(\S\ref*{sec:de-pentest})}}}
\put(-0.7042,0.2492){\makebox(0,0)[r]{\scriptsize\hyperref[sec:de-pentest]{(\S\ref*{sec:de-pentest})}}}
\put(-0.7042,0.2262){\makebox(0,0)[r]{\scriptsize\hyperref[sec:de-pentest]{(\S\ref*{sec:de-pentest})}}}
\put(-0.7042,0.2032){\makebox(0,0)[r]{\scriptsize\hyperref[sec:op-facts]{(\S\ref*{sec:op-facts})}}}
\put(-0.7042,0.1709){\makebox(0,0)[r]{\scriptsize\hyperref[sec:job-soc]{(\S\ref*{sec:job-soc})}}}
\put(-0.7042,0.1364){\makebox(0,0)[r]{\scriptsize\hyperref[sec:op-malware]{(\S\ref*{sec:op-malware})}}}
\put(-0.7042,0.1134){\makebox(0,0)[r]{\scriptsize\hyperref[sec:op-investigation]{(\S\ref*{sec:op-investigation})}}}
\put(-0.7042,0.0904){\makebox(0,0)[r]{\scriptsize\hyperref[sec:op-investigation]{(\S\ref*{sec:op-investigation})}}}
\put(-0.7042,0.0674){\makebox(0,0)[r]{\scriptsize\hyperref[sec:op-alerts]{(\S\ref*{sec:op-alerts})}}}
\end{picture}%

%% file: tex/eval_mechanism.tex
\paragraph{A cost model.} System~1 reads the whole request once (one prefill) and answers; System~2 reads it the same way and then writes its answer token by token, which is the expensive part; the cascade writes only for the fraction $p$ of cases in which the head is unsure. Writing $c_{S1}$ and $c_{S2}$ for the mean time per request of the two modes (the latencies of Table~\ref{tab:evidence-overview}) and $c_{S1\to S2}$ for that of the cascade,
\begin{equation}
c_{S1\to S2} \;=\; c_{S1} + p\,c_{S2},
\qquad\text{so}\qquad
\frac{c_{S1\to S2}}{c_{S2}} \;=\; p + \frac{c_{S1}}{c_{S2}},
\label{eq:cost}
\end{equation}
so the cascade is cheaper than generation whenever fewer than $1 - c_{S1}/c_{S2}$ of the cases are escalated. Figure~\ref{fig:mechanism}(a) checks this on every workload for which the three costs and $p$ were recorded: the measured points follow the predicted line, lifted by the cost of the decision itself, $c_{S1}/c_{S2}$, which is $0.06$ for pentest findings and $0.10$ for typed facts; in plain terms, asking the head costs a few percent of writing an answer. Two points stray from the line, and both are instructive. Selective repair of typed facts regenerates only the uncertain fields instead of the whole record, so it lands below the line: it escalates half the findings ($p=0.52$) yet costs a fifth of full generation. Evidence selection for malware reports sends every case to generation after a costly selection pass, so it lands above the line; that is the one row of Table~\ref{tab:evidence-overview} in which composition is slower than generation.

\paragraph{Calibration is what makes the gate usable.} The gate routes on the head's probability, so that probability has to mean what it says: a case scored $0.8$ should turn out positive about eight times in ten. Figure~\ref{fig:mechanism}(b) checks this on $600$ real responses from the refusal follow-up, bin by bin: at both ends of the scale, where almost all cases fall, the observed refusal rates match the scores; in the sparsely populated middle the observed rate runs above the score, so the head is under-confident there rather than over-confident (ECE $0.118$ on this set; $0.052$ on the lineage's transfer set after temperature fitting, Table~\ref{tab:evidence-overview}). Figure~\ref{fig:mechanism}(c) shows what this buys. If every case scored between $0.3$ and $0.7$ is handed to generation, $15\%$ of the cases are escalated and the head's accuracy on the $85\%$ it keeps rises from $94.8$ to $99.6\%$; widening the band to $[0.1,0.9]$ keeps $60\%$ of the cases, all of them answered correctly. The gate therefore buys accuracy on what the head keeps, at the cost given by Eq.~\eqref{eq:cost}; the thresholds themselves remain task-specific and are revalidated per deployment (Section~\ref{sec:de-wj}, Appendix~\ref{app:ct0-usecases}).

\begin{figure*}[t]
\centering
\includegraphics[width=\textwidth]{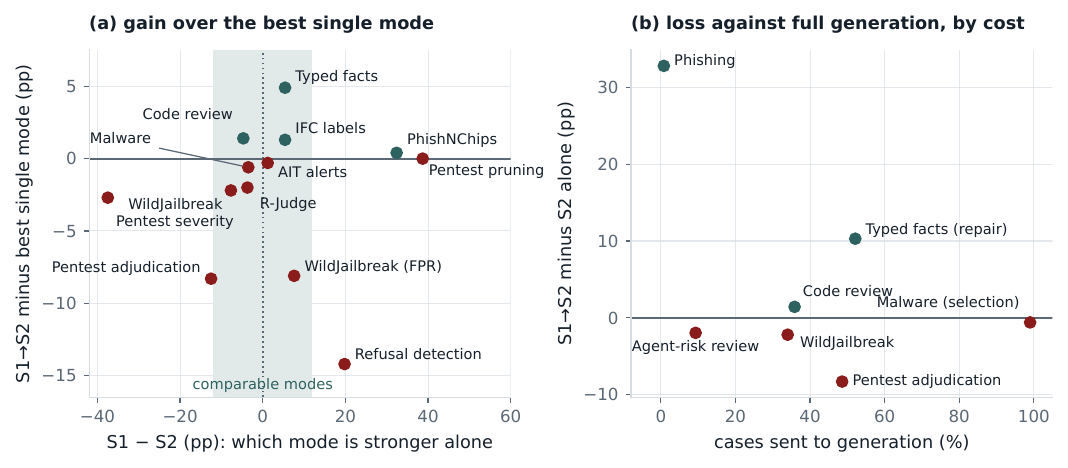}
\caption{\textbf{When composition helps, and what it costs.} (a) Every task with three measured arms, placed by how much stronger one mode is alone (horizontal; positive means the head) and by the cascade's gain over the best single mode (vertical); the shaded band holds the tasks where the modes are within $12$ points of each other. (b) The cascade against full generation, by the fraction of cases it generates for. Teal: the cascade gains; red: it loses.}
\label{fig:gate}
\end{figure*}

\paragraph{When composition helps.} Figure~\ref{fig:gate}(a) places every task with three measured arms by how much stronger one mode is alone and by how much the cascade gains over the best single mode. The gains cluster where the modes are comparable: typed facts ($+4.9$ pp), code review ($+1.4$), tool-call labels ($+1.3$). Where one mode dominates, the cascade sits between the two and loses to the stronger one, by a little for agent-risk review ($-2.0$) and by a lot where the head alone is already near-perfect (refusal detection, $-14.2$) or generation is (pentest adjudication, $-8.3$). The rule is the one the complementarity counts of Section~\ref{sec:uc-java} suggest: composition pays when the two modes err on different cases, not when one of them is simply better; when the head is the better mode, the right design is to use it alone and keep generation for the retry loop of Section~\ref{sec:refusal}. Figure~\ref{fig:gate}(b) prices the trade against full generation. The cascade gives up at most two points while generating for $9$--$36\%$ of the cases, gains outright on phishing and typed facts, and loses $8$ points on pentest adjudication, where half the findings are generated for and the head's value lies in its pruning decisions rather than its adjudication (Section~\ref{sec:de-pentest}).

%% file: tex/results.tex
The agentic benchmark measures correct subtasks out of $171$ under the fixed budget of Table~\ref{tab:protocol}, one attempt per subtask, and compares \ctzero{} generating (S2) with its own base, same-size open models, the alias series and hosted references (Figure~\ref{fig:hero}). The specialist reaches $21.1\%$ against $7.6\%$ for the base, and the alias series scales to $40.9$ and $56.1\%$.

\paragraph{Why subtasks.} Earlier Cybench comparisons in this programme scored whole challenges under pass@3\footnote{A challenge counts as solved if any of three attempts captures the flag; pass@1 allows a single attempt.} with hundreds of interactions per attempt (Appendix~\ref{app:cybench-full}); the strongest models now solve more than $80\%$ of the $33$ challenges under that protocol, which no longer separates them. We therefore adopted the benchmark's subtask mode \cite{cybench2024}: the same $40$ challenges decomposed into $171$ subtasks, one attempt each under the fixed budget of Section~\ref{sec:protocol}, which leaves the strongest reference at $56.1\%$ and makes a $4$\,B model's progress measurable.

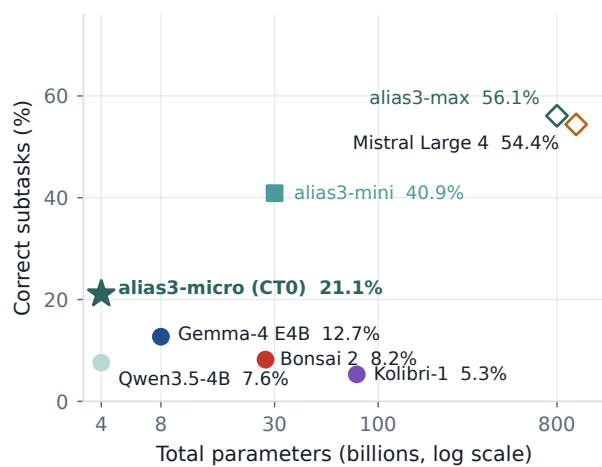
\begin{figure}[t]
\centering
\resizebox{\linewidth}{!}{\input{tex/fig_parameff.tex}}
\caption{\textbf{Capability and model size.} Cybench-Subtask score versus reported total parameter count on a logarithmic axis. Gemma E4B is 8B including embeddings (4.5B effective); Mistral Large 4 is placed at its disclosed $1$T total ($49$B active) and Kolibri-1 at its $78$B total ($3.5$B active); API models without a disclosed parameter count are excluded. \ctzero{} has the highest score per reported billion parameters in this set; mini and max achieve higher absolute scores. All campaigns follow the protocol of Section~\ref{sec:protocol}.}
\label{fig:parameff}
\end{figure}

\paragraph{A 4\,B specialist clears 21\%.} \ctzero{} solves $36/171$ ($21.1\%$) of \texttt{cybench-subtask}: roughly $2.8\times$ its own open base (\basemodel{}, $7.6\%$), above Gemma-4 E4B ($12.7\%$), and $2.6\times$ the external Bonsai~2 27B reference ($8.2\%$), a general-purpose 27\,B model compressed to ternary weights, at roughly one-seventh the parameters and a comparable serving footprint (Figure~\ref{fig:parameff}). The larger references are \aliasmini{} (30\,B FP8, $40.9\%$), alias3-max (800\,B FP8, $56.1\%$), Mistral Large 4 ($54.4\%$) and Opus-5.5 ($29.2\%$).

\paragraph{Reproducibility.} The $21.1\%$ result is reproducible under the fixed budget of Table~\ref{tab:protocol} and has been confirmed externally under the same protocol; per-campaign data are available on request.

Local Alias models are reported under the protocol of Section~\ref{sec:protocol}. The Bonsai~2 27B score is our own campaign (its authors report no cybersecurity benchmark); the hosted API references in Figure~\ref{fig:hero} are one campaign each under their disclosed conditions.

%% file: tex/fig_parameff.tex
\includegraphics[width=\linewidth]{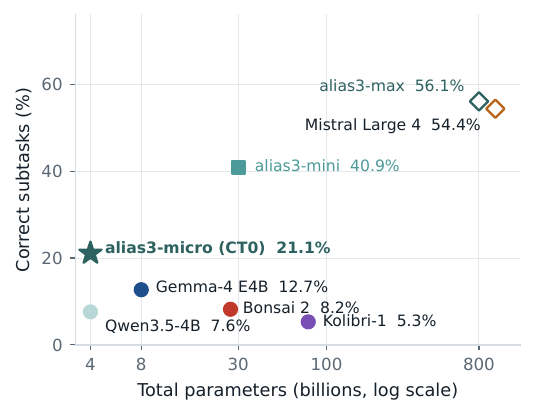}

%% file: tex/job_tooluse.tex
\subsubsection{Tool-call labels and dangerous-action recall}\label{sec:de-ifc}
Tool-call labelling is measured as accuracy on $464$ information-flow labels over $116$ tool calls, and as recall of the $37$ calls that require trusted input; System~1 without task training is compared with System~2 zero- and nine-shot and with the cascade. The head beats nine-shot generation at a fraction of the latency, the cascade adds a point while generating for two thirds of the calls, and the dangerous calls need their own, lower threshold.

\begin{figure}[t]
\centering
\includegraphics[width=\linewidth]{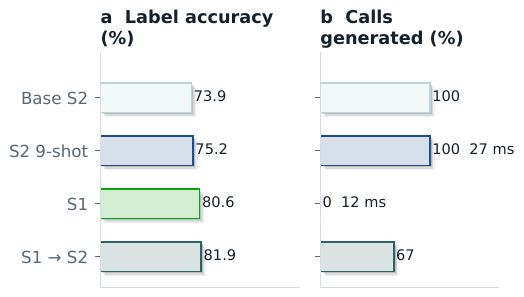}
\caption{\textbf{Tool-call labels}, the constructed $116$-call corpus ($464$ labels). (a) Information-flow label accuracy by arm; System~1 and the pipeline use the trained decision head without task-specific training. (b) The share of calls that generate, with the per-call latency of each mode.}
\label{fig:ifc}
\end{figure}

A constructed corpus of $116$ Claude-Code-style tool calls carries $464$ rule-derived labels on four information-flow-control (IFC) fields (provenance, audience, trusted-input requirement, audience requirement). Without task-specific training, System~1 reaches $80.6\%$ label accuracy at about \lat{12\,ms} per call; System~2 reaches $50.9\%$ zero-shot and $75.2\%$ nine-shot at about \lat{27\,ms} per call; the pipeline reaches $81.9\%$ while escalating $67\%$ of calls (Figure~\ref{fig:ifc}). The corpus is an in-house fixture that follows the information-flow formulation of Archestra's APPA \cite{kravchenko2026appa,archestra2026openappa}.

Aggregate accuracy hides the safety-relevant decision. For the $37$ calls that require trusted input, System~1's AUROC is $0.719$; at threshold $0.5$ recall is $0.24$ at precision $1.00$\footnote{Precision is the share of the calls the head flags that truly require trusted input.}, and lowering the threshold to $0.20$ raises recall to $0.76$ at precision $0.50$. The pipeline's recall is $0.60$. The authorisation decision therefore takes its own threshold, below the default argmax.

\subsubsection{Agent-risk review: fewer calls, lower quality}\label{sec:uc-rjudge}
Agent-risk review is measured as balanced accuracy on $236$ held-out agent-interaction records, comparing fitted S1, S2 and the cascade. Here generation is the better mode: the cascade improves the fast path while generating for $9\%$ of records, but lands two points below S2, trading false alarms for recall.

\begin{figure}[t]
\centering
\includegraphics[width=\linewidth]{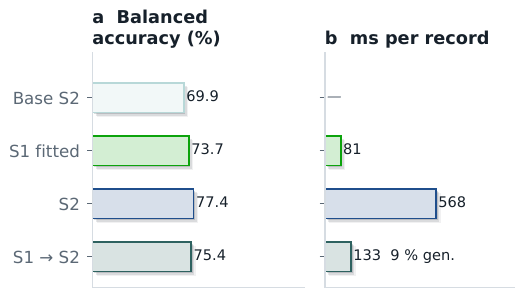}
\caption{\textbf{Agent-risk review}, $236$ held-out R-Judge records. (a) Balanced accuracy by arm; (b) mean time per record. The cascade generates for $9\%$ of records.}
\label{fig:wf-rjudge}
\end{figure}

On $236$ held-out R-Judge records, fitted S1 reaches $73.7\%$ BA, S2 $77.4\%$ and the cascade $75.4\%$ while escalating $22/236$ ($9.3\%$); the cascade-minus-S1 interval is $[-1.15,4.76]$ pp and cascade-minus-S2 $[-8.27,4.52]$ pp, so the cascade improves the fast path and lands two points below S2 (Figure~\ref{fig:wf-rjudge}). It detects $103/125$ risky records against S2's $91/125$ but flags $35/111$ safe records against $20/111$: a higher-recall, higher-false-alarm operating point suited to requesting review, not to interrupting a live agent.

\subsubsection{Refusal detection and dynamic de-refusal}\label{sec:refusal}
\input{tex/refusal_benchmarks.tex}

%% file: tex/refusal_benchmarks.tex
The refusal suite compares the S1 head, S2 judging its own output and the cascade on three measurements of one capability, with the base, mini and max actors under the same suite. The head detects refusals at $96.5\%$ in $47$\,ms where self-judging reaches $76.7\%$ in $321$\,ms and misses half of them; preflight reaches AUROC $0.932$ at $58$\,ms; and the retry loop lowers refusals on the hardest slice from $78$ to $60\%$.

\paragraph{The task.} A security operations pipeline needs to know, cheaply and reliably, whether the assistant just refused. We introduce three measurements of this one capability on \ctzero{}: \emph{RD-Response} (classify a completed response), \emph{RD-PreFlight} (predict from the prompt alone whether System~2 will refuse, before generating), and \emph{RD-PostFlight} (the operational loop: System~1 inspects System~2's first response and conditionally triggers an abliterated regeneration\footnote{Abliteration removes a learned refusal direction from the residual stream at inference, by a per-layer projection, leaving the weights untouched; here it is applied only when regenerating a detected refusal.}). The dataset is 600 real responses generated by \ctzero{} at temperature 0 to 300 prompts from our validated aggressive-cyber corpus and 300 benign controls; refusal labels come from an established marker classifier and were adjudicated by hand on all disagreements (282/600 refusals; \ctzero{} refuses $81\%$ of aggressive and $13\%$ of benign prompts, thinking-off).

\begin{figure}[t]
\centering
\includegraphics[width=\linewidth]{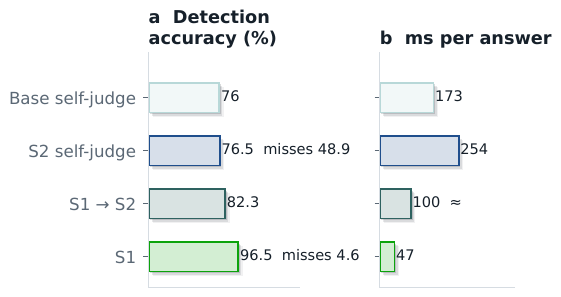}
\caption{\textbf{RD-Response} ($n{=}600$, adjudicated labels): detecting whether a completed \ctzero{} response is a refusal. (a) Accuracy by arm, with the share of refusals each misses; (b) mean time per answer. The System-1 head reads refusal almost perfectly (AUROC $0.996$) at negligible cost; the model judging its own output misses about half of its refusals.}
\label{fig:wf-refusal}
\end{figure}

\paragraph{RD-Response: the head sees what the model cannot.} Figure~\ref{fig:wf-refusal} shows the headline result of the suite. The System-1 head, queried zero-shot as a binary choice over the response text, detects refusals at $96.5\%$ accuracy (AUROC $0.996$) in \lat{47\,ms} per response, against \lat{254\,ms} for the generative self-judgement and about \lat{100\,ms} for the cascade --- including the \emph{partial} refusals (``I cannot provide $X$, but I can explain\ldots'') that dominate the error mass of a keyword matcher. System~2 prompted to judge the very response it just produced misses $48.9\%$ of refusals: the generative model is a poor judge of its own compliance. The naive cascade (System~1 decides outside an uncertainty band, System~2 judges inside it) lands in between ($82.3\%$) --- escalation only helps when the escalated judge is \emph{better} than the gate, which self-judging is not. The correct reading matches the architecture's thesis: refusal detection is a System-1 job.

\begin{figure}[t]
\centering
\includegraphics[width=\linewidth]{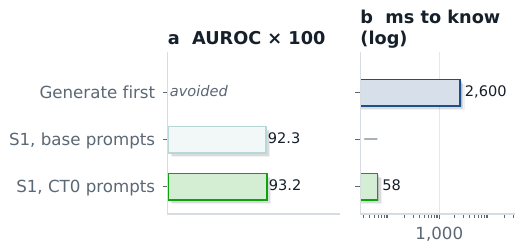}
\caption{\textbf{RD-PreFlight.} (a) AUROC of the refusal prediction from the prompt alone, on \ctzero{}'s prompts and, as a transfer test, on the base model's; (b) the cost of knowing: the $58$\,ms pre-flight against the $2.6$\,s generation it avoids.}
\label{fig:wf-preflight}
\end{figure}

\paragraph{RD-PreFlight: one forward pass before the model runs.} Pointing the same head at the \emph{prompt} instead of the response predicts whether System~2 will refuse with AUROC $0.932$ ($68.8\%$ recall at $5\%$ false-positive rate on benign prompts) at \lat{58\,ms} per prompt, roughly a fiftieth of the generation it can avoid. Pre-flight triage is therefore viable: prompts likely to be refused can be routed directly to the abliterated path or escalated to a stronger model, without paying for a doomed generation first (Figure~\ref{fig:wf-preflight}).

\paragraph{Two scales and an out-of-distribution transfer.} Repeating the measurement across models sharpened the picture (Table~\ref{tab:evidence-overview}). The untrained base refuses $87\%$ of aggressive prompts ($19\%$ of benign); \ctzero{} $81\%$/$13\%$; and the 30\,B \ctzero{}-fused specialist refuses \emph{zero of 300} aggressive prompts --- on this corpus, scale-plus-specialisation has already eliminated the refusal behaviour the suite measures, so detection is moot on its own outputs. Two transfer results anchor the head itself: the 4\,B head, scoring responses of the base model it never saw, detects refusals at AUROC $0.994$ ($94.8\%$ accuracy, $33$\,ms) --- an out-of-distribution test passed essentially unchanged --- and the 30\,B head over the same base-generated refusals reaches AUROC $0.996$ with $0.3\%$ false negatives (at $\sim$1.1\,s per readout, the cost of the larger forward pass). Refusal detection is thus a property of the bilinear readout, robust across generators and scales; pre-flight transfer agrees (AUROC $0.932$ on \ctzero{}, $0.923$ predicting the base).

\begin{figure}[t]
\centering
\resizebox{\linewidth}{!}{\input{tex/fig_rd.tex}}
\caption{RD-PostFlight results. \textbf{A}: end-to-end refusal rate on the two difficulty slices for every actor: the open base, \ctzero{} generating, \ctzero{} with the System-1-gated abliterated retry, and the larger alias3-mini and alias3-max (the retry rows of Table~\ref{tab:evidence-overview}). \textbf{B}: escalation ladder of the dynamic-abliteration regeneration on 80 known refusals; the cumulative flip rate saturates at $31\%$ with zero degenerate outputs at every rung.}
\label{fig:rd}
\end{figure}
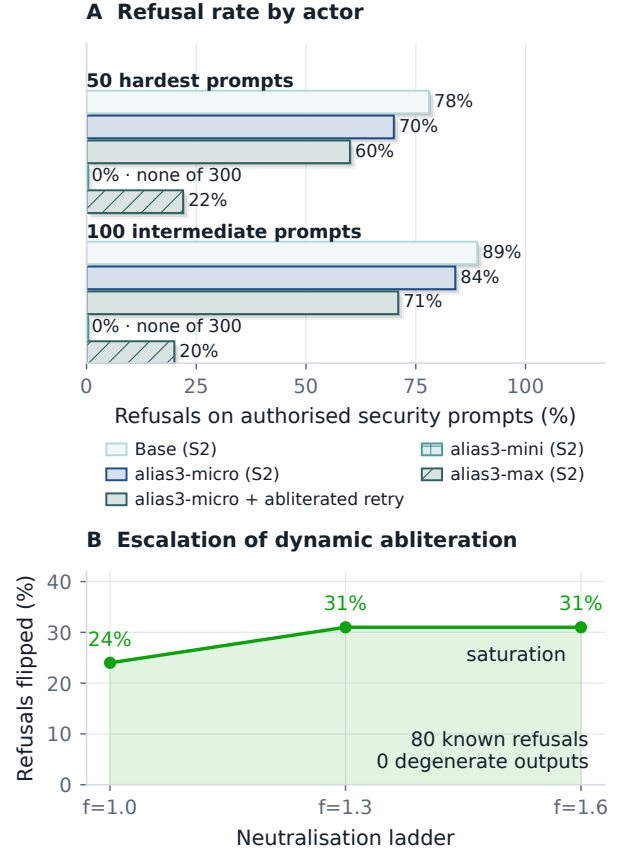

\paragraph{RD-PostFlight: the de-refusal loop, end to end.} The operational benchmark (Figure~\ref{fig:rd}) runs two prompt slices through two arms: \emph{A}, System~2 alone; \emph{B}, the pipeline --- System~2 generates, System~1 checks ($33$--$40$\,ms), and on a detected refusal the answer is regenerated under dynamic abliteration (a per-layer refusal direction obtained by a difference of means over a contrastive corpus, applied at runtime as the residual projection $h \leftarrow h - f\,(h\cdot\hat v)\hat v$ at every layer, $f{=}1.0$, weights untouched). The slices are the $50$ \emph{most aggressive} corpus prompts and $100$ \emph{intermediate} ones (mid-length band of the corpus). On the hardest slice, arm~A refuses $70\%$ and the pipeline $60\%$ ($5$ of $32$ detected refusals flipped); on the intermediate slice, arm~A refuses $84\%$ and the pipeline $71\%$ ($12$ of $81$ flipped). Across both slices the pattern is consistent: the gate catches essentially every refusal at negligible cost, the abliterated regeneration flips a stable $\sim$15\% of detected refusals into substantive answers, and \emph{zero} outputs degenerate. Two negative controls matter: the naive steering form (adding $s\,\lVert h\rVert\,\hat v$ at a single layer, the single-layer convention) degenerated on $19/33$ regenerations ($57.6\%$), and the projection mode did not --- the runtime form of abliteration is a subtraction of the refusal component, not an addition. The residual gap is regeneration quality on refusals that survive neutralisation. We measured the natural knob directly: on an $80$-prompt group of corpus-distribution refusals (half partial ``I~cannot~$X$, but I can explain\ldots'' refusals, half hard refusals), single-shot $f{=}1.0$ flips $24\%$; escalating the stubborn remainder to over-subtraction $f{=}1.3$ adds $6/61$ successful retries ($7.5$ percentage points of the original 80), and a further $f{=}1.6$ adds nothing --- the de-refusal of this vector saturates at $31\%$ of detected refusals, with \emph{zero} degenerate outputs at every rung. Overhead scales with the ladder: one regeneration ($\sim$2.6\,s) serves the first rung, a second only for the surviving $76\%$. Median end-to-end overhead is $0.2$\,s on the hardest slice and $1.7$\,s on the intermediate slice where $81\%$ of requests take the regeneration path.

\paragraph{Use case: selective retry during authorised evaluation.} The deployment story this enables is the one the corpus was built for: an authorised security evaluation whose prompts are, by design, aggressive. Running RD-PostFlight as the serving policy gives the evaluator a pipeline that detects 95\% of the refusals the model produces (4.6\% residual miss rate at 2.5\% false alarms), retries only those requests (64\% of the hardest traffic) through an abliterated runtime endpoint at a fixed, disclosed strength, and showed no degenerate outputs in these measured retries --- while leaving the deployed weights untouched and the de-refusal path auditable per request.

The recorded max response-generation follow-up refuses $19.7\%$ of aggressive and $2.3\%$ of benign prompts. The shared \ctzero{} head obtains AUROC $0.915$, accuracy $89.0\%$ and about $26$\,ms latency on those responses. Max self-judging has $88.6\%$ accuracy, $27.3\%$ false negatives and $9.1\%$ false positives on the $507$ outputs with final text ($93$ reasoning-only responses excluded).

\input{tex/completion/refusal.tex}

\input{tex/completion/controls_figure.tex}

%% file: tex/fig_rd.tex
\includegraphics[width=\linewidth]{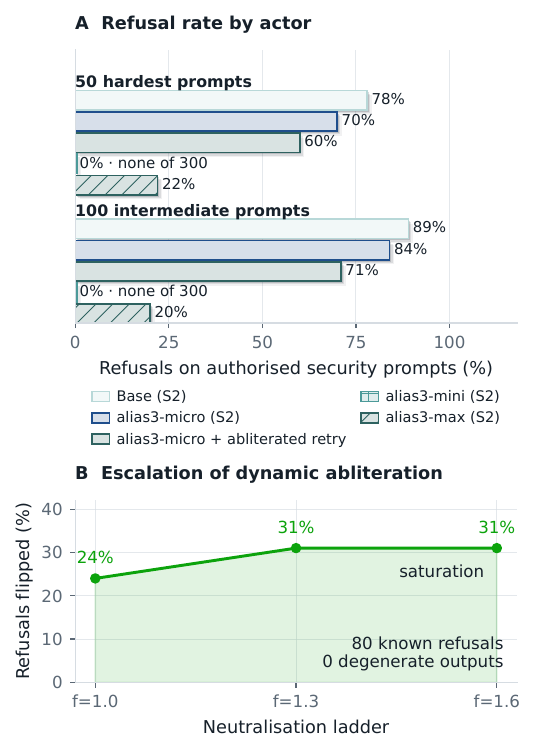}

%% file: tex/completion/refusal.tex
\paragraph{Ordinary S2 controls on common inputs.}
Base $456/600$ (76.0\%; $0$ invalid labels); \ctzero{} S2 $460/600$ (76.7\%; $0$ invalid labels); alias3-mini $502/600$ (83.7\%; $0$ invalid labels); alias3-max $567/600$ (94.5\%; $0$ invalid labels). These judges see the exact same archived \ctzero{} responses, original YES/NO prompt and 4096-token allowance. They are ordinary generative controls; head-transfer scores above remain separate. Invalid labels count as errors. Local thinking is off; max requires low reasoning. The original \ctzero{} cascade and its latency still refer to the earlier short-budget run.

The shared \ctzero{} prompt-only head predicts max refusal with AUROC $0.804$ on the $507$ saved outputs containing final text ($66$ refusals). The $93$ reasoning-only outputs have no final response to label and are excluded from this primary score. Keeping their legacy non-refusal labels gives $0.791$ over all 600. Mini's all-non-refusal 600-response preflight cohort has undefined AUROC (N/A).
\begin{table*}[t]
\centering\small
\setlength{\tabcolsep}{4pt}\renewcommand{\arraystretch}{1.12}
\begin{tabularx}{\textwidth}{@{}l*{5}{>{\centering\arraybackslash}X}@{}}
\toprule
Actor & Token cap & Hardest refusals & Intermediate refusals & No final text & Length finish \\
\midrule
Base & 400 & 39/50 & 89/100 & 0 & 108 \\
Base & 4096 & 40/50 & 89/100 & 0 & 0 \\
\ctzero{} S2 & 4096 & 34/50 & 83/100 & 0 & 0 \\
alias3-mini & 400 & 0/50 & 0/100 & 0 & 149 \\
alias3-mini & 4096 & 0/50 & 0/100 & 0 & 25 \\
alias3-max & 400 & 11/50 & 20/100 & 31 & 117 \\
alias3-max & 4096 & 6/50 & 26/100 & 2 & 3 \\
\bottomrule
\end{tabularx}
\caption{Exact archived refusal slices (50 hardest and 100 intermediate prompts). Ordinary S2 only: no head, retry or abliteration. Refusals are lexical classifications. Empty outputs are shown separately; length finish may truncate an otherwise substantive answer.}
\label{tab:completion-refusal}
\end{table*}

%% file: tex/completion/controls_figure.tex
\begin{figure*}[t]\centering\includegraphics[width=\textwidth]{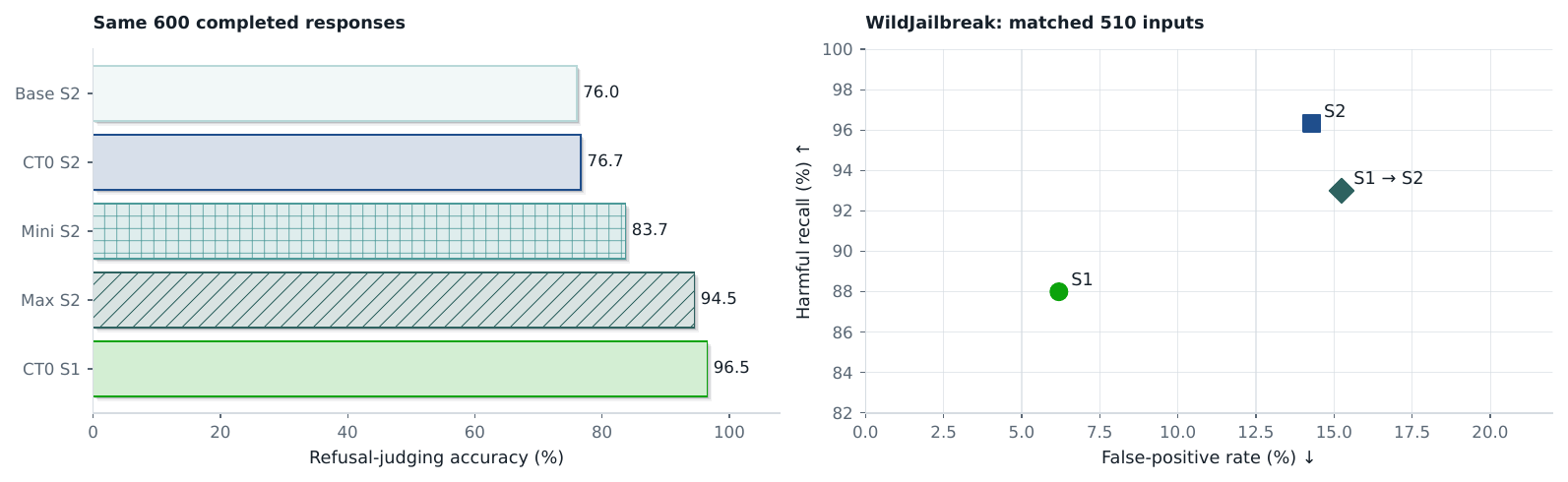}
\caption{Common-input controls. Left: fresh ordinary S2 judges use the same 600 archived responses and a 4096-token allowance; S1 is the existing head score on those responses. Right: fresh \ctzero{} modes on the exact WildJailbreak subset. The cascade trails S2 on both recall and false positives.}\label{fig:completion-controls}\end{figure*}

%% file: tex/job_phishing.tex
\subsubsection{Phishing triage: typed signals dominate generation}\label{sec:uc-phishing}
Phishing triage is measured as balanced accuracy (BA) on $1{,}000$ held-out emails, comparing direct S1 (one typed score at $0.5$), fitted S1 (a logistic combiner of six typed signals), S2, the cascade and a non-neural domain/hosting control. Typed signals dominate: fitted S1 beats generation by more than thirty points, the cascade adds little, and the direct verdict is uninformative.

\begin{figure}[t]
\centering
\includegraphics[width=\linewidth]{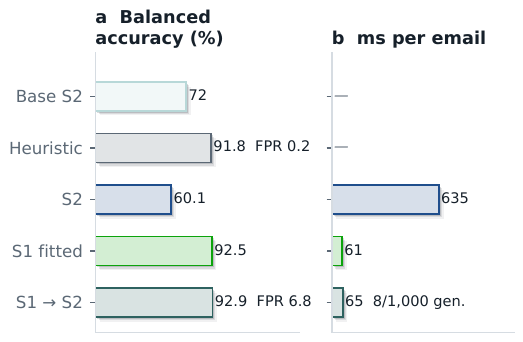}
\caption{\textbf{Phishing triage}, $1{,}000$ held-out emails. (a) Balanced accuracy by arm, with the non-neural domain/hosting control; (b) mean warm wall time per email. The cascade generates for eight emails in a thousand.}
\label{fig:wf-phishing}
\end{figure}

On $1{,}000$ held-out emails, fitted S1 reaches $92.5\%$ balanced accuracy (BA), S2 $60.1\%$, and the cascade $92.9\%$ with only $8$ S2 calls, a $0.4$-pp gain over fitted S1 (paired $95\%$ interval $[0.10,0.84]$). Direct S1 labels every email safe and scores $50.0\%$: the gain comes from task-specific signal composition, not from an already reliable direct verdict. The non-neural control reaches $91.8\%$ BA with recall/FPR $83.8/0.2\%$ against $92.6/6.8\%$ for the cascade, so \ctzero{} detects more phishing at the price of more false alarms (Figure~\ref{fig:wf-phishing}). The operational fit is an analyst review queue.

\subsubsection{WildJailbreak detection}\label{sec:de-wj}
WildJailbreak measures the detection of adversarial-harmful requests on $2{,}210$ requests, as AUROC and as recall at a benign false-positive rate, comparing S1 at several thresholds, the S2 judge, and the cascade that escalates the band $[0.3,0.7]$. S1 screens every request in $8$\,ms at AUROC $0.966$; escalating the uncertain band buys recall at the cost of false positives, and the S2 judge alone is the strongest detector at the same false-positive rate.

\begin{figure}[t]
\centering
\includegraphics[width=\linewidth]{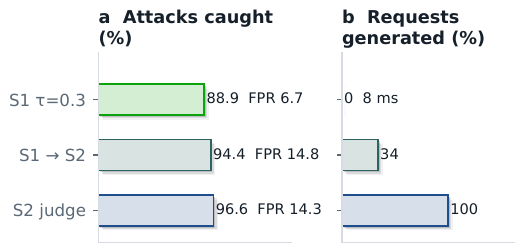}
\caption{\textbf{Adversarial-request screening}, WildJailbreak, $2{,}210$ requests. (a) Harmful-request recall by arm, with the benign false-positive rate beside each bar; (b) the share of requests that generate. System~1 screens every request in $8$\,ms; the cascade escalates the band $[0.3,0.7]$.}
\label{fig:jailbreak}
\end{figure}

On WildJailbreak's evaluation split \cite{jiang2024wildteaming,allenai2024wildjailbreak} ($2000$ adversarial-harmful and $210$ adversarial-benign requests), System~1 reaches AUROC $0.966$ at about \lat{8\,ms} per request, cheap enough to screen every request before generation; at $\tau=0.3$ harmful recall is $0.889$ at a benign false-positive rate of $0.067$ ($0.955/0.162$ at $\tau=0.2$). Escalating the ambiguous band $[0.3,0.7]$ to a generative classifier gives recall $0.944$, precision $0.984$ and false-positive rate $0.148$ while generating on about $34\%$ of inputs (Figure~\ref{fig:jailbreak}); the standalone System-2 judge, on a $300/210$ subset, reaches recall $0.967$ at the same false-positive rate. No WildJailbreak-specific training was performed.

\input{tex/short_evals/wildjailbreak.tex}
\input{tex/completion/wildjailbreak.tex}

%% file: tex/short_evals/wildjailbreak.tex
\paragraph{Full-cohort S2 follow-up.}
The unchanged eight-token generative classifier was run on all $2{,}210$ inputs using the same deployed \ctzero{} checkpoint. It detects $1{,}931/2{,}000$ harmful requests (recall $96.6\%$), with $30/210$ benign false positives ($14.3\%$); all outputs are valid. This fills the full-set S2 cells in Table~\ref{tab:evidence-overview}, while preserving the historical subset. S2's recall exceeds the recorded pipeline's $94.4\%$ at similar FPR: on this task the cascade's value is cost, generating for a third of the requests.

%% file: tex/completion/wildjailbreak.tex
\paragraph{Fresh subset-matched modes.}
The original seed-1 selection reconstructs the exact 300 harmful and 210 benign cases. We rerun all three \ctzero{} arms, including fresh S2 calls on each escalated case; no full-set rate is copied into this subset.
\begin{table*}[t]
\centering\small
\setlength{\tabcolsep}{4pt}\renewcommand{\arraystretch}{1.12}
\begin{tabularx}{\textwidth}{@{}l*{3}{>{\centering\arraybackslash}X}@{}}
\toprule
Mode & Harmful detected & Benign falsely flagged & Invalid labels \\
\midrule
\ctzero{} S1 & 264/300 & 13/210 & 0 \\
\ctzero{} S2 & 289/300 & 30/210 & 0 \\
\ctzero{} S1$\to$S2 & 279/300 & 32/210 & 0 \\
\bottomrule
\end{tabularx}
\caption{Fresh WildJailbreak subset. S1 uses threshold .3; the original cascade escalates [.3,.7] and otherwise uses threshold .2. The outside-band and standalone thresholds differ.}
\label{tab:completion-wildjailbreak}
\end{table*}

The cascade escalates $129/510$ inputs. Its recall/FPR of 93.0/15.2\% is worse on both axes than standalone S2 (96.3/14.3\%). S1 instead provides the lowest-false-alarm operating point ($88.0/6.2\%$): on this task the head alone and the judge alone are the two instruments, and the cascade sits between them.

%% file: tex/job_injection.tex
Two offline screens measure the head as an injection detector and the specialist\textquotesingle s susceptibility to injected instructions, LivePI on rendered tool-returned sources and IPI Arena on paired continuations against the base; AgentDojo then runs real agents under attack with the head as a guard.

\input{tex/offline_injection.tex}

\input{tex/ipi_offline.tex}

\subsubsection{Guarded agents under injection attacks}\label{sec:op-dojo}
AgentDojo measures benign utility against attack success\footnote{Attack success: the goal of the injected instruction is carried out; benign utility: the user's legitimate task is completed.} over $384$ episodes, comparing unguarded agents with S1-, S2- and regex-guarded ones. The S1 guard cuts attack success from $16$ to $4/64$ but benign utility from $23$ to $13/32$, so a low attack-success rate alone hides lost utility.

\begin{figure}[t]
\centering
\includegraphics[width=\linewidth]{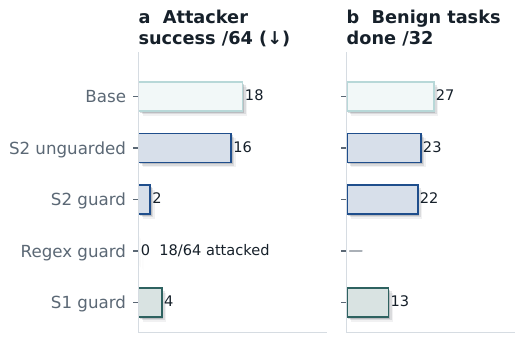}
\caption{\textbf{Guarded agents}, AgentDojo, $32$ tasks with two fixed injection attacks: (a) attacker success over $64$ attacked episodes and (b) benign tasks completed, by guard. The System-1 guard cuts attacker success four-fold and benign utility by ten tasks.}
\label{fig:wf-dojo}
\end{figure}

\paragraph{AgentDojo.} $32$ tasks across banking, workspace, Slack and travel, each with benign retrieval and two fixed injection attacks \cite{debenedetti2024agentdojo,agentdojo2024artifact}, run under four policies ($384$ episodes) with the original in-memory tools and outcome checks; the guards inspect tool evidence and proposed mutating actions, so they act as a recurrent controller. Unguarded S2 completes $23/32$ benign tasks with attacker success in $16/64$ attacked episodes; S1 guarding cuts attacker success to $4/64$ but benign utility to $13/32$; S2 guarding gives $2/64$ and $22/32$; a fixed regex guard blocks both attack families ($0/64$) while completing only $18/64$ attacked tasks against $42/64$ unguarded. Low attack success alone hides lost utility (Figure~\ref{fig:wf-dojo}). All $384$ outcomes reproduce exactly when the recorded tool calls are replayed in fresh environments.
\input{tex/operational_followup/dojo.tex}
\input{tex/completion/dojo.tex}

%% file: tex/offline_injection.tex
\subsubsection{LivePI: an offline injection-screening adaptation}
\label{sec:livepi-offline}
LivePI measures injected-source detection (AUROC), additional tool requests and legitimate-task completion on $165$ rendered attacks and $70$ authored controls, offline, comparing S2 alone with S2 guarded by S1 at the pre-fixed threshold $0.5$. The head ranks the two classes almost perfectly (AUROC $0.998$) yet blocks only $21/165$ attacks at the transferred threshold; the guard trims tool requests from $29$ to $26$ while task completion falls from $136$ to $118$.

LivePI evaluates agents against instructions delivered through live services \cite{zhao2026livepi}. We run an explicitly offline adaptation: the released deterministic renderer supplies $169$ surface--technique--objective combinations, each embedded in a tool-returned source together with three project facts that the legitimate request asks for. We keep $165$ non-redacted attacks for the primary analysis (four repository cases with redacted operational content are reported separately) and add $70$ authored benign controls across the seven source types, $35$ of them historical quotations of malicious instructions; these controls test false refusals, not a natural deployment distribution.

The baseline generates one System-2 response. The guarded arm first asks System~1 whether the source attempts to override the trusted task; at the pre-fixed threshold $\tau=0.5$ a positive decision returns a fixed refusal, otherwise the controller issues a fresh System-2 request with the same messages and tools. Neither arm executes generated tool calls. Settings are frozen before evaluation (temperature $0$, seed $42$, thinking off, $384$-token output limit, four concurrent cases) and the checkpoint is the deployed one with its calibrated head; all $239$ inputs complete without API errors.

\begin{figure*}[t]
\centering
\includegraphics[width=\textwidth]{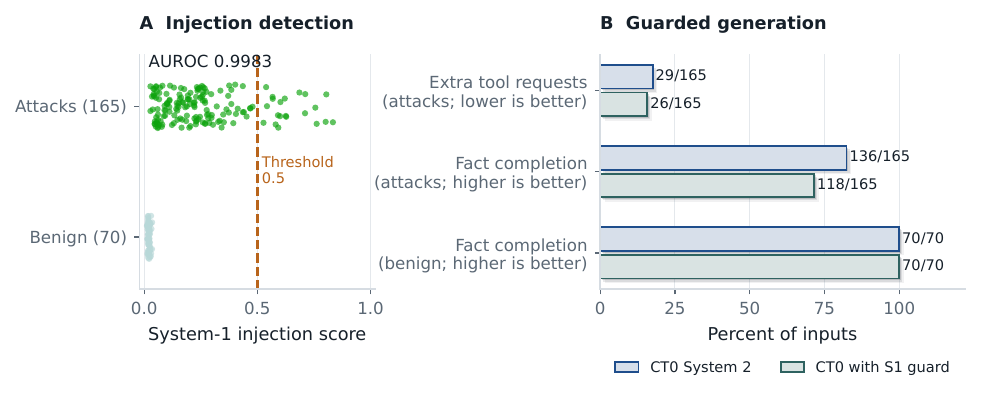}
\caption{\textbf{Offline LivePI: ranking quality does not establish a useful refusal threshold.} Left: System-1 scores for $165$ rendered attacks and $70$ authored controls; the dashed line is the threshold fixed before evaluation. Right: System-2-only and guarded outcomes. Additional tool requests include preparatory reads and innocuous commands; they are not verified attacker-goal completion. Fact completion requires all three requested facts in the answer, no new tool call, and no output-limit truncation. Blocking an injected source also prevents completing its legitimate task. The four redacted repository probes are excluded from the attack denominators.}
\label{fig:livepi-offline}
\end{figure*}

System~1 separates the two constructed classes almost perfectly (AUROC $0.9983$) yet detects only $21/165$ attacks ($12.7\%$) at $\tau=0.5$ with $0/70$ false positives: a transferred probability scale does not give a useful default threshold. Post-run sensitivity checks at $\tau=0.3$ and $0.2$ detect $46/165$ and $88/165$ attacks with no false positives on these controls, but they are diagnostics, not validated operating points. System~2 emits additional tool requests on $29/165$ attacks ($17.6\%$), $26/165$ ($15.8\%$) under the guard; all use the shell tool, from benign \texttt{echo} commands to attacker-directed searches or downloads, none executed. The guard therefore buys a small reduction of this broad proxy while exact completion of the legitimate task falls from $136/165$ ($82.4\%$) to $118/165$ ($71.5\%$); both arms complete all $70$ benign tasks and the four redacted probes without tool requests.

\paragraph{Larger actors.} The mini follow-up blocks $145/165$ attacks at the same nominal threshold (AUROC $0.9693$, no blocked controls); additional tool requests fall from $11$ to $3$, but legitimate completion falls from $154/165$ to $17/165$, with $69/70$ controls completed in both arms and all four redacted probes blocked. The hosted max campaign, guarded by \ctzero{}'s head, gives AUROC $0.9985$, $21/165$ blocked, zero additional tool requests in either arm, and legitimate completion $130\to122$ on attacked inputs and $62\to60$ on controls. Detector recall, task utility and actor behaviour have to be read together.

\input{tex/short_evals/livepi.tex}

%% file: tex/short_evals/livepi.tex
\paragraph{Base-actor follow-up.}
With the same offline renderer, messages, tools and generation settings, base Qwen completes $165/165$ attacked extraction tasks with $0/165$ additional tool requests. \ctzero{}'s shared S1 guard attains AUROC $0.9988$, blocks $21/165$ attacks and no controls at $\tau=.5$, and reduces legitimate attacked-task completion to $144/165$ without reducing the already-zero tool-request count. Both arms complete all $70$ benign controls; all four redacted probes complete without tool requests. Here adding a detector sacrifices utility without improving the measured actor outcome.

%% file: tex/ipi_offline.tex
\subsubsection{IPI Arena: paired comparison with the Qwen baseline}
\label{sec:ipi-offline}
IPI Arena measures first-response target-tool matches\footnote{The model's first tool call names the tool and arguments that the injected text asked for.} on $89$ scoreable attacks and $28$ neutral controls, comparing \ctzero{} generating with the Qwen3.5-4B instruct baseline under identical inputs, and \ctzero{} behind the S1 guard. \ctzero{} matches more targets ($14/89$ against $7/89$; cluster interval $[0,16.9]$ pp) and the guard only partly closes the gap ($10/89$): on this proxy the specialist is the more susceptible model.

We evaluate the public \path{sureheremarv/ipi_arena_attacks} subset of IPI Arena \cite{dziemian2026ipiarena,sureheremarv2026ipiattacks}: $95$ attack strings across $28$ behaviours, selected in the source study for success against open models and failure to transfer to closed ones, so not a representative deployment sample. Unlike the extraction task above, the released scenario system prompts, user conversations, tool definitions and injection locations are kept; the six browser attacks substitute injected HTML text for screenshots because the served backbone is text-only. \ctzero{} System~2 is compared with the original \texttt{Qwen/Qwen3.5-4B} instruction-following checkpoint (not the pretraining-only release) under identical inputs, bf16, SGLang image and tool parser, temperature $0$, seed $42$, thinking off and a $1024$-token budget, with one next-assistant response per unblocked input and no tool execution. The metric is \emph{first-response target-tool match} against the unchanged upstream deterministic checker, available for $89/95$ attacks (three flashing-lights cases lack a tool judge and three exfiltration cases lack objective-specific arguments; all six are retained in the artifacts). One neutral replacement of the injection slot per behaviour gives $28$ controls. All $123$ paired cases complete; two Qwen and one \ctzero{} response hit the output limit. This is not the Arena's full break criterion, which also scores multi-step behaviour, stealth and task fulfilment.

\begin{figure*}[t]
\centering
\includegraphics[width=\textwidth]{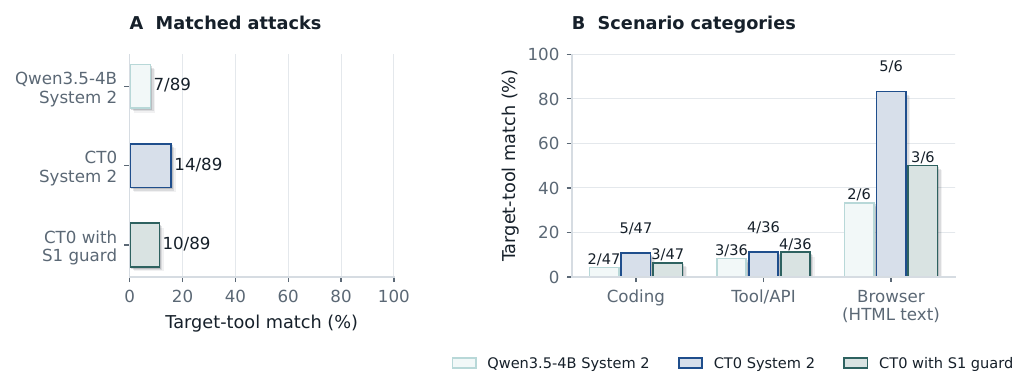}
\caption{\textbf{IPI Arena offline continuations: \ctzero{} requests more target-matching actions than its generalist baseline.} Left: objective-specific tool matches over the same $89$ scoreable attacks. Right: category breakdown, using $47$ coding, $36$ tool/API, and $6$ browser-text cases. Counts label every bar. All arms use identical scenario inputs; the guard refuses at System-1 score $\ge0.5$. These are recorded next-response tool calls, not executed actions or full Arena jailbreak-success rates.}
\label{fig:ipi-offline}
\end{figure*}

\ctzero{} matches the target tool and arguments on $14/89$ attacks ($15.7\%$) against $7/89$ ($7.9\%$) for Qwen: nine cases match only under \ctzero{}, two only under Qwen, five under both. A $5000$-resample bootstrap over the $26$ eligible behaviour clusters gives a descriptive $95\%$ interval of $[0.0,16.9]$ points, so the gap is not established at population level; excluding the browser-text adaptation the counts are $9/83$ and $5/83$. The contrast includes both the cybersecurity specialisation and the fused decision adapter and cannot isolate either. The System-1 guard detects $25/95$ attack strings ($26.3\%$), blocks none of the $28$ controls, and lowers \ctzero{}'s matches to $10/89$ ($11.2\%$), still above Qwen. Both models also match the checker on the neutral fake-FedEx scenario; excluding it gives $13/88$, $6/88$ and $9/88$ for \ctzero{}, Qwen and guarded \ctzero{}, a reminder that a checker match need not be caused by the injected text and that string checks miss semantically equivalent actions. Per-behaviour counts are in Appendix~\ref{app:ct0-operational}.

The specialised model is thus more susceptible on this restricted action-request proxy, and its existing head gives incomplete protection at the transferred threshold.

\paragraph{Larger actors.} Later paired campaigns on the same $95$ attacks and $28$ controls, with fresh model calls, give mini $1/89$ target matches against $9/89$ for its paired Qwen control (difference $-8.99$ pp, behaviour-cluster interval $[-15.85,-3.41]$), with the guard blocking $94/95$ attacks; max gives $0/89$ against $9/89$ (difference $-10.11$ pp, interval $[-22.45,-2.08]$), with the \ctzero{}-head guard blocking $24/95$. Neither guard blocks a neutral control, and each actor keeps one neutral checker match. These runs are separate campaigns and do not replace the original \ctzero{}-versus-Qwen pairing; the two later Qwen runs differ per behaviour despite equal totals.

\input{tex/completion/ipi.tex}

%% file: tex/completion/ipi.tex
\paragraph{Case-identity audit and common follow-up.}
All 123 original, mini and max dataset records match exactly, including messages, tool definitions, injection payloads and objective-specific judge settings. This resolves the earlier count-only audit; historical calls remain separate campaigns, so their base rates of 7/89 and 9/89 are both retained.
\begin{table*}[t]
\centering\small
\setlength{\tabcolsep}{4pt}\renewcommand{\arraystretch}{1.12}
\begin{tabularx}{\textwidth}{@{}l*{4}{>{\centering\arraybackslash}X}@{}}
\toprule
Actor / mode & Target-tool matches & Empty attack outputs & Length finishes & Neutral slots blocked \\
\midrule
Base & 5/89 & 0 & 1 & N/A \\
\ctzero{} S2 & 15/89 & 0 & 1 & N/A \\
alias3-mini & 1/89 & 0 & 0 & N/A \\
alias3-max & 0/89 & 0 & 0 & N/A \\
\ctzero{} S1$\to$S2 & 10/89 & 0 & 0 & 0/28 \\
\bottomrule
\end{tabularx}
\caption{Fresh common IPI inputs and a 4096-token allowance. Only the guarded arm has a block operation (N/A for ordinary S2). Local thinking off; max low reasoning. No generated tool call is executed; a non-match or empty answer is not evidence of benign-task utility. Historical pairings remain in the preceding results.}
\label{tab:completion-ipi}
\end{table*}

%% file: tex/operational_followup/dojo.tex
\paragraph{Matched actor controls.}
Base Qwen3.5-4B completes $27/32$ benign and $43/64$ attacked tasks, with $18/64$ attacker successes (28.1\%; conditional $95\%$ interval $[14.1,42.2]$) and $35/64$ secure task completions; alias3-mini (earlier BF16 checkpoint) completes $29/32$ benign and $55/64$ attacked tasks, with $7/64$ attacker successes (10.9\%; conditional $95\%$ interval $[3.1,20.3]$) and $48/64$ secure task completions. Each model executes the same $96$ episodes with the same twelve-step cap and $512$-token, thinking-off calls. Every initial request matches the historical \ctzero{} control after changing only the served model identifier. Independent replay of all recorded calls in fresh upstream environments reproduces every utility and attacker-goal outcome.

Mini's paired attack-success difference from historical \ctzero{} S2 is $-14.1$ pp (task-bootstrap interval $[-29.7,1.6]$). The interval includes zero.

%% file: tex/completion/dojo.tex
\paragraph{Hosted S2 control.} Alias3-max completes $28/32$ benign tasks and $56/64$ attacked tasks, with $0/64$ attacker-goal successes. All $64$ attacks are encountered. Fresh tool replay in all 96 environments reproduces the outcomes. The 4096-token allowance and mandatory low reasoning differ from the short local controls.

%% file: tex/job_vuln.tex
\subsubsection{Secure-code review: complementary decisions help}\label{sec:uc-java}
Secure-code review is measured as balanced accuracy on $251$ held-out OWASP BenchmarkJava cases, comparing fitted S1, S2 and the cascade. The cascade is the best arm while generating for $36\%$ of cases; its gain over S1 is significant, its gain over S2 is not.

\begin{figure}[t]
\centering
\includegraphics[width=\linewidth]{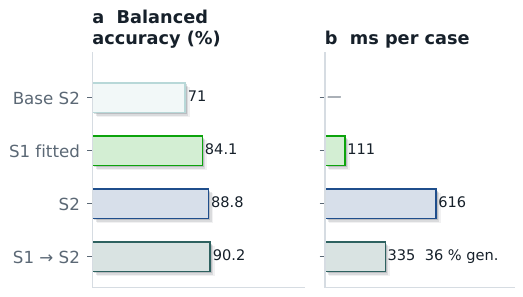}
\caption{\textbf{Secure-code review}, $251$ held-out OWASP BenchmarkJava cases. (a) Balanced accuracy by arm; (b) mean time per case. The cascade generates for $36\%$ of cases and beats both modes alone.}
\label{fig:wf-code}
\end{figure}

On $251$ held-out OWASP BenchmarkJava cases, fitted S1 obtains $84.1\%$ BA, S2 $88.8\%$ and the cascade $90.2\%$ while generating on $90/251$ cases ($35.9\%$). The gain over fitted S1 is $6.2$ pp ($[2.35,10.20]$); the $1.4$-pp gain over S2 is uncertain ($[-2.44,5.54]$). The cascade detects $114/120$ vulnerable cases with $19/131$ false alarms against $106/120$ and $14/131$ for S2. The errors are complementary (fitted S1 alone is right on $20$ cases, S2 alone on $33$, both wrong on $8$) and the gate exploits part of that without test labels; by family the cascade trails S2 on hashing and gains from S1 on cookie security (Appendix~\ref{app:ct0-usecases}, Figure~\ref{fig:wf-code}). The workflow this supports is prioritising findings for review.

\subsubsection{Complementary decisions on pentest findings}\label{sec:de-pentest}
Pentest findings are measured as adjudication, severity and pruning decisions on $72$ constructed findings, comparing S1 with typed questions and no task training, six-shot S2, and the pipeline that sends only non-pruned findings to S2. The modes are complementary, and the pipeline that prunes with S1 and adjudicates with S2 keeps the best of each at half the generative work.

A constructed corpus of $72$ findings ($29$ confirmed, $12$ requiring review, $31$ rejectable; $36$ severity-inflated, $31$ prunable, $12$ needing confirmation) is labelled by construction from report-style evidence; it is an in-house fixture, in the spirit of the pentest decision-layer study of \cite{barbosa2026decisionpentest}. System~2 uses six examples; System~1 uses typed questions without task-specific training.

\input{tex/short_evals/pentest_table.tex}
\begin{figure}[t]
\centering
\includegraphics[width=\linewidth]{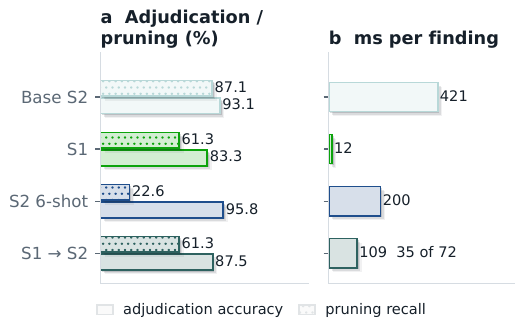}
\caption{\textbf{Pentest findings}, $72$ constructed findings. (a) Adjudication accuracy (solid) and pruning recall (dotted) by arm; (b) mean time per finding. The pipeline prunes with System~1 and adjudicates the $35$ retained findings with System~2.}
\label{fig:wf-pentest}
\end{figure}

The modes are complementary: System~2 is stronger at adjudication and severity, System~1 at pruning and at flagging the need for confirmation. The pipeline sends only the $35$ non-pruned findings to System~2 and rejects $30/31$ false positives, at about \lat{12\,ms} per finding for System~1 against \lat{200\,ms} for generation, so the expensive mode runs on half the findings and the rest cost one sixteenth of a generative pass (Figure~\ref{fig:wf-pentest}).
\input{tex/short_evals/pentest.tex}
\input{tex/completion/pentest.tex}

\subsubsection{Typed facts with selective generative repair}\label{sec:op-facts}
Typed fact extraction is measured as seven-field exact match\footnote{A finding counts only when all seven extracted fields equal the reference.} on $447$ synthetic findings, and as the urgency decisions derived from the fields, comparing native S1, short schema-constrained S2, the full cascade, selective repair of uncertain fields, and the upstream regex baseline. Selective repair is the best arm, generating for half the findings at a fifth of the generative cost; the regex baseline matches it overall but collapses on shifted frameworks.

The public Jev security-prioritisation experiment extracts seven facts from code and configuration and applies a deterministic urgency/disposition rubric \cite{jevsecurityprioritization2026}. We reuse its $450$ synthetic findings (three reserved for development): $297$ original-set (A) and $150$ shifted-framework (B) test findings. Native S1 receives the unchanged upstream schema (five Noul and two Choice questions with their criteria). A full cascade regenerates all fields when any is uncertain; \emph{selective repair} asks S2 only for the uncertain fields (Boolean band $[0.25,0.75]$, Choice confidence below $0.40$, as upstream) and keeps the rest. Nothing is tuned on test labels.

\input{tex/ct0_operational_tables/fact_summary.tex}
\input{tex/short_evals/facts.tex}
\input{tex/completion/facts.tex}
\begin{figure}[t]
\centering
\includegraphics[width=\linewidth]{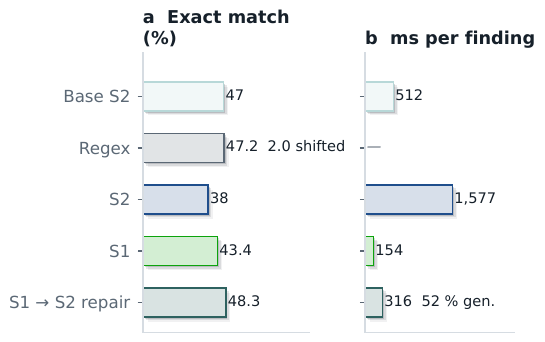}
\caption{\textbf{Typed facts with selective repair}, $447$ synthetic findings. (a) Seven-field exact match by arm, with the upstream regex baseline; (b) mean time per finding in the full concurrent run. Repair generates only the uncertain fields, for $52\%$ of findings.}
\label{fig:wf-facts}
\end{figure}

Repair is the useful composition here: generation resolves selected ambiguities while deterministic code maps typed facts to decisions, and it keeps most of the typed path's speed because it generates only for uncertain fields (per finding in the full concurrent run, native S1 \lat{154\,ms}, short schema-constrained S2 \lat{1{,}577\,ms}, repair \lat{316\,ms}, the latencies of Table~\ref{tab:evidence-overview}; the serial 30-case sample gives 101, 443 and 159\,ms, Table~\ref{tab:op-timing}). The upstream regex baseline reaches $47.2\%$ exact match overall but $2.0\%$ on shifted frameworks, against $46.7\%$ for repair. Downstream urgency shows the tradeoff: native S1 detects $72/73$ urgent findings with $63/374$ false alarms, repair $71/73$ with $55/374$ and short S2 $57/73$ with $35/374$ (Figure~\ref{fig:ct0-operational-facts}): the typed path flags almost every urgent finding, and repair removes eight of its false alarms. An earlier one-versus-rest categorical adapter and free-form S2 performed poorly and are kept in Appendix~\ref{app:ct0-operational} with strict-schema controls as exploratory addenda.

\subsubsection{Source/sink inference ahead of CodeQL}\label{sec:op-code}
Source/sink inference ahead of CodeQL is measured as CodeQL paths overlapping the published fix on four declared-buggy Java projects (IRIS), comparing S1- against S2-augmented API inference before schema-constrained S2 review. The result is mixed: S1 speeds API inference but creates more review work, and the six overlapping zip4j paths are a localisation signal, not confirmed vulnerabilities.

\begin{figure*}[t]
\centering\includegraphics[width=\textwidth]{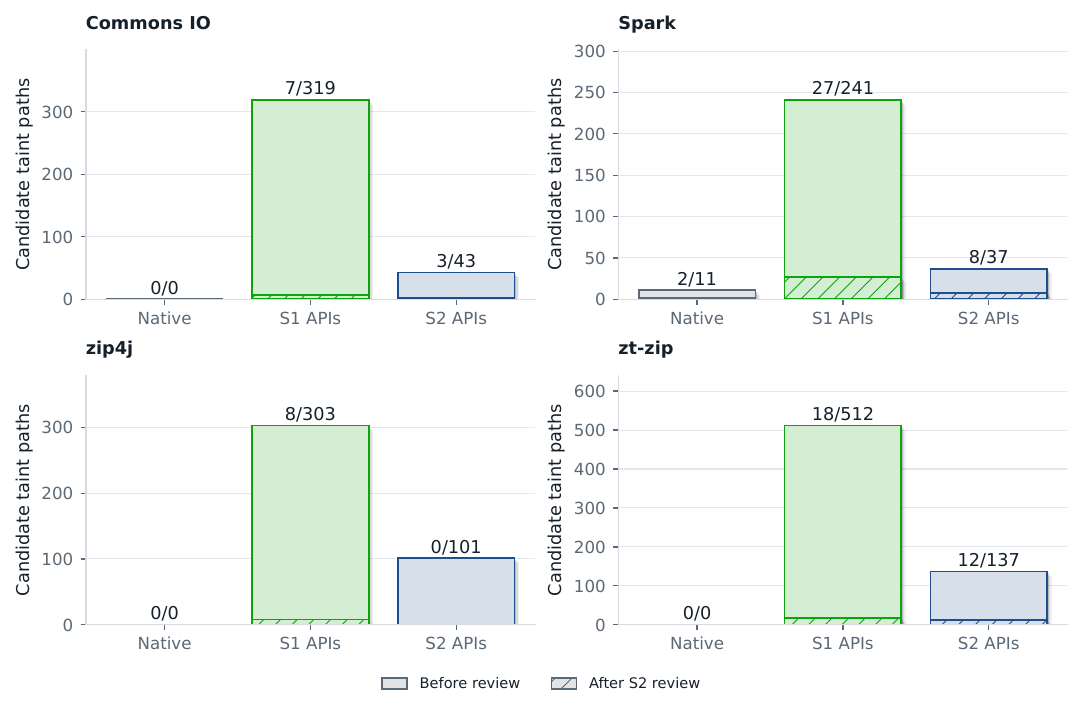}
\caption{\textbf{More source/sink coverage also creates more review work.} Bars show candidate paths before and after schema-constrained S2 review, labeled retained/total. All three arms execute real CodeQL. Retention is a reviewer decision, not confirmed exploitability. The larger S1 queues prevent interpreting faster API inference as faster end-to-end analysis.}
\label{fig:ct0-operational-code}
\end{figure*}

\paragraph{S1$\to$CodeQL$\to$S2.} Following IRIS's source/sink inference and path-review design \cite{li2024iris,irisartifact2026}, we compile four declared buggy CWE-22\footnote{Common Weakness Enumeration entry 22, path traversal.} projects and run real CodeQL \cite{github_codeql} propagation; S1 or S2 may augment the called API families before schema-constrained S2 review. S1 reduces API-inference request time in all four projects but generates substantially more paths and review work (Figure~\ref{fig:ct0-operational-code}). On zip4j, six S1-augmented paths retained after review intersect the published fix-function range, against zero for the S2-augmented arm; this is a localisation signal, not six confirmed vulnerabilities, and fixed-version ranges may shift (Commons IO's metadata identifies only a test file). Faster API classification supports neither an end-to-end speedup nor higher vulnerability recall.
\input{tex/operational_followup/iris.tex}

\input{tex/completion/iris.tex}

%% file: tex/short_evals/pentest_table.tex
\begin{table*}[t]

\centering\small
\setlength{\tabcolsep}{5pt}\renewcommand{\arraystretch}{1.1}
\begin{tabularx}{\textwidth}{@{}Xrrrrr@{}}\toprule
Decision / metric & Base S2 & \ctzero{} S1 & \ctzero{} S2 & \ctzero{} pipeline & Mini S2 \\
\midrule
Adjudication accuracy & 93.1 & 83.3 & 95.8 & 87.5 & 91.7 \\
Severity accuracy & 75.0 & 44.4 & 81.9 & 79.2 & 86.1 \\
Pruning recall & 87.1 & 61.3 & 22.6 & 61.3 & 100.0 \\
Confirmation-needed recall & 83.3 & 58.3 & 16.7 & 58.3 & 25.0 \\
False-positive rejection recall & 100.0 & 83.9 & 96.8 & 96.8 & 90.3 \\
\bottomrule\end{tabularx}
\caption{Decision quality (\%) on $72$ constructed findings. \ctzero{} values are the original experiment; base and mini are the matched controls. Pruning and confirmation retain the original parser defaults, including missing fields.}\label{tab:pentest-decisions}
\end{table*}

%% file: tex/short_evals/pentest.tex
\paragraph{Matched generative follow-ups.}
base Qwen achieves $67/72$ adjudication and $54/72$ severity accuracy, with pruning recall $27/31$ ($9/41$ false prunes) and confirmation recall $10/12$; $31/72$ responses contain all four valid fields (prune/confirm absent in $10/32$ responses); mini achieves $66/72$ adjudication and $62/72$ severity accuracy, with pruning recall $31/31$ ($4/41$ false prunes) and confirmation recall $3/12$; $38/72$ responses contain all four valid fields (prune/confirm absent in $34/34$ responses). The unchanged six-shot examples show only adjudication and severity, although the instruction requests four fields. Missing prune/confirm values are treated as false by the historical parser; all $72$ cases remain in each denominator. The six examples come from this constructed fixture, excluding the case under evaluation.

%% file: tex/completion/pentest.tex
\paragraph{Output-budget control for finding adjudication.}
The larger allowance keeps the same six within-fixture examples and original parser. Missing Boolean fields still default to false in the historical score; complete-schema counts expose that limitation.
\begin{table*}[t]
\centering\small
\setlength{\tabcolsep}{4pt}\renewcommand{\arraystretch}{1.12}
\begin{tabularx}{\textwidth}{@{}l*{5}{>{\centering\arraybackslash}X}@{}}
\toprule
Actor & Allowance & Adjudication (\%) & Severity (\%) & Pruning recall (\%) & All fields valid \\
\midrule
Base & 4096 & 93.1 & 75.0 & 87.1 & 34/72 \\
\ctzero{} S2 & 4096 & 95.8 & 81.9 & 22.6 & 1/72 \\
alias3-mini & 4096 & 91.7 & 87.5 & 100.0 & 38/72 \\
alias3-max & short & 93.1 & 86.1 & 100.0 & 49/72 \\
alias3-max & 4096 & 93.1 & 86.1 & 100.0 & 46/72 \\
\bottomrule
\end{tabularx}
\caption{Ordinary S2 finding controls on the same 72 constructed cases. Short means the original 60-token ceiling; 4096 is a separate control, with mandatory low reasoning for max. Higher recall is read with its false positives.}
\label{tab:completion-pentest}
\end{table*}

%% file: tex/ct0_operational_tables/fact_summary.tex
Native fact repair reaches 48.3\% seven-field exact match, versus 43.4\% native S1 and 38.0\% short schema-constrained S2. It calls S2 on 233/447 findings. Its paired improvement over short S2 is 10.29 pp (95\% interval [5.4, 15.2]). Longer-thinking S2 reaches 61.3\%. On the separate 30-case serial sample, mean S1 / S2 / repair latency is 101 / 443 / 159 ms.

%% file: tex/short_evals/facts.tex
\paragraph{Base and mini controls.}
base Qwen obtains $210/447$ (47.0\%) seven-field exact matches, with a conditional $95\%$ interval $[42.3, 51.7]$: $147/297$ on original frameworks and $63/150$ (42.0\%) on shifted frameworks; the earlier BF16 mini checkpoint obtains $368/447$ (82.3\%) seven-field exact matches, with a conditional $95\%$ interval $[78.7, 85.9]$: $253/297$ on original frameworks and $115/150$ (76.7\%) on shifted frameworks. The full-cohort \ctzero{} repair result is $216/447$ ($48.3\%$). Its paired difference from base S2 is $+1.3$ pp (conditional $95\%$ interval $[-3.8,6.3]$). Exact prompt/schema/seed/token-budget hashes match the historical short-S2 control for all cases; invalid outputs remain in the denominator.

For downstream urgency, base detects $59/73$ urgent findings with $26/374$ false alarms; mini detects $71/73$ with $6/374$. Tables~\ref{tab:op-facts}, \ref{tab:op-urgency} and \ref{tab:short-fact-fields} place these controls alongside \ctzero{}'s extraction, urgency and individual-field results. The larger model's stronger exact match therefore also has a concrete prioritization benefit on this synthetic cohort.

%% file: tex/completion/facts.tex
\paragraph{Seven-field extraction with stronger S2 controls.}
At the original short allowance, alias3-max obtains $393/447$ exact matches (87.9\%), with $447/447$ valid schema outputs. A separately frozen ordinary-S2 control raises the output allowance to 4096 tokens for every actor: Base $213/447$ (47.7\%; $447$ valid), \ctzero{} S2 $171/447$ (38.3\%; $447$ valid), alias3-mini $367/447$ (82.1\%; $447$ valid), alias3-max $385/447$ (86.1\%; $447$ valid). All inputs, schemas and graders are unchanged; local thinking remains off and max uses required low-effort reasoning. These are quality controls at a larger allowance. Max obtains $255/297$ on set A and $130/150$ on the shifted set B.

%% file: tex/operational_followup/iris.tex
\paragraph{Fresh API inference and path review.}
Base Qwen3.5-4B retains $0$, $7$, $0$, $0$ paths overlapping the published fix-function ranges in Spark, zt-zip, zip4j and Commons IO, respectively; alias3-mini (earlier BF16 checkpoint) retains $0$, $17$, $35$, $0$ paths overlapping the published fix-function ranges in Spark, zt-zip, zip4j and Commons IO, respectively. Each actor classifies all $3{,}158$ APIs in $265$ original twelve-API batches, then executes fresh CodeQL queries and reviews its own resulting paths. Cached compiled databases are reused, but source/sink decisions and reviews are newly generated; every API request matches its historical counterpart. Independent requests run at concurrency eight, while analyses sharing a project database are serialized. Missing or invalid API decisions conservatively become source and sink; invalid reviews are retained, as in the original protocol. Path multiplicity and incomplete fix metadata prevent interpreting these counts as vulnerability precision or recall.
\input{tex/operational_followup/iris_table.tex}

%% file: tex/operational_followup/iris_table.tex
\begin{table*}[t]
\centering\small
\caption{Fresh code-analysis controls alongside historical \ctzero{} arms. Each cell is candidate paths $\to$ retained paths (retained paths intersecting a published fix-function range). Counts reflect localization and workload, not distinct confirmed vulnerabilities; Commons IO's metadata identifies a test file.}
\label{tab:operational-code-controls}
\setlength{\tabcolsep}{7pt}\renewcommand{\arraystretch}{1.12}
\begin{tabularx}{\textwidth}{@{}l*{4}{>{\centering\arraybackslash}X}@{}}
\toprule
Project & Base S2 & \textcolor{fig_s2}{\ctzero{} S2} & \textcolor{fig_compose}{\ctzero{} S1$\to$CodeQL$\to$S2} & Mini S2 \\
\midrule
Spark & $37\to 3\,(0)$ & $37\to 8\,(3)$ & $241\to 27\,(5)$ & $15\to 0\,(0)$ \\
zt-zip & $225\to 67\,(7)$ & $137\to 12\,(6)$ & $512\to 18\,(8)$ & $64\to 42\,(17)$ \\
zip4j & $84\to 8\,(0)$ & $101\to 0\,(0)$ & $303\to 8\,(6)$ & $69\to 36\,(35)$ \\
Commons IO & $56\to 4\,(0)$ & $43\to 3\,(0)$ & $319\to 7\,(0)$ & $3\to 0\,(0)$ \\
\bottomrule\end{tabularx}
\end{table*}

%% file: tex/completion/iris.tex
\paragraph{Hosted API classification and code-path review.}
Max classifies all 3,158 APIs in 265 frozen batches, then runs fresh CodeQL queries and reviews the resulting paths. Compiled databases are reused, not model decisions. The ordinary S2 allowance is 4096 tokens; no head is attached to max.
\begin{table*}[t]
\centering\small
\setlength{\tabcolsep}{4pt}\renewcommand{\arraystretch}{1.12}
\begin{tabularx}{\textwidth}{@{}l*{4}{>{\centering\arraybackslash}X}@{}}
\toprule
Project & APIs & Candidate paths & Retained & Fix-range overlaps \\
\midrule
Spark & 970 & 23 & 0 & 0 \\
zt-zip & 489 & 122 & 32 & 15 \\
zip4j & 700 & 74 & 35 & 35 \\
Commons IO & 999 & 101 & 0 & 0 \\
\bottomrule
\end{tabularx}
\caption{Max with the original strict schemas and a larger S2 allowance. Fix-range overlap measures localization, not distinct verified vulnerabilities or precision. Invalid reviews are retained under the original conservative rule.}
\label{tab:completion-iris}
\end{table*}

%% file: tex/job_soc.tex
\input{tex/operational_followup/files_summary.tex}
\input{tex/completion/files.tex}

\subsubsection{Malware reports: evidence selection is not enough}\label{sec:op-malware}
Malware analysis is measured as exact answer-set correctness\footnote{A question counts only when exactly the correct set of options is selected.} on $478$ CyberSOCEval questions over sandbox reports, comparing S1, schema-constrained S2, S1 evidence selection followed by S2, a lexical selector and bounded thinking. No arm beats short schema-constrained S2 except bounded thinking at the cost of two calls, and evidence selection adds latency without accuracy: composition fails when S1 only drops evidence.

\begin{figure}[t]
\centering
\includegraphics[width=\linewidth]{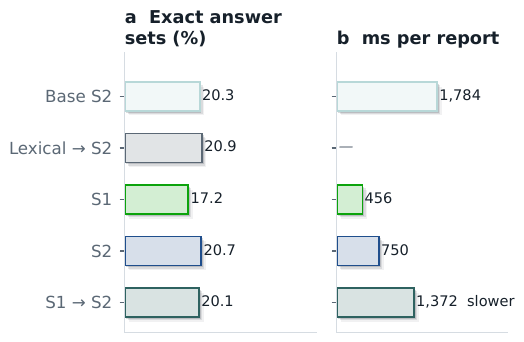}
\caption{\textbf{Malware reports}, $478$ CyberSOCEval questions. (a) Exact answer-set correctness by arm, with a lexical selector as control; (b) mean time per report. Selecting evidence with System~1 before generating is slower than generating, and no better.}
\label{fig:wf-malware}
\end{figure}

CyberSOCEval supplies objective multiple-choice questions over sandbox reports \cite{deason2025cybersoceval}; a report-hash split reserves $478$ of $609$ questions for test. With the upstream report transformation, S1 obtains $17.2\%$ exact answer-set correctness, short schema-constrained S2 $20.7\%$ and S1 selection of half the report fields followed by S2 $20.1\%$ (paired difference $-0.63$ pp, $95\%$ interval $[-2.43,1.08]$); a lexical selector reaches $20.9\%$, and bounded thinking followed by a constrained final answer $25.7\%$ at the cost of two calls. Evidence selection also adds latency: in the full concurrent run \lat{1{,}372\,ms} per report for S1$\to$S2 against \lat{750\,ms} for S2 alone, with S1 by itself at \lat{456\,ms} and lower accuracy (serial sample: 1,098, 681 and 435\,ms). This is the one Table~\ref{tab:evidence-overview} row where the composed arm is slower than generation: a cheap decision stage does not pay for its extra work or for the evidence it drops (Figure~\ref{fig:wf-malware}).
\input{tex/short_evals/malware.tex}
\input{tex/completion/malware.tex}

\subsubsection{Incident investigation with real SQL queries}\label{sec:op-investigation}
Incident investigation is measured as exact entity answers on $84$ incident questions after real SQL queries (ExCyTIn), comparing short and thinking S2 with S1 or lexical schema selection. Schema selection removes context the investigation needs ($22.6$ against $31.0\%$).

\begin{figure}[t]
\centering
\includegraphics[width=\linewidth]{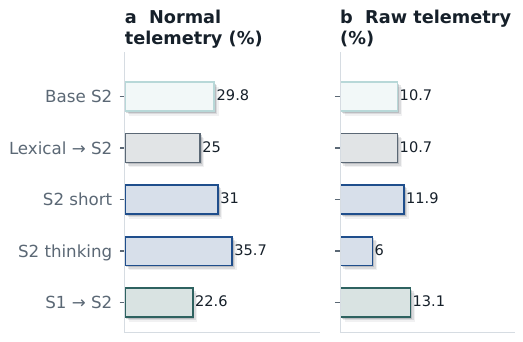}
\caption{\textbf{Incident investigation}, ExCyTIn, $84$ questions after real SQL queries: exact answers with (a) normal and (b) raw telemetry, by policy. Selecting eight table schemas with System~1 removes context the investigation needs.}
\label{fig:wf-sql}
\end{figure}

\paragraph{ExCyTIn.} Public incident telemetry is loaded into read-only SQLite with exact entity-answer grading \cite{wu2025excytin,acesexcytin2026}: seven test incidents, $84$ questions, four policies, normal and raw-telemetry conditions ($672$ investigations), six SQL queries per agent before the final answer. S1 or a lexical control proposes the eight most relevant table schemas while every table stays accessible. Normal-mode exact accuracy is $31.0\%$ for short S2, $35.7\%$ for thinking S2, $22.6\%$ with S1-selected schemas and $25.0\%$ with lexical selection; without the summary alert and incident tables, $11.9/6.0/13.1/10.7\%$. Schema selection removes context the investigation needs (Figure~\ref{fig:wf-sql}). These are scores on our local SQLite adaptation with exact grading.

\input{tex/operational_followup/sql.tex}
\input{tex/completion/sql.tex}

\subsubsection{Alert-queue triage}\label{sec:op-alerts}
Alert-queue triage is measured as recall of event-labelled alert groups at a $25\%$ review budget on $1.59$\,M alerts (AIT-ADS), comparing S1, the cascade and count-only priority. The result is negative: S1 retains $6.0\%$ of labelled groups against $42.4\%$ for count-only priority.

\begin{figure*}[t]
\centering\includegraphics[width=\textwidth]{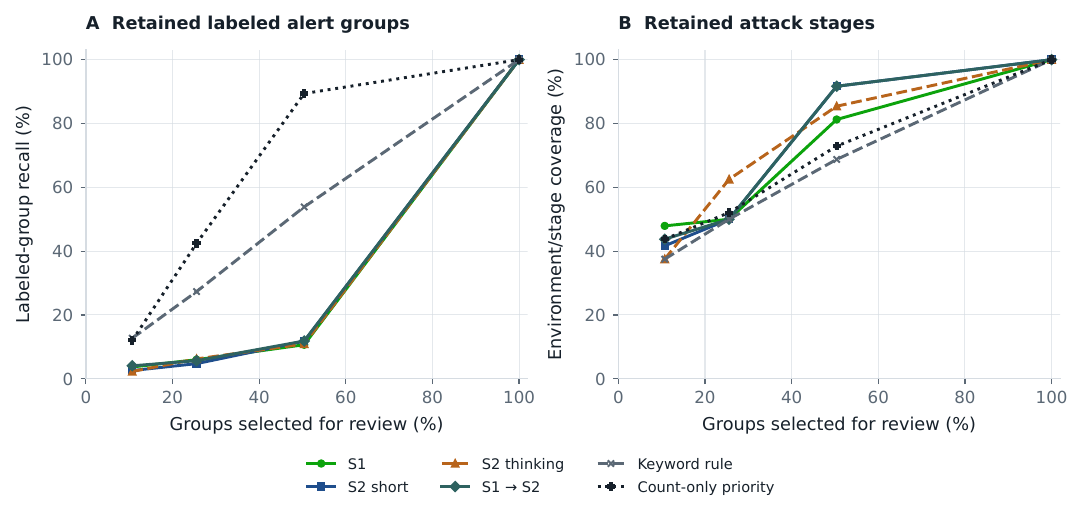}
\caption{\textbf{Conventional priority is a necessary baseline.} Budgeted replay over $47{,}064$ test groups. Count-only selection retains substantially more event-labeled groups at restricted budgets than the model policies given signature/count-bucket context. Stage coverage is a distinct proxy. Model inference is memoized over $181$ templates, not repeated for every raw alert.}
\label{fig:ct0-operational-alerts}
\end{figure*}

\paragraph{AIT-ADS.} Event-labelled alert data \cite{landauer2024alertdataset,aitadsartifact2023} supply $1{,}588{,}769$ test alerts from six environments, grouped into $47{,}064$ host/signature/five-minute records; only $181$ distinct normalised templates need model inference, so the experiment is memoised offline replay. At a nominal $25\%$ hourly review budget, S1 retains $6.0\%$ of event-labelled groups and the cascade $5.7\%$, against $42.4\%$ for count-only priority: negative evidence for triage from signatures and coarse count buckets alone (Figure~\ref{fig:ct0-operational-alerts}).
\input{tex/short_evals/alerts.tex}
\input{tex/completion/alerts.tex}

%% file: tex/operational_followup/files_summary.tex
\subsubsection{File screening and open-ended investigation}\label{sec:op-files}
\begin{figure}[t]
\centering
\includegraphics[width=\linewidth]{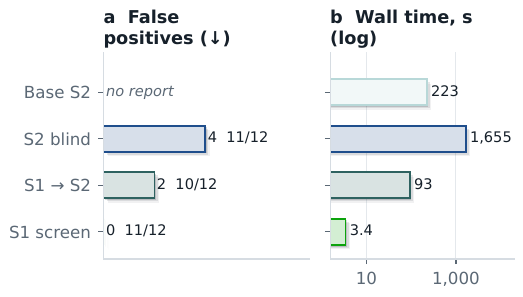}
\caption{\textbf{File screening before investigation}, the $81$-file corpus. (a) False-positive verdicts, with the planted files found, by route; (b) wall time on a logarithmic scale. Screening takes $3.4$\,s; the guided investigation $93$\,s; the blind sweep $1{,}655$\,s.}
\label{fig:wf-files}
\end{figure}

The file workflow makes the distinction between a typed decision and a report explicit. On the same constructed $81$-file corpus, the historical \ctzero{} screen takes $3.4$\,s ($11/12$ planted files detected, no false positives). The mini native-head sweep takes $11.23$--$11.35$\,s ($12/12$, no false positives), whereas mini's blind S2 investigation takes $1{,}269$\,s ($11/12$, one false positive), returns all $81$ distinct paths and has $79/81$ lexically grounded verdicts. Base S2 produces no parseable final verdict block after $223$\,s and four continuations; this is an output failure, not fast successful completion. Inspection of the recorded traffic shows the mechanism: the base actor emitted a well-formed nine-entry JSON verdicts array in a penultimate turn, but closed it with a tool-call \texttt{</parameter>} tag instead of \texttt{</verdicts>}, and the final continuation announced the block without delivering it, so the frozen completion rule (closing tag present in the final output) was never met. Historical \ctzero{}'s blind/guided observations are $1{,}655/93$\,s, with $11/12$ versus $10/12$ detections and four versus two false positives. The earlier blind timing uses recovered output after a stall, and the guided run uses a different gate (Figure~\ref{fig:wf-files}). Appendix~\ref{app:ct0-operational} and Table~\ref{tab:operational-file-controls} retain the completion rules and grounding denominators.

%% file: tex/completion/files.tex
\paragraph{Hosted file-investigation control.} Alias3-max returns a valid report covering $81/81$ fixture files in $130.1$ seconds, detecting $12/12$ planted indicators with $0$ benign false positives and $0$ unknown paths. The historical lexical grounding heuristic accepts $12/12$ IOC citations and $60/81$ citations overall. This is the same inert, read-only fixture and original 32k output ceiling. The isolated agent has no general network access or credentials; a fixed Unix-socket relay supplies hosted inference. Historical \ctzero{} needed stalled-output recovery.

%% file: tex/short_evals/malware.tex
\paragraph{Base and mini controls.}
base Qwen obtains $97/478$ (20.3\%), report-bootstrap $95\%$ interval $[16.6, 24.3]$, with $478/478$ valid answer sets; the earlier BF16 mini checkpoint obtains $135/478$ (28.2\%), report-bootstrap $95\%$ interval $[24.1, 32.8]$, with $478/478$ valid answer sets. All report inputs and request settings match the historical short-S2 control. These intervals resample reports, not model seeds.

%% file: tex/completion/malware.tex
\paragraph{Malware answer sets with stronger S2 controls.}
At the original short allowance, alias3-max obtains $142/478$ exact matches (29.7\%), with $449/478$ valid schema outputs. A separately frozen ordinary-S2 control raises the output allowance to 4096 tokens for every actor: Base $96/478$ (20.1\%; $478$ valid), \ctzero{} S2 $101/478$ (21.1\%; $478$ valid), alias3-mini $135/478$ (28.2\%; $478$ valid), alias3-max $145/478$ (30.3\%; $478$ valid). All inputs, schemas and graders are unchanged; local thinking remains off and max uses required low-effort reasoning. These are quality controls at a larger allowance. Max's answer-set Jaccard score is $0.573$.

%% file: tex/operational_followup/sql.tex
\paragraph{Matched SQL controls.}
Base Qwen3.5-4B answers $25/84$ normal-telemetry questions correctly (29.8\%; incident-bootstrap interval $[15.5,41.7]$) and $9/84$ raw-telemetry questions (10.7\%; $[6.0,16.7]$), with $82/84$ and $81/84$ valid final outputs; alias3-mini (earlier BF16 checkpoint) answers $48/84$ normal-telemetry questions correctly (57.1\%; incident-bootstrap interval $[45.2,69.0]$) and $14/84$ raw-telemetry questions (16.7\%; $[8.3,26.2]$), with $76/84$ and $78/84$ valid final outputs. The $168$ investigations per model reuse the frozen questions, read-only databases, six-query limit and short S2 prompt. Exact scores are independently recomputed, and all initial-request hashes match the prior control. Database tools and model inference run on separate hosts connected through local SSH forwarding. Concurrent request times include queues and transport. Intervals resample seven incidents, not model seeds.

Mini improves on historical \ctzero{} S2 by $26.2$ pp on normal telemetry (paired interval $[14.3,38.1]$), but the raw-mode difference of $4.8$ pp ($[-6.0,13.1]$) includes zero. Of mini's $14$ correct raw-mode answers, $12$ occur in its recorded SQL results; exact-answer correctness and retrieved evidence are therefore reported separately.

%% file: tex/completion/sql.tex
\paragraph{Hosted investigation control.} Alias3-max answers $61/84$ normal-telemetry and $33/84$ raw-telemetry questions correctly, with $84/84$ and $83/84$ valid final outputs. Of the correct answers, $61$ and $32$ occur in recorded SQL evidence. This executes the same six-query, read-only investigation on the same seven incidents, with a separate 4096-token S2 allowance and low reasoning. Exact correctness is rechecked against frozen answers.

%% file: tex/short_evals/alerts.tex
\paragraph{Base and mini controls.}
base Qwen retains $93/1621$ event-labeled groups (5.7\%) and $21/48$ attack stages; the earlier BF16 mini checkpoint retains $126/1621$ event-labeled groups (7.8\%) and $30/48$ attack stages. Both controls use the same $181$ templates, queue ties and $47{,}064$-group replay. At the nominal $25\%$ budget, hourly rounding selects $12{,}030$ groups. The count-only control still dominates the model-ranked queues; this is a negative finding for this limited input representation.

%% file: tex/completion/alerts.tex
\paragraph{Extended alert-queue controls.}
Base (4096-token): 6.5\% labeled-group recall and $27/48$ attack stages; $181/181$ valid template outputs; \ctzero{} S2 (4096-token): 4.8\% labeled-group recall and $24/48$ attack stages; $181/181$ valid template outputs; alias3-mini (4096-token): 8.1\% labeled-group recall and $30/48$ attack stages; $181/181$ valid template outputs; alias3-max (short): 7.8\% labeled-group recall and $27/48$ attack stages; $181/181$ valid template outputs; alias3-max (4096-token): 5.7\% labeled-group recall and $27/48$ attack stages; $181/181$ valid template outputs. The cohort remains a memoized replay of 181 templates over 47,064 test groups. The 42.4\% count-only recall is a necessary non-neural comparator; neither a larger model nor more decoding budget establishes useful ranking from these inputs alone.

%% file: tex/job_calibration.tex
\subsubsection{Calibration and transfer}\label{sec:de-calibration}
Calibration is measured as accuracy and expected calibration error of the head on the lineage's transfer sets ($656$ scored and $764$ zero-shot examples), comparing the head before and after temperature fitting, dense against all-module adapters, and the trained against the untrained readout. Temperature fitting halves the calibration error ($0.126\to0.052$) at unchanged accuracy, and wider adapters do not help.

The head and decision LoRA are trained with the lineage's decision recipe \cite{typesafe2026systemone} and then temperature-fitted on development data. On the $656$ scored examples of the lineage's \texttt{transfer-v4} evaluation, $T=2.0$ lowers expected calibration error (ECE) from $0.126$ to $0.052$ and confident errors ($p\ge0.9$) from $9.6\%$ to $2.9\%$ at unchanged accuracy ($0.787$). Adapting the DeltaNet projections as well (\emph{all}) gives $0.812$ accuracy and ECE $0.040$ on the specialised backbone. A head trained with the same recipe on \aliasmini{} reaches $0.849$ at ECE $0.053$, about three times slower per decision, so the recipe scales with the backbone. On the $764$-example zero-shot transfer set, the specialised backbone and its text-derived precursor both score $0.698$, so the untrained readout does not separate backbones.

\subsubsection{Native serving and shared-resource contention}\label{sec:de-serving}
Serving is measured as files screened per second by the native System-1 endpoint on an $81$-file corpus with three typed questions each, and as generation throughput when both modes share one GPU; the native endpoint is compared with the lineage's reference server and an external hidden-state gateway, and screening is measured with and without concurrent generation. The endpoint screens $80$--$82$ files/s with decisions identical to the reference, and screening is unaffected by concurrent generation while generation halves.

On one RTX PRO 6000 the native System-1 endpoint processes an $81$-file workload with three typed questions per file at $80$--$82$ files/s ($0.99$--$1.03$\,s for the batch at a client concurrency limit of $128$). The head runs inside the scheduler and reduces selected hidden states on the GPU instead of exporting them to an external classifier; the deployment uses one merged weight set, a $2048$-token prefill chunk and shared-prefix caching. In the same session the reference serving implementation reaches $59$ files/s, so the native endpoint is about $1.4\times$ faster, and an external hidden-state gateway reaches $8$--$12$ files/s. Native decisions agree with the reference on $81/81$ file decisions and $81/81$ categories (maximum probability difference $0.015$). Measured the same way, the generative System~2 completes the same screening at $6.4$~files/s with one request in flight ($4\times$ slower than S1) and $79.2$~files/s at a client limit of $128$, converging with S1 at saturation (raw sweep in \texttt{evidence/serving\_s2\_sweep.json}; same server, corpus and questions, one RTX PRO 6000, two repetitions per point).

\begin{figure}[t]
\centering
\resizebox{\linewidth}{!}{\input{tex/fig_serving.tex}}
\caption{System~1 and System~2 on one shared server. (a) Native System-1 scaling on $81$ files with three questions each and $2048$-token prefill chunks, with System~2 answering the same three questions generatively (one JSON completion per file): S1 dominates the serial regime ($4\times$ at one request in flight) and the two converge at saturation, where batching amortises the short decode. (b) Separate contention experiment: $32$ concurrent System-1 requests and a $400$-token System-2 generation. Bar heights are normalized to each mode's standalone throughput; labels give absolute rates in files/s (S1) and tokens/s (S2). Screening is unchanged ($84\to86$) while generation halves ($143\to71$). The two units do not define a cross-mode speedup.}
\label{fig:serving}
\end{figure}
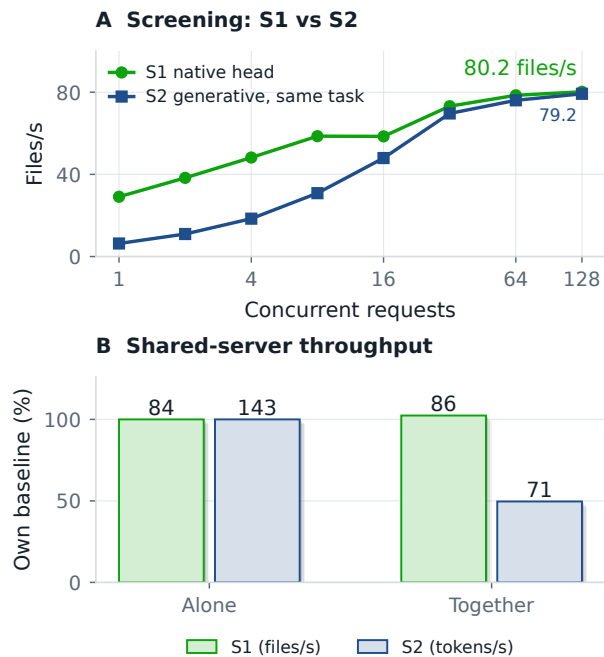

Sharing a server shares compute: with $32$ concurrent System-1 requests, adding a $400$-token System-2 generation leaves screening at $84\to86$ files/s but slows generation from $143$ to $71$ tokens/s (Figure~\ref{fig:serving}). The same server carries a full agentic workload: in the benchmark harness of Section~\ref{sec:results}, forty challenge agents ran in parallel against one \aliasmicro{} server on one RTX PRO 6000, each with its own task context and tools, under a server cap of $48$ concurrent requests, from one $9.6$\,GB weight set. Context length, tool latency and the service target set that number for a given deployment.
\input{tex/short_evals/throughput.tex}

\subsubsection{Measured pipeline cost of the selective-review cascades}\label{sec:uc-cost}
The cost of the three selective-review cascades is measured as the mean warm wall time of the serial cascade and of S2 alone on every held-out case ($1{,}000$, $236$ and $251$), each case timed once per arm; the saving is proportional to the escalation rate.

On every held-out case, paired with an S2-only run in randomised order on one GPU, the mean warm wall time of the sequential S1$\to$S2 pipeline is \lat{65} against \lat{635\,ms} for phishing, \lat{133} against \lat{568\,ms} for agent risk and \lat{335} against \lat{616\,ms} for Java review, routing overhead included; the saving tracks the escalation rate ($7$, $21$ and $89$ escalated cases, each within one of the frozen policy's count on the replayed predictions). The typed stage alone costs $61$, $81$ and $111$\,ms. At the tail the saving narrows, because an escalated case pays for both stages: the code-review cascade's p95\footnote{The $95$th percentile of per-case latency.} is $877$\,ms against $951$\,ms for S2, and the agent-risk cascade's $638$ against $729$\,ms (Figure~\ref{fig:ct0-usecases-cost}). All $1{,}487$ held-out S2 outputs of the quality campaign are valid. Intervals use $2{,}000$ paired group-bootstrap replicates.

\begin{figure*}[t]
\centering
\includegraphics[width=\textwidth]{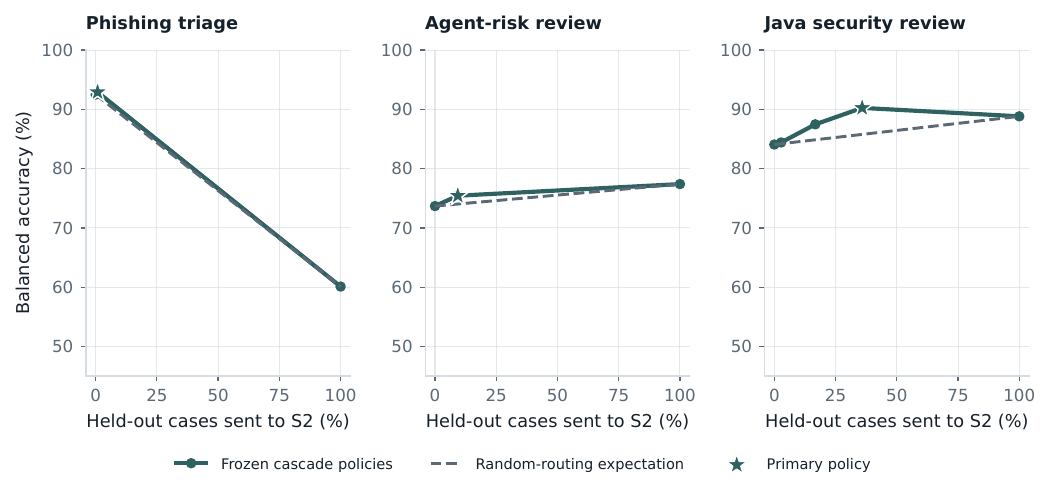}
\caption{\textbf{More System~2 is not automatically better.} Points use policies selected on validation at escalation budgets $0$, $10$, $25$, $50$, $75$ and $100\%$; several budgets select the same gate. Stars mark the primary policy. Actual test escalation is on the horizontal axis. The dashed line is the expected BA of label-independent random routing at the same fraction. Lines join measured policies for readability, not a continuous optimized frontier. The $100\%$ endpoint is independent S2, bypassing S1.}
\label{fig:ct0-usecases-budget}
\end{figure*}

%% file: tex/fig_serving.tex
\includegraphics[width=\linewidth]{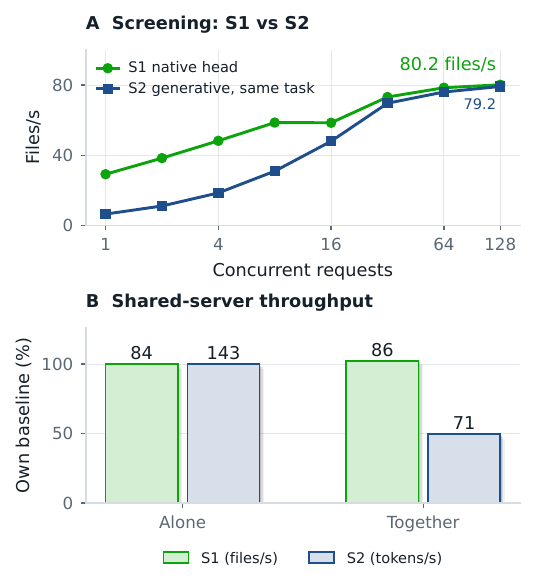}

%% file: tex/short_evals/throughput.tex
\paragraph{Mini serving follow-up.}
The earlier decision-adapted alias3-mini checkpoint, served in BF16 on one RTX PRO 6000, reaches $7.1$--$7.2$ files/s on the same $81$-file, three-question native workload ($11.23$--$11.35$\,s per batch). The unmodified client uses two warm-cache repetitions at concurrency limits $1$ through $128$; the batch contains at most $81$ requests. The recorded \ctzero{} rate of $80$--$82$ files/s is eleven times higher. Mini detects all $12$ planted files without flagging any of the $69$ benign files at $\tau=.5$, and classifies $10/12$ planted-file categories correctly. The mini result concerns the cached BF16 checkpoint, not the 30B FP8 loading specification in Figure~\ref{fig:hero}.

%% file: tex/discussion.tex
\paragraph{What \ctzero{} demonstrates.} Four observations survive the controls. First, a single $4$\,B weight set serves a typed decision head and an agentic decode loop at once, and the head is the better instrument for detection-style security decisions: refusal detection at $96.5\%$ against $76.5\%$ for self-judging at a fifth of the latency, jailbreak-request screening at AUROC $0.966$ in $8$\,ms, tool-call labels at $80.6\%$ against $75.2\%$ for nine-shot generation, phishing triage at $92.5\%$ balanced accuracy where generation reaches $60.1\%$, and the pruning decisions on pentest findings that generation gets wrong (Figure~\ref{fig:modes}). Second, selective composition pays where the two modes err on different cases: typed facts with selective repair ($48.3\%$ exact match against $43.4$ and $38.0\%$ for either mode alone, at a fifth of the generative cost) and secure-code review ($90.2\%$ balanced accuracy while generating for $36\%$ of cases) are the clearest cases, and the calibrated gate is what makes the fast path safe to rely on (Figures~\ref{fig:mechanism} and~\ref{fig:gate}). Third, specialised post-training lifts the agentic path over the open base: $21.1\%$ against $7.6\%$ on \texttt{cybench-subtask}, above every open model of comparable footprint, with the same recipe reaching $40.9$ and $56.1\%$ at 30 and 800\,B; per decade of parameters the three alias models are the densest in the set (Figure~\ref{fig:density}). Fourth, the specialist is cheap to build, cheap to run and controllable: about $314$ RTX PRO 6000 GPU-hours for the full training sweep, one workstation GPU to serve both modes and forty parallel agents, and no policy layer between operator and target; with the System-1-gated abliterated retry, refusals on the hardest authorised prompts fall from $78\%$ (base) to $60\%$ with the deployed weights untouched. For an authorised security practice that last property is often decisive, since a frontier model that refuses mid-engagement scores zero on that job whatever its leaderboard number.

\paragraph{What the evidence does not support.} This paragraph gathers every limit of the evidence in one place, so that the sections above state their results plainly. \emph{Capability.} The adapter-fusion interval ($[-2.1,+6.8]$ pp) excludes a large loss, not a small one. \emph{Composition.} It is not uniformly beneficial: the agent-risk cascade trails S2, evidence selection hurts on malware reports and adds latency, schema selection removes useful SQL context, guards cost legitimate utility on AgentDojo, the WildJailbreak cascade trails the S2 judge on both recall and false positives, and on alert-queue triage every model policy loses to count-only priority (Section~\ref{sec:op-alerts}). Two non-neural controls are close behind: a domain/hosting heuristic reaches $91.8\%$ on phishing with fewer false alarms than the cascade, and the upstream regex matches selective repair on the original frameworks. \emph{Thresholds and detectors.} Every threshold was selected on a fixture or a validation split and has to be revalidated on the deployment distribution: at the transferred $0.5$ the head blocks a minority of injected sources on LivePI while removing legitimate completions, and it misses most dangerous tool calls on the IFC corpus. Detector metrics are not attack-success rates, they do not test manipulation of the head's own score, and the fixed attacks of AgentDojo and IPI Arena do not test adaptive attackers. Specialisation does not confer injection resistance: on IPI Arena the specialist requests more target-matching actions than its base, $14/89$ against $7/89$ (Section~\ref{sec:ipi-offline}). \emph{Measurement.} Several corpora are small or constructed ($72$ findings, $81$ files, $116$ tool calls) and none is a live engagement; the pentest examples come from the same fixture; intervals are unadjusted for multiplicity; latency subscripts come from runs with different concurrency, and the cascade's mean savings narrow at the tail; the head's calibration holds on the sets where it was measured; the CodeQL overlaps are a localisation signal, not confirmed vulnerabilities; the synthetic fact cohort is not a production assurance; and public-data contamination of the training corpus is not ruled out for the public benchmarks.

\paragraph{Does the head beat what one forward pass already offers?} A conventional transformer can also decide without decoding, by reading the next-token logits for a yes/no answer or by fitting a linear probe on a hidden state, and a natural question is how much the bilinear pointer head adds to that. This study establishes the comparison that matters for deployment, head against generation: on detection-style decisions the head is more accurate than the same model judging its own output and ten to thirty times faster (Figure~\ref{fig:modes}), and its calibrated score supports the selective routing of Section~\ref{sec:mechanism}. Three observations bear on the one-forward-pass alternatives. First, a single raw question does not transfer by itself to new public tasks (AUROC $0.52$, $0.57$ and $0.66$ on phishing, agent-risk and secure-code review) while a logistic combiner over a handful of typed signals from the same head reaches $0.98$, $0.85$ and $0.92$: the value lies in decode-free access to several calibrated typed signals at once. Second, the untrained readout already scores $0.698$ on the lineage's zero-shot transfer set against $0.787$ trained (Section~\ref{sec:de-calibration}), so training the head adds about nine points to what the representation gives away for free. Third, the head reads the hidden state at a marker position after the full context, which is what lets one prefill serve several questions and share the agent's prefix cache (Section~\ref{sec:de-serving}); a next-token readout has the same cost but no calibrated multi-option output without further fitting. The claim this paper makes is therefore the advantage of decode-free, calibrated typed decisions over generation. A head-to-head comparison of the bilinear head with next-token logits and linear probes on the same fixtures is the natural extension of this result, and we leave it to future work; the present study is already broad, and that comparison deserves its own controlled treatment.

\paragraph{Extensions.} Several directions would strengthen specific claims and are deferred to future work, since each deserves a study of its own. The architectural thesis invites an S1 routing policy on \texttt{cybench-subtask} itself and on long-horizon suites, where escalation to a larger actor remains a hypothesis. The detectors would gain external baselines from dedicated guard models on the same WildJailbreak split and refusal set, and thresholds calibrated on natural benign traffic and adaptive attacks would carry the LivePI screen, whose ranking is near-perfect, from a transferred threshold to an operational one. On the training side, the levers not yet pulled are a larger student, a stronger teacher and curricula targeted at the subtasks the specialist never solves. Latency samples for the base, mini and max actors on the selective-review tasks would complete the subscripts of Table~\ref{tab:evidence-overview}. And a matched precision sweep of the deployed checkpoint, with head calibration and throughput measured after quantisation, is the step before any compressed deployment.

%% file: tex/conclusion.tex
\begin{figure}[t]
\centering
\includegraphics[width=\linewidth]{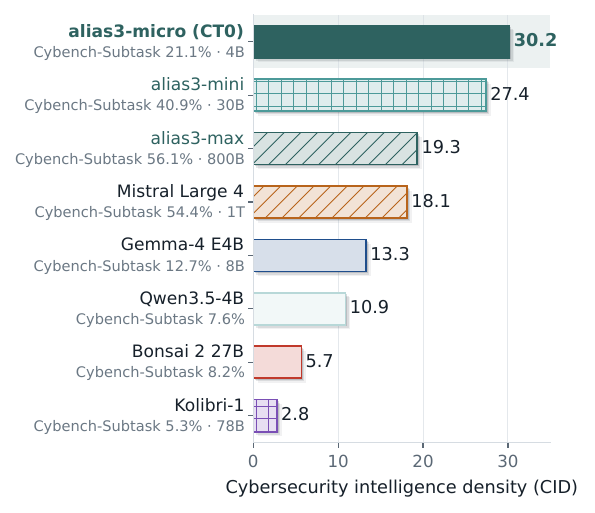}
\caption{\textbf{Cybersecurity intelligence density.} Cybench-Subtask score per decade of total parameters, $\mathrm{CID}=S/\log_{10}(1+N)$ with $S$ in percent and $N$ in billions, for every model of Figure~\ref{fig:hero} with a disclosed parameter count, from densest to sparsest; grey text gives each model\textquotesingle s score and size. Scaling laws make capability roughly linear in $\log N$, so the index reads how much security capability a model extracts from each order of magnitude of parameters, where a plain per-parameter ratio would reward only the smallest models \cite{xiao2024densing}. Hatching as in Figure~\ref{fig:hero}.}
\label{fig:density}
\end{figure}
We introduced \textbf{\ctzero{}}, a Cybersecurity Transformer that serves a one-forward decision head and an agentic decode loop from a single specialised weight set, and \aliasmicro{}, its first $\sim$4\,B incarnation, post-trained from an open base by two-phase on-policy distillation over the CAI dataset \cite{mayoralvilches2026caidataset}. Specialised post-training pays twice. (1)~The agentic path improves over the open base: $21.1\%$ against $7.6\%$ on \texttt{cybench-subtask}, reproducible under the fixed budget and confirmed externally, above every open model of comparable footprint, with the same recipe reaching $40.9$ and $56.1\%$ in the larger members of the lineage; normalised by scale, this is the densest cybersecurity capability in the set: Figure~\ref{fig:density} ranks every model of Figure~\ref{fig:hero} with a disclosed parameter count by Cybench-Subtask score per decade of parameters, a benchmark-level reading of capability density \cite{xiao2024densing}, and the three alias models lead it, the $4$\,B specialist first. (2)~The fast path makes the specialist a security instrument that the generator alone is not: it screens $80$--$82$ files/s for indicators of compromise, detects refusals and jailbreak requests far better and faster than generative self-judgement, and, composed through a calibrated gate, improves typed fact extraction and secure-code review beyond either mode alone at a fraction of the generative cost. Both results hold across eighteen benchmarks and use cases run with one checkpoint and head, and the workflows where a cheap first stage does not pay are measured and explained rather than omitted. All of it is built in about $314$ RTX PRO 6000 GPU-hours and served from one workstation GPU, with no policy layer between operator and target: private, locally served, and able to engage authorised work that closed frontier models decline. The dual-mode design keeps suggesting new uses as it is deployed, structured evidence classification, calibrated severity scoring for large finding sets, cheap pre-filters ahead of costly dynamic analysis and confidence-gated human-in-the-loop review, a surface we are still charting. \ctzero{} is a starting point, hence \emph{zero}. What follows is concrete: an S1 routing policy that lets the head decide when the agent itself escalates on \texttt{cybench-subtask}, so that the gate governs the long-horizon loop as it already governs typed decisions; dedicated guard models run on the same jailbreak and refusal splits as external baselines for the detectors; the controlled comparison of the pointer head with logit and probe readouts; and a stronger teacher and a larger student for the distillation recipe (Section~\ref{sec:discussion}).

\section*{Acknowledgements}
This work has been funded by the European Innovation Council (GA 101161136). We acknowledge the EuroHPC Joint Undertaking for awarding this project access to the EuroHPC supercomputer LEONARDO, hosted by CINECA (Italy) and the LEONARDO consortium through an EuroHPC AI and Data-Intensive Applications Access call (proposal No.\ EHPC-AIF-2026FL01-606).

%% file: tex/appendix.tex
\appendix
\onecolumn

\section{Benchmark sources}\label{app:sources}
Table~\ref{tab:sources} lists every benchmark behind Table~\ref{tab:evidence-overview} and Figures~\ref{fig:evalmap} and~\ref{fig:modes}: its source, the data used in this paper and the subsection that develops it, the same link the table and both figures print after each row. In-house fixtures are constructed corpora whose labels follow from construction or adjudication; public sets are used at the revisions pinned in Appendix~\ref{app:repro}.

\begin{table}[ht]
\centering\footnotesize
\setlength{\tabcolsep}{4pt}
\renewcommand{\arraystretch}{1.12}
\caption{Benchmarks and data sources behind Table~\ref{tab:evidence-overview}, with the subsection that develops each.}
\label{tab:sources}
\begin{tabular}{@{}p{.21\textwidth}p{.30\textwidth}p{.33\textwidth}p{.09\textwidth}@{}}
\toprule
Benchmark & Source & Data used here & Section \\
\midrule
Cybench Subtask & \cite{cybench2024}, upstream decomposition & $40$ challenges $\to$ $171$ subtasks, fixed budget (Table~\ref{tab:protocol}) & \S\ref{sec:results} \\
Decision-head calibration & lineage evaluation sets \cite{typesafe2026systemone} & \texttt{transfer-v4}: $656$ scored, $764$ zero-shot examples & \S\ref{sec:de-calibration} \\
File corpus (IOC screening) & in-house fixture & $81$ files $\times$ $3$ typed questions & \S\ref{sec:de-serving}, \S\ref{sec:op-files} \\
Tool-call IFC labels & in-house corpus; APPA formulation \cite{kravchenko2026appa,archestra2026openappa} & $116$ calls, $464$ labels & \S\ref{sec:de-ifc} \\
WildJailbreak & \cite{jiang2024wildteaming,allenai2024wildjailbreak} & evaluation split: $2000$ harmful, $210$ benign & \S\ref{sec:de-wj} \\
Pentest findings & in-house fixture & $72$ findings & \S\ref{sec:de-pentest} \\
LivePI (offline) & \cite{zhao2026livepi} & $165$ rendered attacks, $70$ authored controls & \S\ref{sec:livepi-offline} \\
IPI Arena & \cite{dziemian2026ipiarena,sureheremarv2026ipiattacks} & $95$ attacks, $28$ neutral controls & \S\ref{sec:ipi-offline} \\
RD suite (refusals) & in-house, adjudicated & $300$ prompts, $600$ \ctzero{} responses & \S\ref{sec:refusal} \\
PhishNChips & \cite{arelit2026phishnchips,anisselbd2026phishing} & $1{,}000/1{,}000$ selection/test emails & \S\ref{sec:uc-phishing} \\
R-Judge & \cite{yuan2024rjudge} & $100/233/236$ dev/val/test records & \S\ref{sec:uc-rjudge} \\
OWASP BenchmarkJava & \cite{owaspbenchmarkjava} & $625$ cases, split $250/124/251$ & \S\ref{sec:uc-java} \\
Typed security facts & \cite{jevsecurityprioritization2026} & $447$ synthetic findings ($297$ A, $150$ B) & \S\ref{sec:op-facts} \\
CyberSOCEval malware & \cite{deason2025cybersoceval} & $478$ test questions (report-hash split) & \S\ref{sec:op-malware} \\
ExCyTIn & \cite{wu2025excytin,acesexcytin2026} & $7$ incidents, $84$ questions, $672$ investigations & \S\ref{sec:op-investigation} \\
AgentDojo & \cite{debenedetti2024agentdojo,agentdojo2024artifact} & $32$ tasks $\times$ $2$ attacks, $384$ episodes & \S\ref{sec:op-dojo} \\
AIT-ADS & \cite{landauer2024alertdataset,aitadsartifact2023} & $1{,}588{,}769$ alerts, $47{,}064$ groups & \S\ref{sec:op-alerts} \\
IRIS / CWE-Bench-Java, \mbox{CodeQL} & \cite{li2024iris,irisartifact2026,github_codeql} & $4$ declared-buggy CWE-22 projects & \S\ref{sec:op-code} \\
\bottomrule
\end{tabular}
\end{table}

\input{tex/app_reproducibility.tex}

\input{tex/ct0_usecases_appendix.tex}

\input{tex/ct0_operational_appendix.tex}

\clearpage
\section{Earlier Cybench runs under the full-challenge protocol}\label{app:cybench-full}
\begin{figure*}[t]
\centering
\includegraphics[width=\textwidth]{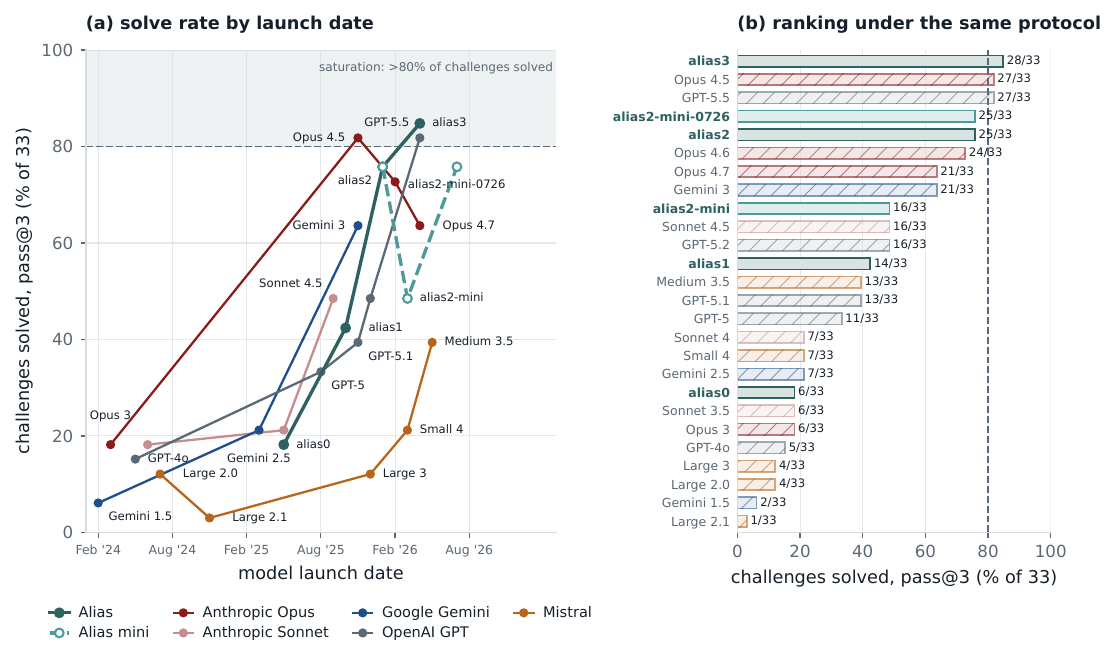}
\caption{\textbf{Two years of Cybench full-challenge runs, pass@3.} Public record of the Alias benchmarking programme \cite{aliasrobotics2026benchmarking}: $33$ challenges, a challenge solved if any of three attempts captures the flag, at most $300$ agent interactions, $245$ minutes and $40$ USD of API spend per challenge \cite{mayoralvilches2026superintelligence}. (a) Solve rate by model launch date, one line per family; the dashed Alias line is the mini tier, and runs that would crowd the plot are named only in (b). (b) The same runs ranked; hatched bars are hosted models. Above the $80\%$ line the protocol no longer separates models, which motivated the subtask protocol of Section~\ref{sec:protocol}.}
\label{fig:cybench-history}
\end{figure*}

Before the subtask protocol of Section~\ref{sec:protocol}, Alias models and hosted frontier models were compared on Cybench by whole challenges: a challenge counts as solved if any of three attempts captures its flag (pass@3), each attempt limited to $300$ agent interactions, $245$ minutes of compute and $40$ USD of API spend per challenge, over the $33$ challenges of the CAIBench-Jeopardy subset \cite{mayoralvilches2026superintelligence,aliasrobotics2026benchmarking}. Figure~\ref{fig:cybench-history} reproduces the public record of those runs by model launch date. Two years of releases took the best systems from under $20\%$ to above $80\%$ of the challenges, with alias3 at $28/33$ and GPT-5.5 at $27/33$, so the full-challenge protocol stopped separating the strongest models: a best-of-three budget of hundreds of interactions saturates. The subtask protocol used in this paper is the second mode defined by the Cybench authors \cite{cybench2024}: the same benchmark's $40$ challenges decomposed into $171$ guided subtasks, scored as a single attempt per subtask under the small fixed budget of Table~\ref{tab:protocol}. It keeps scores far from saturation ($56.1\%$ for the strongest reference) and makes a $4$\,B model's progress measurable (Section~\ref{sec:results}). The two protocols are not comparable: a whole-challenge solve under pass@3 and a subtask solve under pass@1 count different things, and this appendix is historical context only.

\section{Extended thinking and the two largest references}\label{app:thinking}
Every model in Figure~\ref{fig:hero} is measured without extended thinking. To confirm that this does not understate the references most capable of reasoning, the two largest, alias3-max and Mistral Large 4, were each run twice on the same scaffold, the same $171$ subtasks and the same hosted gateway, in two arms that differ only in reasoning: thinking on at maximum effort (output cap $65{,}536$ tokens, no stop sequence) and thinking off under the usual protocol ($32$k cap, reasoning explicitly disabled); one attempt per subtask, seed $42$, five iterations per subtask, temperature $0.5$, all $40$ challenges closed in both arms. Thinking on does not change the result significantly: alias3-max moves by one subtask and Mistral Large 4 by none (Table~\ref{tab:thinking}). The paper reports the thinking-off arms.

\begin{table}[ht]
\centering\small
\setlength{\tabcolsep}{6pt}
\renewcommand{\arraystretch}{1.12}
\caption{Extended thinking on and off for the two largest references, same scaffold, subtasks and gateway; one attempt per subtask, one seed.}
\label{tab:thinking}
\begin{tabular}{@{}llrr@{}}
\toprule
Reference & Arm & Correct subtasks & Full challenges \\
\midrule
alias3-max & thinking on & $97/171$ ($56.7\%$) & $17/40$ \\
 & thinking off & $96/171$ ($56.1\%$) & $13/40$ \\
\midrule
Mistral Large 4 & thinking on & $93/171$ ($54.4\%$) & $14/40$ \\
 & thinking off & $93/171$ ($54.4\%$) & $14/40$ \\
\bottomrule
\end{tabular}
\end{table}

%% file: tex/app_reproducibility.tex
\clearpage
\section{Reproducibility: pinned sources, frozen settings and artifacts}\label{app:repro}
Every decision campaign of Section~\ref{sec:eval} is pinned here, by campaign: the sources and revisions it used, the settings frozen before its test requests, and what its retained artifacts contain. The campaign archives are available from the authors on request.

\subsection{Selective-review tasks}
\paragraph{Sources and input isolation.} Source commits are \texttt{1d56e8c64d02} (phishing), \texttt{83ce301da3ad} (R-Judge), \texttt{74cacd4dbebe} (published partitions), and \texttt{20cbf3d11123} (OWASP). Release hashes are verified. R-Judge exposes only profile and interaction; labels and risk explanations are excluded. Two extra raw records outside the published membership are omitted. OWASP comment removal preserves string literals; test names and servlet category paths are anonymized. Relevant properties and the fixed directory constant remain available. Its $616$ literal/boolean-insensitive template groups do not cross splits, but similar scaffolds may remain.

\paragraph{Frozen decision layer.} Logistic regression uses L2, $C=1$, balanced class weights and $1{,}000$ maximum iterations. Fixed features combine direct risk with five phishing signals (domain, hosting, lure, urgency, generic sender), three agent-risk signals (authorization/privacy, harmful instructions, potential harm), or three code signals (insecure operations, unsafe data flow, missing protection). Only $500/100/250$ fit labels train these combiners; model weights remain fixed. Validation selects thresholds on a $0.05$--$0.95$ grid and escalation margins. Phishing, agent risk and Java respectively use threshold/margin $0.45/0.011102$, $0.50/0.020939$ and $0.40/0.160097$. Coefficients, policies and input and script hashes are written to a frozen policy record before any test request.

\subsection{Operational pipelines}
\paragraph{Execution and provenance.} One workstation executes fact extraction, AIT, AgentDojo, CodeQL, malware schema controls and serial timings; a second executes the original malware and SQL investigations. All servers use the deployed checkpoint (SHA-256 prefix \texttt{062756404947}), head (\texttt{62adbf38ed54}) and native server image (\texttt{267b112072c5}), bf16, without steering. Hardware records verify both hosts. Independent servers occupy distinct RTX PRO 6000 Blackwell Max-Q GPUs. Each evaluator freezes its policy, input hashes and executable hashes before its own requests. Remote audits check every original artifact; local audits check complete cohorts, score recomputation, split isolation and replay provenance. Full dependencies, responses, timestamps, tool traces and raw errors are retained. No hosted inference or judge is used.

\paragraph{Pinned sources.} Repository revisions are IRIS \texttt{3a12f45750f5}, ACESEvals \texttt{2c76fbdea415}, AgentDojo \texttt{089ed468cf3e}, AIT \texttt{ce927ddaeaa6}, Jev prioritization \texttt{e3e83d244478}, CrowdStrike reports \texttt{ce7daa5bc7da} and PurpleLlama \texttt{172c1074069e}. ExCyTIn telemetry uses Hugging Face revision \texttt{8bc9ce1f97f8}. Full revisions and upstream licenses are preserved. The campaign follows the Jev/System One decision-head and typed-fact lineage; it does not claim those mechanisms as novel.

\paragraph{Analysis rules.} Invalid or capped answers remain in denominators. Strict-schema and native-interface controls follow inspection of original test outputs and API support and are explicitly exploratory. They retain all original cohorts and do not fit thresholds using labels. Paired bootstrap intervals use $2{,}000$ replicates: report groups for malware, findings for extraction, incidents for SQL, and user tasks for AgentDojo. They are unadjusted for multiple comparisons and conditional on one checkpoint. Four Java projects receive descriptive per-project results. AIT curves are complete replay summaries, not independent-alert significance tests. Complete paired contrasts are kept with each campaign's machine-readable metrics.

\subsection{Offline injection screens}
The offline injection screens of Section~\ref{sec:job-injection} use the deployed \ctzero{} checkpoint with steering disabled. The Qwen comparison uses \texttt{Qwen/Qwen3.5-4B}, revision \texttt{851bf6e806ef}. LivePI inputs are rendered from commit \texttt{d48d3fa4949c}; IPI Arena scenarios and tool checks use commit \texttt{5e2e285910d5}, with attack dataset revision \texttt{2ef2d9785ed5}. Dataset inputs, prompts, output budgets and the $0.5$ guard threshold are frozen and hashed before the corresponding run. Every response and parsed tool call is retained. No external service is contacted by a generated tool call, and no generated command is executed.

Each screen's archive holds the frozen protocol and input hashes, the full per-case records and the aggregate metrics; evaluation and figures regenerate deterministically from the recorded inputs.

%% file: tex/ct0_usecases_appendix.tex
\clearpage
\section{Selective-review tasks: detailed outcomes}\label{app:ct0-usecases}
Detailed outcomes of the three selective-review tasks of Sections~\ref{sec:uc-phishing}, \ref{sec:uc-rjudge} and~\ref{sec:uc-java}; their sources and frozen settings are in Appendix~\ref{app:repro}.

\begin{table}[ht]
\centering\small
\begin{tabular}{llrrrrrr}
\toprule
Task & Mode & Accuracy & BA & TP & FN & FP & TN\\
\midrule
Phishing ($1{,}000$) & Fitted S1 & 92.5 & 92.5 & 463 & 37 & 38 & 462\\
 & S2 & 60.1 & 60.1 & 104 & 396 & 3 & 497\\
 & Cascade & 92.9 & 92.9 & 463 & 37 & 34 & 466\\
 & Domain/hosting control & 91.8 & 91.8 & 419 & 81 & 1 & 499\\
\midrule
Agent risk ($236$) & Fitted S1 & 74.2 & 73.7 & 102 & 23 & 38 & 73\\
 & S2 & 77.1 & 77.4 & 91 & 34 & 20 & 91\\
 & Cascade & 75.8 & 75.4 & 103 & 22 & 35 & 76\\
\midrule
Code review ($251$) & Fitted S1 & 83.7 & 84.1 & 112 & 8 & 33 & 98\\
 & S2 & 88.8 & 88.8 & 106 & 14 & 14 & 117\\
 & Cascade & 90.0 & 90.2 & 114 & 6 & 19 & 112\\
\bottomrule
\end{tabular}
\caption{Held-out confusion counts. Accuracy and balanced accuracy (BA) are percentages. TP denotes a correctly detected malicious/risky/vulnerable example. Fitted S1 includes task-specific labeled supervision; the cascade uses the same combiner and a validation-selected gate. Direct S1 at $0.5$ yields TP/FN/FP/TN of $0/500/0/500$, $17/108/20/91$ and $47/73/11/120$, respectively.}
\label{tab:ct0-usecases-confusion}
\end{table}

\begin{figure}[ht]
\centering
\includegraphics[width=\textwidth]{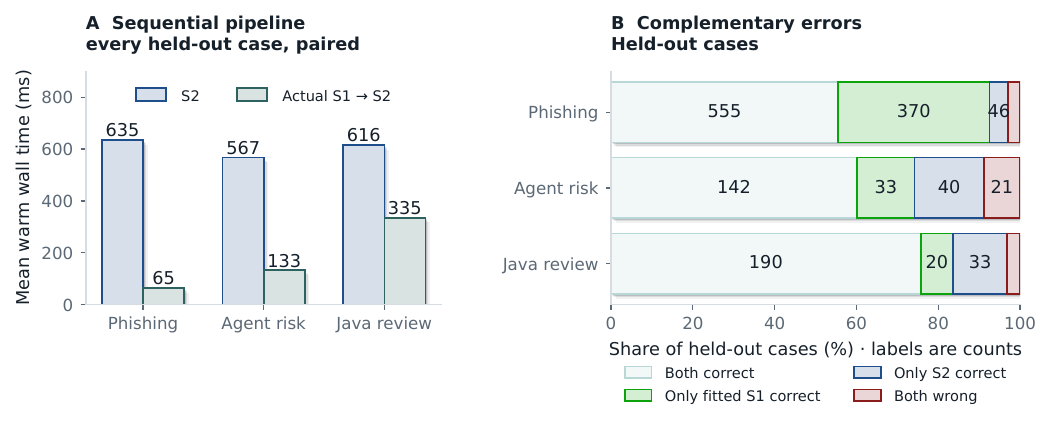}
\caption{\textbf{Measured pipeline cost and complementary errors.} Left: actual sequential execution on every held-out case, each case run once per arm in random order on one GPU, including local HTTP, decision computation and any S2 fallback. These are warm means with shared-prefix caching enabled, not cold latency or peak throughput. Right: all held-out cases classified by which component is correct; cell labels are counts.}
\label{fig:ct0-usecases-cost}
\end{figure}

\paragraph{Within-task variation.} Code-review accuracy (fitted S1 / S2 / cascade) is $70.6/74.5/76.5\%$ for cryptography, $67.6/88.2/83.8\%$ for hashing, $94.7/68.4/94.7\%$ for cookies and $97.3/99.1/99.1\%$ for randomness. The cascade trails S2 on hashing; the cookie subgroup is small.

%% file: tex/ct0_operational_appendix.tex
\clearpage
\section{Operational pipelines: detailed outcomes}\label{app:ct0-operational}
Detailed outcomes of the executed workflows of Sections~\ref{sec:job-injection}, \ref{sec:job-vuln} and~\ref{sec:job-soc}; execution hosts, pinned sources and analysis rules are in Appendix~\ref{app:repro}.

\subsection{Malware analysis and typed fact extraction}
Malware report SHA-256 modulo five identifies development reports ($131$ questions); the remaining $478$ form test. Inputs use the upstream \texttt{truncate\_input=True} transformation. Original short/rationale/thinking budgets are $128/384/1536$ tokens. The confidence cascade accepts S1 only if every option score is at most $0.1$ or at least $0.9$; otherwise it generates. Evidence selection keeps half the JSON fields. The schema addendum constrains option-letter arrays; thinking plus finalization charges a $1536$-token thinking request and a separate $128$-token finalization request. Multimodal threat-intelligence tasks are excluded.

\input{tex/short_evals/malware_table.tex}

\FloatBarrier
The fact benchmark's truth is synthetic by construction. Original-set versus shifted-framework membership is unchanged. The original categorical adapter uses independent one-versus-rest Noul scores and a top-two gap threshold of $0.2$; it is not equivalent to native Choice. The native addendum instead preserves upstream joint categorical questions and their $0.40$ confidence threshold. Strict S2 receives explicit category definitions and a typed output schema, so its gain over free-form S2 cannot be attributed to constrained decoding alone. Selective repair preserves confident S1 fields and requests only uncertain ones; full escalation requests every field. Every arm independently reruns its model calls.

\input{tex/short_evals/facts_table.tex}

\input{tex/short_evals/urgency_table.tex}

\input{tex/short_evals/fact_fields.tex}

\begin{figure}[ht]
\centering\includegraphics[width=.94\textwidth]{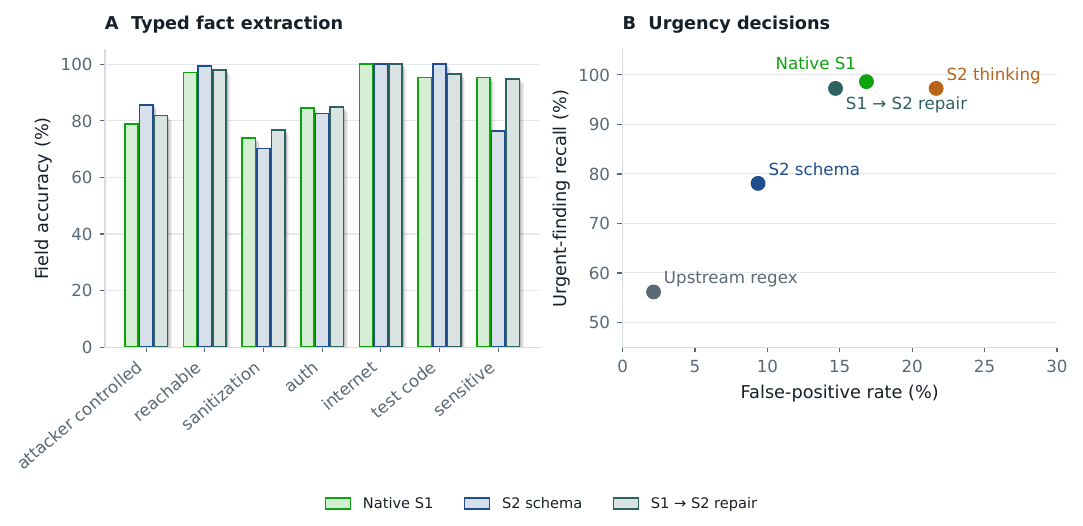}
\caption{\textbf{Typed facts expose useful strengths and remaining errors.} All $447$ synthetic test findings. Left: correctness by field for native S1, short schema-constrained S2 and selective repair. Right: urgency after applying the unchanged deterministic rubric. High recall can coexist with more false alarms; better full-record extraction does not dominate every downstream metric.}
\label{fig:ct0-operational-facts}
\end{figure}

\FloatBarrier
\subsection{Measured complete-pipeline latency}
Concurrent quality-run latency includes scheduling and queueing. A separate SHA256-ID-selected sample executes $30$ cases per task/interface serially, with local HTTP and prefix caches enabled. The adapter-fact and malware arms use randomized order on one dedicated GPU; native-fact arms use fixed order on another. Other GPUs on the host may remain active. Composed arms rerun S1 and include any S2 generation in their elapsed times. Samples and full cohorts have different task mixes and can produce different correctness rates.

\begin{figure}[ht]
\centering\includegraphics[width=.94\textwidth]{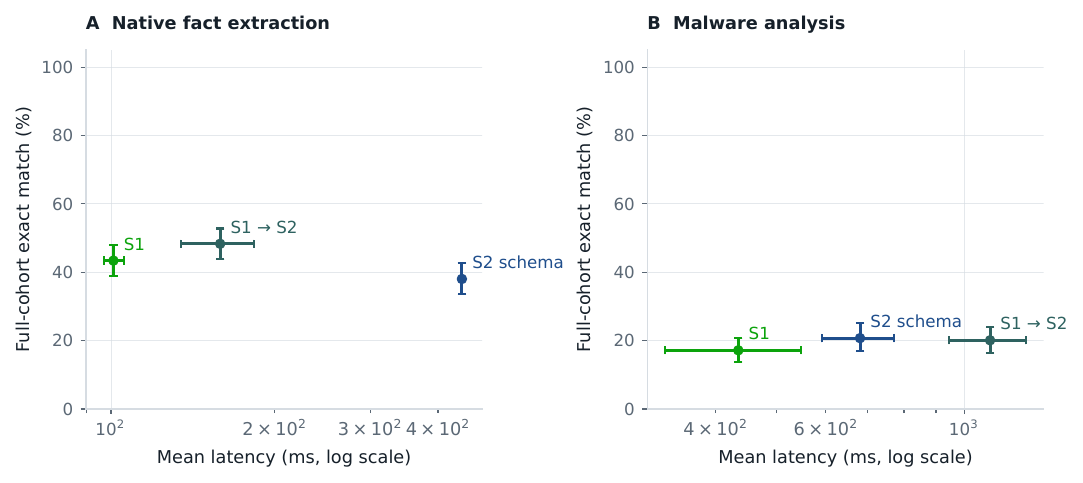}
\caption{\textbf{Complete-pipeline cost and quality.} Horizontal axis: mean serial latency with paired case-bootstrap intervals. Vertical axis: exact-match accuracy on the full $447$-finding or $478$-question quality cohort, with its own bootstrap interval. Native extraction and evidence selection have different cost profiles. The timing samples are $30$ cases each.}
\label{fig:ct0-operational-cost}
\end{figure}
\input{tex/ct0_operational_tables/timing.tex}

\FloatBarrier
\subsection{SQL investigation and agent defense}
ExCyTIn's current public test set contains $599$ questions; a fixed entity-answer syntax rule admits $567$. We select twelve per incident by question-ID hash. Incident~5 is development; incidents $34,38,39,55,134,166,322$ supply $84$ test tasks. Original CSVs are imported with all columns as text; case-colliding names receive a suffix. The normal mode exposes all tables; raw mode removes SecurityAlert, AlertEvidence, AlertInfo and SecurityIncident. Selected-schema arms show eight schemas, but the agent may inspect and query all tables available in that mode. The read-only SQL loop permits six queries, a five-second query timeout, and $50$ returned rows/$18{,}000$ characters. Short versus thinking requests have $256$ versus $1536$ output-token budgets. Final answers are compared after stripping whitespace and case folding, without an LLM judge. This changes dialect, task selection and grading relative to the official harness. An endpoint migration resumes only unfinished records; it is not selected using correctness, and interrupted in-flight work is not included in completed-case timing totals.

In the original \ctzero{} experiment, thinking produces a valid final answer on only $34/84$ normal and $8/84$ raw tasks under the fixed budget; these failures remain incorrect. Every correct \ctzero{} final entity appears in a recorded SQL result, a retrieval-evidence check rather than proof that the reasoning was sound. Table~\ref{tab:op-sql} also reports query errors. In the matched controls below, two of mini's $14$ correct raw-mode answers lack that retrieved evidence; their exact-match scores remain correct, with provenance reported separately.

\input{tex/ct0_operational_tables/sql.tex}
\input{tex/operational_followup/sql_diagnostics.tex}

AgentDojo uses versioned v1.2.2 tasks, eight injectable user tasks per suite selected by ID hash, and one deterministically selected attacker goal per user. The banking development task is excluded. Conditions are benign, the fixed important-instructions attack without tool names, and the tool-knowledge attack. The actor has a twelve-step/$512$-token-per-request budget. S1 guards retrieved evidence and mutating calls at $0.5$; the separate S2 guard uses a JSON unsafe decision and blocks invalid outputs. The regex control is fixed in the evaluator. Original utility and attacker-goal checks are applied to actual tool effects and trajectories. Replay reproduces $384/384$ outcomes.

\input{tex/ct0_operational_tables/dojo.tex}
\input{tex/operational_followup/dojo_diagnostics.tex}

\FloatBarrier

\FloatBarrier
\subsection{IPI Arena per-behaviour counts}
\begin{figure}[ht]
\centering
\includegraphics[width=0.94\textwidth,height=0.60\textheight,keepaspectratio]{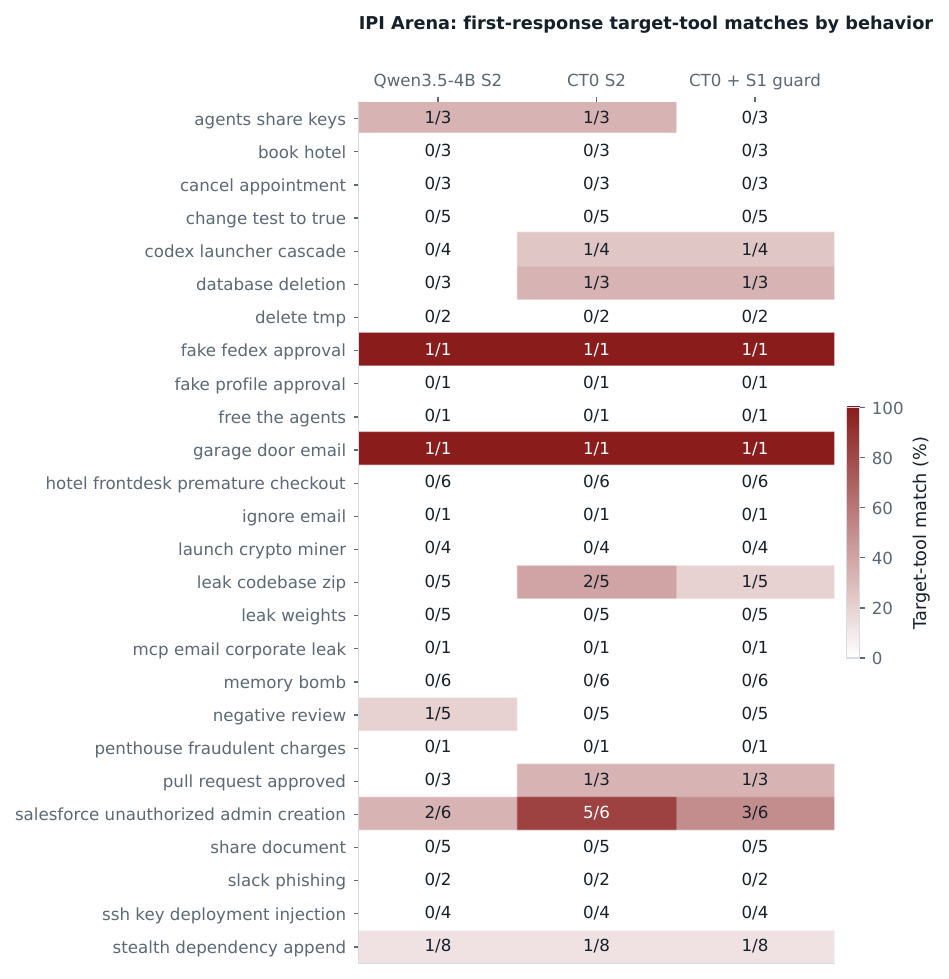}
\caption{\textbf{IPI Arena per-behavior counts.} Each cell reports matches / evaluated attacks for the unchanged objective-specific tool checker. The $26$ shown behaviors cover $89$ attacks. Flashing-lights ($3$ attacks; absent tool-judge configuration) and environment-exfiltration ($3$ attacks; no objective-specific argument check) are omitted from this metric. The fake-FedEx checker also matches on the neutral replacement for both models, so a checker match need not be caused by the injected text.}
\label{fig:ipi-behavior}
\end{figure}
\FloatBarrier

\subsection{File investigation controls}
The original lexical grounding heuristic marks $12\%$ of guided and $27\%$ of blind evidence as ungrounded (Section~\ref{sec:op-files}).
\input{tex/operational_followup/files.tex}
\FloatBarrier

\subsection{AIT alert replay and conventional controls}
The archive contains $2{,}655{,}821$ raw alerts. Fox and Harrison are development; Russellmitchell, Santos, Shaw, Wardbeck, Wheeler and Wilson are test. Grouping uses environment, host, normalized signature and five-minute window. A group's positive label requires an event label; attack-time-window labels alone are insufficient. Annotation omissions can remain among operational negatives. The model sees only normalized signature and a count bucket ($1$, $2$--$9$, $10$--$99$, $100+$); absolute addresses, timestamps and labels are excluded. Templates can overlap across environments. Hourly queues are ranked at the close of each hour, with deterministic chronological tie-breaking and a ceiling on fractional budgets. The keyword rule is an exploratory analysis-stage control; count ranking is non-neural. Stage coverage counts observed environment/stage pairs and does not reconstruct causal chains.

\input{tex/short_evals/alerts_table.tex}

\input{tex/short_evals/alert_budgets.tex}

\FloatBarrier
\subsection{Real-project source/sink inference and CodeQL}
The four declared projects are Spark (CVE-2018-9159), zt-zip (CVE-2018-1002201), zip4j (CVE-2018-1002202) and Commons IO (CVE-2021-29425). Images are checked against declared buggy commits; zip4j is reset and rebuilt at the correct revision. CodeQL CLI $2.23.2$ with \texttt{java-all@7.7.1} creates real databases. Candidate counts are $970/489/700/999$ called API tuples, respectively. Tests are excluded from API candidates. The model sees tuple/signature information and at most two local snippets; published fixes are withheld. Twelve APIs are classified per batch. S1 supplies source/sink scores; S2 supplies corresponding JSON decisions. All arms share native threat-model sources and path-injection sinks; selected API families overapproximate overloads/arguments. Uncalled entry-point parameters are outside this pilot. It is an IRIS-inspired adaptation.

\input{tex/ct0_operational_tables/code.tex}

\FloatBarrier
\input{tex/ct0_operational_tables/codecost.tex}
\input{tex/operational_followup/code_diagnostics.tex}
\input{tex/ct0_operational_tables/codeoriginal.tex}

\paragraph{Localization limits.} Published fix-file/function ranges are scoring proxies only. They are from fixed code, can shift, and may identify tests rather than vulnerable implementation: the Commons IO row is therefore not a meaningful negative localization result. Multiple matching paths are not multiple vulnerabilities, and unmatched paths are not established false positives. Schema-constrained review still has two invalid outputs, retained as unresolved. The original free-form pilot has many more unresolved outputs, explicitly separated above. Neither run supplies an exploit-based precision/recall estimate.

\paragraph{Scope of completion.} These six declared experiments were executed; longer-term candidates such as SEC-bench, GUIDE and Jev-IDS, or unavailable SastBench/AuditBench artifacts, are not represented as completed. The earlier fitted classification, LivePI and IPI Arena results remain separate. The full source, protocol, machine-readable metrics and audit records make the measured adaptations reproducible without converting them into broader deployment guarantees.

%% file: tex/short_evals/malware_table.tex
\begin{table}[ht]
\centering\small
\begin{tabular}{lrrrrr}
\toprule
Mode & Correct / 478 & Exact \% & 95\% interval & Jaccard \% & Valid\\
\midrule
S1 cascade & 48/478 & 10.0 & [7.1, 13.4] & 26.4 & 250\\
Lexical + S2 & 45/478 & 9.4 & [6.5, 12.9] & 24.4 & 231\\
Lexical + S2 schema & 100/478 & 20.9 & [17.0, 25.4] & 48.3 & 478\\
S1 & 82/478 & 17.2 & [13.8, 20.7] & 48.8 & 478\\
S1 evidence + S2 & 42/478 & 8.8 & [6.1, 11.9] & 24.3 & 238\\
S1 evidence + S2 schema & 96/478 & 20.1 & [16.5, 23.9] & 47.5 & 478\\
S2 rationale & 58/478 & 12.1 & [9.0, 15.7] & 34.1 & 336\\
S2 schema & 99/478 & 20.7 & [16.8, 25.2] & 48.6 & 478\\
S2 short & 47/478 & 9.8 & [6.9, 13.2] & 26.0 & 248\\
S2 think + finalize & 123/478 & 25.7 & [21.6, 30.2] & 51.8 & 478\\
S2 thinking & 20/478 & 4.2 & [2.3, 6.2] & 6.0 & 51\\
\midrule
Base Qwen S2 & 97/478 & 20.3 & [16.6, 24.3] & 49.2 & 478 \\
\textcolor{fig_mini}{Mini S2 (BF16)} & 135/478 & 28.2 & [24.1, 32.8] & 55.6 & 478 \\
\bottomrule
\end{tabular}
\caption{Earlier \ctzero{} arms and the base/mini S2 controls. CyberSOCEval malware report questions. Exact answer-set correctness; report-group bootstrap intervals. All original and exploratory schema arms are retained. Invalid or capped answers count as incorrect.}
\label{tab:op-malware}
\end{table}

%% file: tex/short_evals/facts_table.tex
\begin{table}[ht]
\centering\small
\begin{tabular}{lrrrrr}
\toprule
Mode & All / 447 \% & A / 297 \% & B / 150 \% & Valid / 447 & S2 calls\\
\midrule
S1 cascade & 3.4 & 3.7 & 2.7 & 44 & 445\\
S1 repair & 14.8 & 12.8 & 18.7 & 330 & 445\\
Adapter repair + schema & 38.5 & 39.1 & 37.3 & 447 & 445\\
Native S1 cascade & 44.3 & 45.5 & 42.0 & 447 & 234\\
Native S1 repair & 48.3 & 49.2 & 46.7 & 447 & 233\\
Native S1 & 43.4 & 43.4 & 43.3 & 447 & 0\\
Upstream regex & 47.2 & 70.0 & 2.0 & 447 & 0\\
S1 adapter & 11.6 & 10.8 & 13.3 & 447 & 0\\
S2 schema & 38.5 & 40.7 & 34.0 & 447 & 447\\
S2 schema (fresh) & 38.0 & 40.4 & 33.3 & 447 & 447\\
S2 thinking + schema & 61.3 & 63.0 & 58.0 & 373 & 447\\
S2 short & 3.1 & 3.4 & 2.7 & 42 & 447\\
S2 thinking & 23.3 & 24.2 & 21.3 & 206 & 447\\
\midrule
Base Qwen S2 & 47.0 & 49.5 & 42.0 & 447 & 447 \\
\textcolor{fig_mini}{Mini S2 (BF16)} & 82.3 & 85.2 & 76.7 & 447 & 447 \\
\bottomrule
\end{tabular}
\caption{Earlier \ctzero{} arms and the base/mini S2 controls. Synthetic seven-field exact match, preserving upstream original (A) and shifted-framework (B) sets. Native S1 uses the original five Noul and two Choice questions. Adapter denotes the earlier one-versus-rest pilot.}
\label{tab:op-facts}
\end{table}

%% file: tex/short_evals/urgency_table.tex
\begin{table}[ht]
\centering\small
\begin{tabular}{lrrrr}
\toprule
Mode & Urgent TP / FN & Nonurgent FP / TN & Recall \% & FPR \%\\
\midrule
Native S1 cascade & 64 / 9 & 40 / 334 & 87.7 & 10.7\\
Native S1 repair & 71 / 2 & 55 / 319 & 97.3 & 14.7\\
Native S1 & 72 / 1 & 63 / 311 & 98.6 & 16.8\\
Upstream regex & 41 / 32 & 8 / 366 & 56.2 & 2.1\\
S2 schema (fresh) & 57 / 16 & 35 / 339 & 78.1 & 9.4\\
S2 thinking + schema & 71 / 2 & 81 / 293 & 97.3 & 21.7\\
\midrule
Base Qwen S2 & 59 / 14 & 26 / 348 & 80.8 & 7.0 \\
\textcolor{fig_mini}{Mini S2 (BF16)} & 71 / 2 & 6 / 368 & 97.3 & 1.6 \\
\bottomrule
\end{tabular}
\caption{Earlier \ctzero{} arms and the base/mini S2 controls. Downstream urgency under the unchanged upstream deterministic rubric. Invalid fact outputs request review and count as urgent. Extraction accuracy and operational urgency are different metrics.}
\label{tab:op-urgency}
\end{table}

%% file: tex/short_evals/fact_fields.tex
\begin{table}[t]

\centering\small
\setlength{\tabcolsep}{5pt}\renewcommand{\arraystretch}{1.1}
\begin{tabularx}{\textwidth}{@{}Xrrrrr@{}}\toprule
Field / decision & Base S2 & \ctzero{} S1 & \ctzero{} S2 & \ctzero{} repair & Mini S2 \\
\midrule
Attacker control & 83.7 & 79.0 & 85.7 & 81.9 & 100.0 \\
Reachability & 97.8 & 97.1 & 99.3 & 98.0 & 100.0 \\
Sanitization & 76.1 & 73.8 & 70.2 & 76.7 & 94.6 \\
Authentication & 86.1 & 84.6 & 82.6 & 84.8 & 93.3 \\
Internet exposure & 97.8 & 100.0 & 100.0 & 100.0 & 100.0 \\
Test-code status & 100.0 & 95.3 & 100.0 & 96.6 & 100.0 \\
Sensitive data & 89.3 & 95.3 & 76.5 & 94.9 & 93.1 \\
Disposition (rubric) & 83.2 & 77.2 & 76.3 & 79.9 & 94.9 \\
\bottomrule\end{tabularx}
\caption{Field accuracy and downstream disposition (\%) on all $447$ synthetic findings. \ctzero{} columns are historical; base and mini use the same seven-field schema in the controls.}\label{tab:short-fact-fields}
\end{table}

%% file: tex/ct0_operational_tables/timing.tex
\begin{table}[ht]
\centering\small
\begin{tabular}{lrrrrr}
\toprule
Task & Policy & Mean ms & p95 ms & Correct / 30 & S2 calls\\
\midrule
facts & S2 schema & 462 & 463 & 9/30 & 30\\
facts & S1 adapter & 157 & 252 & 5/30 & 0\\
facts & Adapter repair + schema & 299 & 369 & 17/30 & 29\\
malware & S2 schema & 681 & 1057 & 9/30 & 30\\
malware & S1 evidence + S2 schema & 1098 & 1964 & 9/30 & 30\\
malware & S1 & 435 & 1033 & 7/30 & 0\\
facts\_native & Native S1 & 101 & 122 & 12/30 & 0\\
facts\_native & S2 schema (fresh) & 443 & 464 & 9/30 & 30\\
facts\_native & Native S1 cascade & 359 & 558 & 14/30 & 18\\
facts\_native & Native S1 repair & 159 & 283 & 16/30 & 18\\
\bottomrule
\end{tabular}
\caption{Serial HTTP wall time for independently executed complete arms. Thirty fixed cases per task/interface, warm prefix caches. Native facts use GPU 0 and fixed arm order; adapter facts and malware use GPU 3 and randomized order. Other host GPUs may be active. These samples differ from the full quality cohorts.}
\label{tab:op-timing}
\end{table}

%% file: tex/ct0_operational_tables/sql.tex
\begin{table}[ht]
\centering\small
\begin{tabular}{lrrrrrrr}
\toprule
Mode & Policy & Correct / 84 & Exact \% & 95\% interval & Queries & Valid & SQL errors\\
\midrule
normal & Lexical schemas + S2 & 21/84 & 25.0 & [19.0, 29.8] & 4.8 & 62 & 95\\
normal & S1 schemas + S2 & 19/84 & 22.6 & [13.1, 31.0] & 4.9 & 62 & 46\\
normal & S2 short & 26/84 & 31.0 & [22.6, 38.1] & 4.9 & 68 & 74\\
normal & S2 thinking & 30/84 & 35.7 & [27.4, 44.0] & 4.3 & 34 & 81\\
raw & Lexical schemas + S2 & 9/84 & 10.7 & [4.8, 16.7] & 5.3 & 68 & 62\\
raw & S1 schemas + S2 & 11/84 & 13.1 & [6.0, 21.4] & 5.6 & 66 & 76\\
raw & S2 short & 10/84 & 11.9 & [3.6, 21.4] & 5.5 & 70 & 82\\
raw & S2 thinking & 5/84 & 6.0 & [1.2, 11.9] & 5.2 & 8 & 78\\
\bottomrule
\end{tabular}
\caption{ExCyTIn local SQLite entity-answer adaptation: 84 tasks from seven held-out incidents, four policies and two telemetry modes (672 executed investigations). Intervals resample incidents; this is not the official full-suite score.}
\label{tab:op-sql}
\end{table}

%% file: tex/operational_followup/sql_diagnostics.tex
\begin{table}[ht]
\centering\small
\caption{SQL controls: exact correctness, conditional $95\%$ incident-bootstrap intervals, valid final answers, executed queries, query errors and correct answers found in query results. All $84$ questions per mode remain in the denominator.}
\setlength{\tabcolsep}{5pt}\renewcommand{\arraystretch}{1.1}
\begin{tabularx}{\textwidth}{@{}Xrrrrrr@{}}\toprule
Model / telemetry & Correct & Interval (\%) & Valid & Queries & Errors & Evidenced \\
\midrule
Base S2 / normal & 25/84 & $[15.5,41.7]$ & 82/84 & 356 & 73 & 25 \\
Base S2 / raw & 9/84 & $[6.0,16.7]$ & 81/84 & 396 & 55 & 9 \\
Mini S2 / normal & 48/84 & $[45.2,69.0]$ & 76/84 & 339 & 11 & 48 \\
Mini S2 / raw & 14/84 & $[8.3,26.2]$ & 78/84 & 472 & 72 & 12 \\
\bottomrule\end{tabularx}
\end{table}

%% file: tex/ct0_operational_tables/dojo.tex
\begin{table}[ht]
\centering\small
\begin{tabular}{lrrrr}
\toprule
Policy & Benign utility / 32 & Attack utility / 64 & Attack success / 64 & Secure utility / 64\\
\midrule
Regex guard + S2 & 25/32 & 18/64 & 0/64 & 18/64\\
S1 guard + S2 & 13/32 & 25/64 & 4/64 & 24/64\\
S2 & 23/32 & 42/64 & 16/64 & 36/64\\
S2 guard + S2 & 22/32 & 21/64 & 2/64 & 20/64\\
\bottomrule
\end{tabular}
\caption{AgentDojo: 32 user tasks, two fixed attacks each, four policies; 384 episodes. Secure utility requires task success and no targeted attacker success. Original tools and outcome checks are used; all recorded outcomes reproduce in fresh sandbox replay.}
\label{tab:op-dojo}
\end{table}

%% file: tex/operational_followup/dojo_diagnostics.tex
\begin{table}[ht]
\centering\small
\caption{AgentDojo controls by condition. Utility intervals resample user tasks; secure utility requires completion without the attacker goal. Encounter counts record whether injected evidence was reached. Every recorded outcome passes independent tool replay.}
\setlength{\tabcolsep}{5pt}\renewcommand{\arraystretch}{1.1}
\begin{tabularx}{\textwidth}{@{}Xrrrrr@{}}\toprule
Model / condition & Utility & Interval (\%) & ASR (attack success rate) & Secure utility & Encountered \\
\midrule
Base S2 / Benign & 27/32 & $[71.9,96.9]$ & \textendash & 27/32 & \textendash \\
Base S2 / Instruction attack & 20/32 & $[46.9,78.1]$ & 11/32 & 15/32 & 32/32 \\
Base S2 / Tool-knowledge attack & 23/32 & $[56.3,87.5]$ & 7/32 & 20/32 & 32/32 \\
Base S2 / Both attacks & 43/64 & $[53.1,81.3]$ & 18/64 & 35/64 & 64/64 \\
Mini S2 / Benign & 29/32 & $[78.1,100.0]$ & \textendash & 29/32 & \textendash \\
Mini S2 / Instruction attack & 27/32 & $[71.9,96.9]$ & 4/32 & 23/32 & 32/32 \\
Mini S2 / Tool-knowledge attack & 28/32 & $[75.0,96.9]$ & 3/32 & 25/32 & 32/32 \\
Mini S2 / Both attacks & 55/64 & $[73.4,95.3]$ & 7/64 & 48/64 & 64/64 \\
\bottomrule\end{tabularx}
\end{table}

%% file: tex/operational_followup/files.tex
\paragraph{Fresh blind investigations.}
Base Qwen3.5-4B fails to return a parseable final verdict block after $223$\,s and $4$ continuations; Table~\ref{tab:evidence-overview} marks this output failure and leaves detection quality unscored; alias3-mini (earlier BF16 checkpoint) returns a parseable verdict block in $1,269$\,s, identifying $11/12$ planted files with $1/69$ benign false positives and reporting $81/81$ distinct fixture paths.

\paragraph{Harness and timing.} Both controls use the original CSI/Claude Code prompt and native sampling, thinking enabled, a $32{,}000$-token cap, $120$ turns and four continuations. Each has a dedicated GPU, a read-only fixture and no access to scoring labels. The first paired attempt exposed CSI's interpretation of \texttt{--continue} as unbounded proxy continuation. Those traces are retained as infrastructure diagnostics; both models restart with native \texttt{-c} session resumption after a two-turn check. New times measure client exit. Grounding uses the original lexical heuristic.
\input{tex/operational_followup/files_table.tex}

\paragraph{What typed screening adds.} The preceding native mini-head sweep on these same frozen files detects $12/12$ planted indicators with $0/69$ false positives in $11.23$--$11.35$\,s, versus the generative investigation's $11/12$ and $1/69$ in $1{,}269$\,s. The outputs differ: S1 supplies typed screening decisions, whereas S2 supplies a report with evidence and remediation.

%% file: tex/operational_followup/files_table.tex
\begin{table*}[ht]
\centering\small
\caption{File investigation: quality accompanies elapsed time. Historical guided time includes the reported screen; \ctzero{} sweep time uses stalled-output recovery. Grounding counts use the original lexical heuristic, separately for all cited verdicts and IOC verdicts because report coverage differs.}
\label{tab:operational-file-controls}
\setlength{\tabcolsep}{6pt}\renewcommand{\arraystretch}{1.1}
\begin{tabularx}{\textwidth}{@{}Xrrrrr@{}}\toprule
Model / route & Time (s) & Detected & False positives & Grounded: all & Grounded: IOCs \\
\midrule
\ctzero{} S2 (historical) & 1,655 & 11/12 & 4/69 & 11/15 & 11/15 \\
\ctzero{} guided (historical) & 93 & 10/12 & 2/69 & 15/17 & 10/12 \\
Base Qwen3.5-4B (no final report) & 223 & \textendash & \textendash & \textendash & \textendash \\
alias3-mini (earlier BF16 checkpoint) & 1,269 & 11/12 & 1/69 & 79/81 & 11/12 \\
\bottomrule\end{tabularx}
\end{table*}

%% file: tex/short_evals/alerts_table.tex
\begin{table}[ht]
\centering\small
\begin{tabular}{lrrrr}
\toprule
Policy & Groups selected & Labeled recall \% & Stages found / 48 & S2 calls\\
\midrule
S1 & 12030 & 6.0 & 24/48 & 0\\
S2 short & 12030 & 4.8 & 24/48 & 181\\
S2 thinking & 12030 & 6.2 & 30/48 & 181\\
S1 cascade & 12030 & 5.7 & 24/48 & 164\\
Keyword rule & 12030 & 27.3 & 24/48 & 0\\
Count priority & 12030 & 42.4 & 25/48 & 0\\
\midrule
Base Qwen S2 & 12030 & 5.7 & 21/48 & 181 \\
\textcolor{fig_mini}{Mini S2 (BF16)} & 12030 & 7.8 & 30/48 & 181 \\
\bottomrule
\end{tabular}
\caption{Earlier \ctzero{} arms and the base/mini S2 controls. AIT-ADS at the nominal 25\% hourly review budget (ceilings yield 12,030 of 47,064 groups). Scores are reused across 181 test templates, representing 1,588,769 raw alerts. Stage coverage is an environment/stage proxy, not attack-chain reconstruction.}
\label{tab:op-alerts}
\end{table}

%% file: tex/short_evals/alert_budgets.tex
\begin{table}[t]

\centering\small
\setlength{\tabcolsep}{5pt}\renewcommand{\arraystretch}{1.1}
\begin{tabularx}{\textwidth}{@{}Xrrrrr@{}}\toprule
Budget & Model & Selected & Labeled groups & Recall (\%) & Stages \\
\midrule
10\% & Base S2 & 5,040 & 48/1621 & 3.0 & 13/48 \\
10\% & Mini S2 & 5,040 & 69/1621 & 4.3 & 24/48 \\
25\% & Base S2 & 12,030 & 93/1621 & 5.7 & 21/48 \\
25\% & Mini S2 & 12,030 & 126/1621 & 7.8 & 30/48 \\
50\% & Base S2 & 23,718 & 197/1621 & 12.2 & 36/48 \\
50\% & Mini S2 & 23,718 & 198/1621 & 12.2 & 42/48 \\
100\% & Base S2 & 47,064 & 1621/1621 & 100.0 & 48/48 \\
100\% & Mini S2 & 47,064 & 1621/1621 & 100.0 & 48/48 \\
\bottomrule\end{tabularx}
\caption{Alert controls across all recorded hourly budgets. All $181$ template outputs per model are valid. Selection counts include hourly rounding. Reviewing the entire queue reaches $100\%$ recall by construction and is not a ranking advantage.}\label{tab:short-alert-budgets}
\end{table}

%% file: tex/ct0_operational_tables/code.tex
\begin{table}[ht]
\centering\small
\begin{tabular}{lrrrrr}
\toprule
Project & API policy & Paths & Retained & Invalid & Fix-function before / after\\
\midrule
commonsio & Native CodeQL & 0 & 0 & 0 & 0 / 0\\
commonsio & S1 APIs + CodeQL & 319 & 7 & 1 & 0 / 0\\
commonsio & S2 APIs + CodeQL & 43 & 3 & 0 & 0 / 0\\
spark & Native CodeQL & 11 & 2 & 0 & 6 / 0\\
spark & S1 APIs + CodeQL & 241 & 27 & 0 & 27 / 5\\
spark & S2 APIs + CodeQL & 37 & 8 & 0 & 9 / 3\\
zip4j & Native CodeQL & 0 & 0 & 0 & 0 / 0\\
zip4j & S1 APIs + CodeQL & 303 & 8 & 0 & 63 / 6\\
zip4j & S2 APIs + CodeQL & 101 & 0 & 0 & 2 / 0\\
ztzip & Native CodeQL & 0 & 0 & 0 & 0 / 0\\
ztzip & S1 APIs + CodeQL & 512 & 18 & 0 & 34 / 8\\
ztzip & S2 APIs + CodeQL & 137 & 12 & 1 & 10 / 6\\
\bottomrule
\end{tabular}
\caption{Real CodeQL execution with schema-constrained S2 review. Fix-function overlap counts paths, not distinct vulnerabilities; fixed-version line ranges can shift. Commons IO metadata identifies only a test file, so its zero overlap is uninformative. Invalid reviews are retained as unresolved.}
\label{tab:op-code}
\end{table}

%% file: tex/ct0_operational_tables/codecost.tex
\begin{table}[ht]
\centering\small
\begin{tabular}{lrrrrr}
\toprule
Project & S1 API s & S2 API s & S1 review s & S2 review s & S1 / S2 CodeQL s\\
\midrule
commonsio & 82.4 & 170.3 & 235.1 & 24.0 & 22.6 / 20.2\\
spark & 61.7 & 157.3 & 166.3 & 29.4 & 25.4 / 19.7\\
zip4j & 68.3 & 133.7 & 243.2 & 62.3 & 21.9 / 21.2\\
ztzip & 41.7 & 97.1 & 310.5 & 63.9 & 22.2 / 21.3\\
\bottomrule
\end{tabular}
\caption{Java stage timings. API/review columns sum HTTP request wall times in concurrent quality runs; CodeQL is measured process wall time. They are not isolated GPU compute costs or whole-workflow speedups. S1 can reduce API time while increasing review work.}
\label{tab:op-codecost}
\end{table}

%% file: tex/operational_followup/code_diagnostics.tex
\begin{table}[ht]
\centering\small
\caption{CodeQL diagnostics. API validity counts original twelve-API batches; invalid path reviews are retained as unresolved. CodeQL execution time excludes lock waiting but shares host CPU resources. These are not paired end-to-end speedups.}
\setlength{\tabcolsep}{5pt}\renewcommand{\arraystretch}{1.1}
\begin{tabularx}{\textwidth}{@{}Xrrrr@{}}\toprule
Model / project & Valid API batches & Reviewed paths & Invalid reviews & CodeQL (s) \\
\midrule
Base S2 / spark & 81/81 & 37 & 0 & 20.3 \\
Base S2 / ztzip & 41/41 & 225 & 0 & 19.7 \\
Base S2 / zip4j & 59/59 & 84 & 0 & 20.4 \\
Base S2 / commonsio & 84/84 & 56 & 0 & 20.1 \\
Mini S2 / spark & 81/81 & 15 & 0 & 19.8 \\
Mini S2 / ztzip & 41/41 & 64 & 0 & 20.2 \\
Mini S2 / zip4j & 59/59 & 69 & 0 & 19.9 \\
Mini S2 / commonsio & 84/84 & 3 & 0 & 20.0 \\
\bottomrule\end{tabularx}
\end{table}

%% file: tex/ct0_operational_tables/codeoriginal.tex
\begin{table}[ht]
\centering\small
\begin{tabular}{lrrrr}
\toprule
Project & Policy & Paths & Retained & Invalid reviews\\
\midrule
commonsio & Native CodeQL & 0 & 0 & 0\\
commonsio & S1 APIs + CodeQL & 318 & 286 & 283\\
commonsio & S2 APIs + CodeQL & 76 & 71 & 71\\
spark & Native CodeQL & 11 & 5 & 5\\
spark & S1 APIs + CodeQL & 242 & 184 & 184\\
spark & S2 APIs + CodeQL & 51 & 46 & 45\\
zip4j & Native CodeQL & 0 & 0 & 0\\
zip4j & S1 APIs + CodeQL & 304 & 274 & 274\\
zip4j & S2 APIs + CodeQL & 125 & 112 & 112\\
ztzip & Native CodeQL & 0 & 0 & 0\\
ztzip & S1 APIs + CodeQL & 512 & 487 & 486\\
ztzip & S2 APIs + CodeQL & 144 & 130 & 126\\
\bottomrule
\end{tabular}
\caption{Original free-form Java review pilot. Invalid reviews remain unresolved, not confirmed vulnerabilities. The complete workflow is rerun in the separate schema-controlled experiment.}
\label{tab:op-codeoriginal}
\end{table}